\documentclass[11pt,a4paper]{article}
\usepackage{jheppub}
\usepackage[utf8]{inputenc}
\usepackage[T1]{fontenc}
\pdfoutput = 1
\usepackage[noEucal]{main}
\usepackage{aas_macros}
\usepackage{pdfsync}
\usepackage{float}
\usepackage[normalem]{ulem}
\usepackage{url}
\numberwithin{equation}{section}
\usepackage[ragged]{footmisc}
\def\far{{\text{far}}}
\def\intt{{\text{int}}}
\def\near{{\text{near}}}
\def\BPS{{\hbox{\tiny BPS}}}
\def\CLP{{\hbox{\tiny CLP}}}
\def\mix{{mix}}
\def\5{{(5)}}

\renewcommand{\tilde}{\widetilde}
\def\lsim{\mathrel{\rlap{\lower4pt\hbox{\hskip1pt$\sim$}}
    \raise1pt\hbox{$<$}}}
\def\gsim{\mathrel{\rlap{\lower4pt\hbox{\hskip1pt$\sim$}}
    \raise1pt\hbox{$>$}}}
\def\be{\begin{equation}}
\def\ee{\end{equation}}
\def\bea{\begin{eqnarray}}
\def\eea{\end{eqnarray}}

\begin{document}

\title{Charged Rotating Hairy Black Holes in $\ads_5\times S^5$: Unveiling their Secrets}

\author[a]{\'Oscar~J.C.~Dias,}
\author[b,c]{Prahar Mitra}
\author[c]{and Jorge~E.~Santos}

\affiliation[a]{STAG Research Centre and Mathematical Sciences, University of Southampton,\\ Southampton SO17 1BJ, UK}
\affiliation[b]{Institute for Theoretical Physics, University of Amsterdam,
Science Park 904,\\ Postbus 94485, 1090 GL Amsterdam, The Netherlands}
\affiliation[c]{Department of Applied Mathematics and Theoretical Physics, University of Cambridge,\\
Wilberforce Road, Cambridge, CB3 0WA, UK}

\emailAdd{ojcd1r13s@soton.ac.uk}
\emailAdd{p.mitra@uva.nl}
\emailAdd{jss55@cam.ac.uk}

\abstract{Using a mix of analytical and numerical methods, we construct new rotating, charged ``hairy'' black hole solutions of $D=5$, $\CN=8$ gauged supergravity that are dual, via the AdS/CFT correspondence, to thermal states in $D=4$, $\CN=4$ SYM at finite chemical and angular potential, thereby complementing and extending the results of \cite{Bhattacharyya:2010yg, Markeviciute:2018yal, Markeviciute:2018cqs}. These solutions uplift to asymptotically $\ads_5 \times S^5$ solutions of Type IIB supergravity with equal angular momenta along $\ads_5$ ($J=J_1=J_2$) and $S^5$ ($Q=Q_1=Q_2=Q_3$). As we lower the mass $E$ at fixed $Q$ and $J$, the known Cveti\v{c}-L\"u-Pope (CLP) black holes are unstable to scalar condensation and the hairy black holes constructed here emerge as novel solutions associated to the instability. In the region of phase space where the CLP and hairy black holes coexist, the hairy black holes dominate the microcanonical ensemble and, therefore, describe a new thermodynamic phase of SYM. The hairy black holes extend beyond the CLP extremality surface all the way to the BPS surface, defined by $E = 3 Q + 2 J / L$. Through a combination of analytical and numerical techniques, we argue that the BPS limit of the hairy black holes is a singular, horizonless solution, and \emph{not} a new two-parameter family of BPS black holes that extend the known one-parameter Gutowski-Reall (GR) black hole solution, in contradiction with the conjectures of \cite{Bhattacharyya:2010yg, Markeviciute:2018yal}. To further support our conclusions, we perform a near-horizon analysis of the BPS equations and argue that they do not admit any regular solutions with an horizon.
}

\maketitle

%%%%%%%%%%%%%%%%%%%%%%%%%
\section{Introduction}
\label{sec:intro}
%%%%%%%%%%%%%%%%%%%%%%%%%

The AdS/CFT correspondence conjectures an equivalence between quantum gravitational theories in asymptotically anti-de Sitter (AdS) spacetimes and conformal field theories (CFTs) living on the asymptotic boundary of AdS~\cite{Maldacena:1997re, Gubser:1998bc, Witten:1998qj, Aharony:1999ti}. Under this correspondence, classical stationary solutions in AdS are dual to equilibrium phases of the CFT at large central charge, a regime in which the CFT is strongly interacting. A complete understanding of the strongly coupled phases of holographic CFTs can be obtained by thoroughly investigating gravitational solutions in AdS. In this context, a particularly important class of gravitational solutions are black holes because their existence implies an exponentially large number of states in the CFT.
One of the many successes of AdS/CFT has been the microscopic counting of the states that contribute to the Bekenstein-Hawking entropy of asymptotically flat D1-D5-P black holes of Type IIB supergravity from first principles~\cite{Strominger:1996sh, Breckenridge:1996is} (i.e., by counting the D-brane states on the CFT side). A similar programme to complete the entropy microscopic counting of asymptotically AdS$_5 \times S^5$ black holes of Type IIB supergravity is ongoing~\cite{Hosseini:2018tha, Choi:2018hmj, Cabo-Bizet:2018ehj, Benini:2018ywd, Zaffaroni:2019dhb}.

In its original form~\cite{Maldacena:1997re}, the correspondence states that Type IIB superstring theory on $\ads_5\times S^5$ (at equal radii $L$) with string coupling $g_s$ and string length $\ell_s$ with $N$ units of five-form flux on $S^5$ is dual to 4-dimensional $\CN=4$ supersymmetric Yang-Mills theory (SYM) with gauge group $SU(N)$ and 't Hooft coupling $\l = g_{\text{YM}}^2 N$. The parameters on the two sides of the correspondence are related by
\begin{equation}
\begin{split}
g_s = \frac{\l}{2\pi N}  , \qquad \frac{\ell_s}{L} = \frac{1}{(2\l)^{1/4}} .
\end{split}
\end{equation}
The string theory side of this correspondence is best understood in the weakly coupled ($g_s \ll 1$) and low energy ($\ell_s/L \ll 1$) limit. In this regime, Type IIB superstring theory reduces to Type IIB supergravity, and the dual CFT becomes planar and strongly coupled ($N \to \infty$ with $\l \gg 1$). The full ten-dimensional theory -- consisting of 128 on-shell bosonic degrees of freedom and 128 fermionic superpartners -- is rather complicated to work with, and a complete classification of all its stationary solutions appears to be completely intractable. To make progress, it is necessary to consider consistent truncations that retain only a small number of fields.\footnote{A consistent truncation sets fields to zero in a way that is compatible with the equations of motion of the theory. Imposing invariance under a symmetry is a convenient way to ensure that a truncation is consistent.} A particularly nice consistent truncation of Type IIB supergravity on $\ads_5\times S^5$ is $U(1)^3$ gauged supergravity in five-dimensions~\cite{Liu:2007rv}, which retains only the metric $g_{ab}$, three Abelian gauge fields $A_a^K=\{A_a^1, A_a^2, A_a^3\}$, three charged scalars $\Phi_K=\{\Phi_1, \Phi_2, \Phi_3\}$ and two neutral scalar fields $\varphi_r = \{ \varphi_1, \varphi_2\}$ (see~\cite{Dias:2022eyq} and Section~\ref{sec:conclusion} for a brief account, including original references, of consistent truncations relevant for this system). General stationary asymptotically $\ads_5$ solutions in this theory are labelled by their energy $E$, three $U(1)$ electric charges $(Q_1, Q_2, Q_3)$ and two angular momenta $(J_1, J_2)$ along the two independent rotation planes of $\ads_5$.\footnote{These solutions uplift to solutions of the $SL(2,\mrr)$-invariant sector of ten-dimensional Type IIB supergravity, which retains the metric and self-dual five-form field~\cite{Cvetic:2000nc}. The symmetry group of asymptotically $\ads_5 \times S^5$ solutions is $SO(1,5) \times SO(6)$. $E$ is the eigenvalue of $SO(1,1) \subset SO(1,5)$, $(J_L,J_R)$ are the weight vectors of $SO(4)\subset SO(1,5)$ and $(Q_1,Q_2,Q_3)$ are weight vectors of $SO(6)$.} On the SYM side, these solutions are dual to states with $SO(4) \cong SU(2)_L \times SU(2)_R$ spin $J_L = (J_1 + J_2)/2$ and $J_R = ( J_1 - J_2 )/2$ and $R$-charge given by $SU(4)$ weight vector $(Q_1, Q_2, Q_3)$. All solutions satisfy the BPS bound,
\begin{equation}
\begin{split}
\label{BPS_Bound}
E \geq E_\BPS = |Q_1| + |Q_2| + |Q_3| + |J_1|/L + |J_2|/L \, .
\end{split}
\end{equation}
Solutions that saturate this bound are supersymmetric. Generic supersymmetric solutions depend on five independent parameters $(Q_1, Q_2, Q_3, J_1, J_2)$.

When $\Phi_K=0$, $U(1)^3$ gauged supergravity reduces to the STU model~\cite{Behrndt:1998jd} (which contains the $\CN=2$ supergravity multiplet coupled to 2 additional vector multiplets) which has been a major focus of study and exploration in the past two decades~\cite{Cvetic:2004ny, CVETIC2004273, Gutowski:2004ez, Gutowski:2004yv, Chong:2005da, Chong:2005hr, Cvetic:2005zi, Kunduri:2006ek, Chong:2006zx, MEI200764, Wu:2011gq} (see~\cite{Dias:2022eyq} for a complete review of all black hole solutions in the STU model). Of particular significance among these is the 1/16 BPS black hole constructed by Kunduri, Lucietti, and Reall (KLR) in~\cite{Kunduri:2006ek}. A rather surprising feature of this black hole is that it only depends on four independent parameters instead of the five ones one naively expects. This is because the KLR black holes satisfy a non-linear charge constraint $\Delta_{\hbox{\tiny KLR}} = 0$, where\footnote{The most general six parameter {\it non}-supersymmetric black hole is known due to Wu~\cite{Wu:2011gq}. The BPS limit $E \to E_\BPS$ of this solution results in a complex metric. A real solution can be obtained by setting the imaginary part of the metric to zero, which imposes $\Delta_{\hbox{\tiny KLR}}=0$. In other words, in the six-dimensional phase space of solutions of the STU model (i.e., $U(1)^3$ gauged supergravity with $\Phi_K=0$), the five-dimensional hypersurface of extremal ($T=0$) Wu black holes intercepts the five-dimensional BPS hypersurface along a four-dimensional hypersurface that describes the supersymmetric KLR black holes.}
\begin{equation}
\begin{split}
\label{charge_constraint}
\Delta_{\hbox{\tiny KLR}} \! &\equiv \!\left( \frac{1}{2L} \!+ \!\frac{Q_1 + Q_2 + Q_3}{N^2} \right)\!\! \left( \frac{Q_1 Q_2 + Q_2 Q_3 + Q_3 Q_1}{N^4}  - \frac{J_1 + J_2}{2N^2L^2} \right) -  \frac{Q_1 Q_2 Q_3}{N^6} - \frac{J_1 J_2}{2N^4L^3 } \, .
\end{split}
\end{equation}
The KLR black hole was found in 2006, and until recently, the physical interpretation of this charge constraint has remained a mystery. Indeed, a generic 1/16 BPS state in SYM is characterised by five fugacities, so any attempt to count the microstates of the KLR black holes would fail due to the mismatch in the number of parameters on both sides. However, it has recently been shown that the microstates of the KLR black hole can be counted by evaluating a twisted Witten index in SYM~\cite{Hosseini:2018tha, Choi:2018hmj, Cabo-Bizet:2018ehj, Benini:2018ywd, Zaffaroni:2019dhb}. The five chemical potentials that enter the index (one for each of the two angular momenta and three $R$-charges) must satisfy a linear constraint to preserve supersymmetry so that the most general index depends only on four parameters. This linear constraint on the chemical potentials is equivalent to the non-linear constraint \eqref{charge_constraint} on the charges.

Given the results above, it is important to ask the following question: \emph{Do supersymmetric black holes with $\Delta_{\hbox{\tiny KLR}} \neq 0$ exist?} An important hint in this direction arises from the results of~\cite{Bhattacharyya:2010yg}, which studied a restricted class of solutions with $Q_1 = Q_2 = Q_3\equiv Q$ and $J_1 = J_2\equiv J$ (In this case, the STU model \cite{Behrndt:1998jd} reduces to minimal gauged supergravity~\cite{Gunaydin:1983bi}). With this restriction, the KLR black hole reduces to the Gutowski-Reall (GR) black hole~\cite{Gutowski:2004ez}, a one-parameter family of supersymmetric black hole solutions. Using both numerical and perturbative techniques, the authors of~\cite{Bhattacharyya:2010yg} constructed a static ($J=0$), regular, supersymmetric {\it hairy} soliton solution which has $\Phi_1=\Phi_2=\Phi_3\equiv \Phi \neq 0$.\footnote{Here, a `soliton' refers to a solution without a horizon, i.e. it has an entropy of $\CO(N^0)$ entropy at large $N$, and `hairy' means the solution has non-vanishing $\Phi$.} Then, using a non-interacting thermodynamic model,\footnote{Here, the hairy black hole is approximated by a non-interacting equilibrium mix of the supersymmetric GR black hole~\cite{Gutowski:2004ez} and the regular supersymmetric soliton that maximises the entropy of the system.} the authors conjectured the existence of a new supersymmetric {\it hairy} black hole (so with $\Phi \neq 0$) that has finite entropy, is everywhere regular, and does not satisfy the charge constraint \eqref{charge_constraint}.

Motivated by the conjecture of~\cite{Bhattacharyya:2010yg},~\cite{Markeviciute:2018yal, Markeviciute:2018cqs} numerically constructed non-supersymmetric hairy black holes (thus establishing the existence of these solutions) and found solutions with temperatures as low as $TL\sim \CO(10^{-3})$.\footnote{The analysis done in~\cite{Markeviciute:2018yal, Markeviciute:2018cqs} is strictly valid for $T\neq 0$. While the $T\to 0$ limit can be taken, the computational cost gets increasingly higher the smaller $T L$ gets.\label{foot:BPSlimit}} With the numerical data that was collected, the analysis of~\cite{Markeviciute:2018yal, Markeviciute:2018cqs} suggests that in the extremal $T\to 0$ limit, the chemical potentials and angular velocity both limit to 1 ($\mu \to 1^-$ and $\O_H L\to 1^+$) and the entropy remains finite. This was interpreted as evidence that the extremal limit of the hairy black holes of~\cite{Markeviciute:2018yal, Markeviciute:2018cqs} is the conjectured supersymmetric hairy black hole of ~\cite{Bhattacharyya:2010yg}.

In this paper, we revisit the results of~\cite{Bhattacharyya:2010yg, Markeviciute:2018yal, Markeviciute:2018cqs} and supplement them with three new analyses: two analytic calculations and an improved numerical analysis. First, employing methods similar to the ones used in~\cite{Basu:2010uz, Bhattacharyya:2010yg, Dias:2011tj, Dias:2016pma, Dias:2011at, Stotyn:2011ns, Dias:2013sdc, Cardoso:2013pza, Dias:2015rxy, Dias:2022eyq}, we construct the non-supersymmetric hairy black hole solution in a double perturbative expansion in the horizon radius and scalar condensate amplitude (equivalently, in $Q$ and $J$) and then study the BPS limit of these solutions.\footnote{We will find that the perturbative construction is valid in the expected small $E, Q, J$ regime but, surprisingly, only if  $T L$ is not too small. Indeed, the perturbative expansion breaks down when $T L= \CO(Q L/N^2)$.} Secondly, we study the near-horizon structure of the supersymmetric black hole by directly analysing the BPS equations. Finally, we use numerical methods to lower the temperature down to $TL \sim \CO(10^{-7})$ which is, say, four orders of magnitude lower than the typical temperature reached in~\cite{Markeviciute:2018yal, Markeviciute:2018cqs}. With these new pieces of evidence, we will conclude that the scenario conjectured in~\cite{Bhattacharyya:2010yg} for the supersymmetric hairy black hole does {\it not} occur. Namely, in $U(1)^3$ gauged supergravity with three equal charged scalar fields and equal angular momenta (and for the sector of symmetries considered in~\cite{Bhattacharyya:2010yg, Markeviciute:2018yal, Markeviciute:2018cqs}), the supersymmetric limit of the hairy black holes is {\it singular}, in the sense that the entropy vanishes and the scalar field and curvature invariants at the horizon diverge. This also means that the low-temperature numerical data collected in~\cite{Markeviciute:2018yal, Markeviciute:2018cqs} is misleading -- one needs to collect data for temperatures that are several orders of magnitude lower to evaluate the BPS limit of the hairy black holes properly.

The remainder of this paper is organised as follows. Section~\ref{sec:setup} discusses $U(1)^3$ gauge supergravity and the class of solutions we are interested in. In Section~\ref{sec:CLP_bh}, we start by reviewing known solutions within this class, namely the Cveti\v{c}-L\"u-Pope (CLP) black hole without scalar hair (subsection~\ref{sec:CLP_bh_subsec}). We further show that some CLP black holes have a scalar condensation instability (subsection~\ref{sec:Onset3Q}). Finally, we use a non-interacting thermodynamic model (that essentially boils down to the one in~\cite{Bhattacharyya:2010yg}) to find evidence for the existence of hairy black holes in the theory (subsection~\ref{sec:toymodel}). In Section~\ref{sec:hbh-pert}, we use perturbation theory with a matching asymptotic expansion (that is valid for small $E$, $Q$, and $J$) to construct hairy black holes (details of the construction are relegated to Appendix~\ref{app:perturbative_eq}). These hairy black holes of the  $U(1)^3$ theory merge, in a phase diagram of solutions, with the CLP family at the onset of the scalar condensation instability of the latter. In Section~\ref{sec:numerics}, we numerically find the exact hairy black hole solutions of $ U(1)^3$ gauged supergravity.
In Section~\ref{sec:MainResults}, we present and discuss our physical results (subsection~\ref{sec:NumericalResults}). We are particularly interested in the supersymmetric limit of these solutions.  Comparing the numerical results with the perturbative ones, we find that the regime of validity of the latter must be revisited to discover that it breaks down at very low temperatures close to the BPS limit (subsection~\ref{sec:PerturbativeFailing}).
 In Section~\ref{sec:susy-analysis}, we perform a near-horizon analysis of the BPS equations to find directly the regular supersymmetric black holes of the theory. Finally, in Section~\ref{sec:conclusion}, we summarise and discuss further our results, make some conjectures and comment on future directions. The reader that wants to skip the strategy and technical details employed to build up our understanding of the system can skip all other sections and jump immediately to Section~\ref{sec:conclusion} (eventually complemented with a reading of Section~\ref{sec:MainResults}, which contains detailed results and figures).

%%%%%%%%%%%%%%%%%%%%%%%%%%%%%%%%%%%%%%%%%%%%%%%%%%
\section{Setup of the problem}
\label{sec:setup}
%%%%%%%%%%%%%%%%%%%%%%%%%%%%%%%%%%%%%%%%%%%%%%%%%%

%%%%%%%%%%%%%%%%%%%%%%%%%
\subsection{Action and equations of motion}
\label{sec:action}
%%%%%%%%%%%%%%%%%%%%%%%%%

Type IIB supergravity in ten dimensions can be consistently truncated to $U(1)^3$ gauged supergravity in five dimensions~\cite{Liu:2007rv} (see~\cite{Dias:2022eyq} for details on how this truncation is obtained). The bosonic fields of the latter include three $U(1)$ gauge fields $A_a^K$ ($K=1,2,3$),  three charged scalars $\Phi_K$, and two neutral scalar fields which are packaged into three scalars $X_K$ constrained by the relation $X_1 X_2 X_3 = 1$. The bosonic part of the action is 
\begin{equation}
\begin{split}\label{OurCTaction}
S &= \frac{1}{16 \pi G_5} \int \dt^5 x \sqrt{-g} \left[ R - V
- \sum_{K=1}^3 \frac{1}{X_K^2} \left( \frac{1}{2} (\nabla X_K)^2
 + \frac{1}{4}  \left(F^K\right)^2 \right)  \right. \\
&\left. \qquad \qquad \qquad \qquad - \frac{1}{8} \sum_{K=1}^{3} \left( |D \Phi_K|^2 -  \frac{(\nabla \l_K)^2}{4(4+\l_K)} \right) \right] - \frac{1}{16 \pi G_5} \int  F^1 \wedge F^2  \wedge A^3 \, , 
\end{split}
\end{equation}
where we have defined (with no Einstein summation convention over $K=1,2,3$)
\begin{equation}\label{OurCTaction2}
\begin{split}
D_a\Phi_K \equiv \partial_a\Phi_K - i\, \frac{2}{L} A^K_a \Phi_K , \qquad F^K_{a b} \equiv \partial_a A^K_b - \partial_b A^K_a , \qquad \l_K \equiv \Phi^\dagger_K \Phi_K\,,
\end{split}
\end{equation}
where $L$ is the AdS radius. The scalar potential $V$ is given by
\begin{equation}\label{OurCTaction3}
\begin{split}
V &=\frac{1}{2L^2}\bigg[ \sum_{K=1}^3 X_K^2 \l_K  - 2 \prod_{K=1}^3 \sqrt{4+\l_K} \sum_{K=1}^3 \frac{1}{X_K \sqrt{4+\l_K} } \bigg].
\end{split}
\end{equation}
The three charged scalars saturate the Breitenlohner-Freedman (BF) bound~\cite{Breitenlohner:1982jf, Breitenlohner:1982bm} and have charge $q_K L = \D_K = 2$ where $\D_K$ is the dimension of the CFT operator dual to $\Phi_K$. As demonstrated in~\cite{Cardoso:2004hs, Basu:2010uz, Dias:2011tj} (within Einstein-AdS gravity), small near-extremal charged black holes are unstable to condensation of a charged scalar field whenever $q L > \D$ and there are novel hairy black holes (i.e. with charged scalar hair) associated to this instability~\cite{Basu:2010uz, Dias:2011tj}. Since the charged scalar in \eqref{OurCTaction} saturates the instability bound, whether or not they condense requires a more detailed study. For static configurations in the theory \eqref{OurCTaction}, this analysis was done in~\cite{Bhattacharyya:2010yg, Markeviciute:2016ivy, Dias:2022eyq}, where it was shown that small near-extremal charged black holes are indeed unstable to scalar condensation and can decay to a static hairy black hole in the microcanonical ensemble (because the entropy of the latter is higher for given energy and charge of the system).

In this paper, we will generalise the work of~\cite{Bhattacharyya:2010yg, Markeviciute:2016ivy, Dias:2022eyq} to stationary (rotating) configurations with $J_1=J_2=J$, complementing the results of~\cite{Markeviciute:2018yal, Markeviciute:2018cqs}. We restrict ourselves to the truncation studied in~\cite{Bhattacharyya:2010yg, Markeviciute:2016ivy,Markeviciute:2018yal, Markeviciute:2018cqs}, which sets
\begin{equation}
\begin{split}
\label{truncation}
A_a^K = A_a , \qquad \Phi_K = \Phi , \qquad X_K = 1 \qquad (K=1,2,3). 
\end{split}
\end{equation}
This can also be obtained by restricting to a $S_3$-invariant field configurations as shown in~\cite{Bhattacharyya:2010yg}. Stationary generalisations of~\cite{Dias:2022eyq} for truncations with different charges are left for future publications (see~\cite{Dias:2022eyq} for static studies in sectors with different charges). The bosonic part of the action defined by \eqref{truncation} is
\begin{equation}
\begin{split}\label{action}
S_{bulk} &= \frac{1}{16\pi G_5} \int \dt^5 x \sqrt{-g} \bigg[ R + \frac{12}{L^2} - \frac{3}{4} F^2  - \frac{3}{8} \bigg( | D \Phi |^2 -  \frac{4}{L^2} \l   - \frac{ ( \n \l )^2  }{4( 4 +  \l )} \bigg) \bigg]  \\
&\qquad \qquad \qquad \qquad  - \frac{1}{16 \pi G_5} \int  F \wedge F  \wedge A .
\end{split}
\end{equation}
The equations of motion of the theory are
\begin{subequations}
\label{EoM}
\begin{equation}
\begin{split}
& R_{ab} - \frac{1}{2} g_{ab} R - \frac{6}{L^2}  g_{ab}  =  \frac{3}{2} T_{ab}^A  + \frac{3}{8}  T_{ab}^\Phi ,  \\
& \n_b F^{ab} - \frac{1}{2}  ( \star F )^{abc} F_{bc}  = \frac{1}{2L} \text{Im} ( \Phi^\dagger D_a \Phi ), \\
& D^2 \Phi + \frac{4}{L^2} \Phi =  \bigg( \frac{ \n^2 \l }{2(4 + \l )} -  \frac{ ( \n \l )^2 }{4( 4 + \l )^2}  \bigg) \Phi , \\
\end{split}
\end{equation}
where the energy-momentum tensor contributions are:
\begin{equation}
\begin{split}
& T_{ab}^A \equiv F_{ac} F_b{}^c - \frac{1}{4} g_{ab} F^{cd} F_{cd} ,  \\
& T_{ab}^\Phi \equiv D_{(a} \Phi^\dagger D_{b)} \Phi  - \frac{1}{2} g_{ab} \bigg( | D \Phi |^2   - \frac{4}{L^2} \l \bigg) - \frac{1}{4(4 + \l )} \bigg( \n_a \l \n_b \l - \frac{1}{2} g_{ab} ( \n \l )^2 \bigg) .
\end{split}
\end{equation}
\end{subequations}

%%%%%%%%%%%%%%%%%%%%%%%%%
\subsection{Solution ansatz\"e, boundary conditions and thermodynamic quantities}
\label{sec:ansatz}
%%%%%%%%%%%%%%%%%%%%%%%%%

We want to find asymptotically AdS$_5$ rotating black hole solutions of \eqref{action} with spherical horizon topology and a charged scalar field $\Phi$ coupled to the gauge potential $A$. For simplicity, we consider cohomogeneity-1 solutions, for which the angular momenta along the two rotation axes of the $S^3$ are the same, $J_1=J_2=J$. Such solutions and some of their key properties were already studied in~\cite{Bhattacharyya:2010yg, Markeviciute:2016ivy, Markeviciute:2018yal}, and we wish to revisit this problem in this paper to get solutions with temperatures even closer to zero (this is fundamental to address the main question of this paper). To find such solutions, we use the ansatz\"e
\begin{subequations}\label{3Q:ansatz}
\begin{equation}
\begin{split}
\dt s^2 &= - f(r) \dt t^2 + \frac{g(r)}{f(r)}  \dt r^2+ r^2 \bigg[ h(r) \bigg( \dt \psi + \frac{1}{2} \cos \t \dt \phi  - w(r) \dt t \bigg)^2  + \frac{1}{4} \dt \O_2^2 \bigg] , \\
A &= A_t (r) \dt t + A_\psi(r) \bigg( \dt \psi + \frac{1}{2} \cos \t \dt \phi \bigg) , \qquad \Phi = \Phi^\dagger = \Phi(r) , 
\end{split}
\end{equation}
where
\begin{equation}
\begin{split}
\dt \O_2^2 = \dt \t^2 + \sin^2 \t \dt \phi^2 , \qquad \t \in 
(0,\pi) , \qquad \phi,\psi \in [0,2\pi)\,.
\end{split}
\end{equation}
\end{subequations}
Surfaces of constant time $t$ and radial $r$ coordinates have the geometry of a homogeneously squashed $S^3$, written as an $S^1$ Hopf fibred over $S^2$. The coordinate $\psi$ parameterises the fibre, while $\t,\phi$ are the standard spherical coordinates on $S^2$. The two orthogonal planes of rotation of $S^3$ are at $\t=0$ and $\t=\pi$. Note that $\phi$ also has period $2\pi$, but the Hopf fibration requires that the coordinate that parametrises the $S^1$ fibre is such that  $\psi \to \psi + \pi$ when $\phi \to \phi + 2\pi$. Note also that we have used $U(1)$ gauge freedom to eliminate the phase of the scalar field and thus work with a real charged scalar field, $\Phi =\Phi^\dagger$.

Inserting the ansatz \eqref{3Q:ansatz} into the equations of motion \eqref{EoM}, we find seven coupled nonlinear differential equations -- two first-order ODEs for $\{f,g\}$ and five second-order ODEs for $\{h, w, A_t, A_\psi, \Phi\}$. Thus, the solution is determined by twelve integration constants at each of the two boundaries of the system, some of which are fixed by an appropriate physical choice of boundary conditions, and the others are obtained once the solution to the boundary value problem is found, and are related to the physical observables.

At the conformal boundary (i.e., as $r\to\infty$), we require that the solution is asymptotically globally $\ads_5$ and that $\Phi$ and $A_{\psi}$ are not sourced, which implies that the functions appearing in \eqref{3Q:ansatz} have the following asymptotic expansion
\begin{equation}
\begin{split}\label{asymp_exp}
f|_{r\to\infty} &= \frac{r^2}{L^2}  + 1 + \frac{L^2}{r^2} C_f + \CO \bigg( \frac{L^4}{r^4} \bigg) , \qquad \qquad ~~\! h|_{r\to\infty}= 1 + \frac{L^4}{r^4} C_h +  \CO \bigg( \frac{L^6}{r^6} \bigg)    , \\
w|_{r\to\infty} &= \frac{1}{L} \bigg[ w_\infty  + \frac{L^4}{r^4} (-2j) +  \CO \bigg( \frac{L^6}{r^6} \bigg)  \bigg]  , \qquad \quad\!\! g|_{r\to\infty} = 1 +  \CO \bigg( \frac{L^4}{r^4} \bigg) , \\
A_t|_{r\to\infty} &= \mu - \frac{L^2}{r^2} ( e + 2 w_\infty \b )  +  \CO \bigg( \frac{L^4}{r^4} \bigg) , \qquad\! A_\psi|_{r\to\infty} = L \bigg[ \frac{L^2}{r^2}  ( 2\b )  +  \CO \bigg( \frac{L^6}{r^6} \bigg)  \bigg]  , \\
\Phi|_{r\to\infty} &= \frac{L^2}{r^2} \e  +  \CO \bigg( \frac{L^4}{r^4} \bigg)\,,
\end{split}
\end{equation}
where $C_f$, $C_h$, $w_{\infty}$, $j$, $e$, $\beta$ and $\epsilon$ are integration constants. With these boundary conditions, we find that the conformal metric on the boundary is that of $\mrr_t \times S^3$,
\begin{equation}
\begin{split}
\dt s^2 |_\CB = - \dt t^2 + L^2 \bigg[ \bigg( \dt \psi + \frac{1}{2} \cos \t \dt \phi  - \frac{w_\infty}{L} \dt t \bigg)^2  + \frac{1}{4} \dt \O_2^2 \bigg].
\end{split}
\end{equation}
Note that the time and angular coordinates on the Einstein cylinder are not $(t,\psi)$ but $(t',\psi')$ where 
\begin{equation}
\begin{split}
\psi' = \psi - \frac{w_\infty}{L} t , \qquad t' = t \, . 
\end{split}
\end{equation}
In the $(t',\t,\phi,\psi')$ coordinates, the asymptotic boundary is described by the Einstein Static Universe (thermodynamic quantities are necessarily computed in this non-rotating frame~\cite{Caldarelli:1999xj, Gibbons_2005}). Note also that in \eqref{asymp_exp}, we have required that the radius of $S^3$ is asymptotically $r$ by choosing the gauge $h|_{r\to\infty}=1$. The equations of motion then completely fix the asymptotic decay of $g$ as displayed in \eqref{asymp_exp}.

Under the AdS/CFT dictionary, $\mu$ sources the operator dual to $A_{t'}$ (as mentioned before, the source for $A_{\psi'}$ has been set to zero). $e$ and $\b$ are then the expectation values (VEVs) of the operators dual to $A_{t'}$ and $A_{\psi'}$, respectively:
\begin{equation}\label{VEVmaxwell}
\begin{split}
\avg{J_{t'}} = - \frac{N^2 e}{4 \pi^2 L^4} , \qquad \avg{J_{\psi'} } = \frac{N^2 \b}{2 \pi^2 L^3 } , 
\end{split}
\end{equation}
where we used the AdS/CFT dictionary to write $G_5 = \frac{\pi L^3}{2N^2}$. 

The scalar field $\Phi$ has mass $m^2=-2/L^2$ which saturates the BF bound in $\ads_5$~\cite{Breitenlohner:1982bm,Breitenlohner:1982jf}. Therefore, asymptotically it decays as $\Phi|_{r\to\infty}\sim s_{\Phi}\frac{L^2}{r^2}\,\ln r +\frac{L^2}{r^2}\,\e +\cdots$ where $s_{\Phi}$ is the source for the operator dual to $\Phi$ (which has dimension $\D=2$) and the associated VEV is proportional to $\e$. We are interested in solutions dual to CFT states that are not sourced so in \eqref{asymp_exp} we have already set the Dirichlet boundary condition $s_{\Phi}=0$ and the VEV dual to $\Phi$ is
\begin{equation}\label{VEVscalar}
\begin{split}
\avg{\CO_\Phi} = \frac{N^2\e}{\pi^2 L^2}.
\end{split}
\end{equation}
In summary, in the asymptotic analysis of our system, we started with 12 UV parameters. After imposing Dirichlet boundary conditions that fix the sources of all fields except $A_{t'}$ and $w$, we are left with eight free UV parameters: the chemical potential source $\mu$, the angular velocity at the boundary $w_\infty$ and six parameters $\{ C_f, C_h, j, q, \b, \e \}$ that are related to physical observables of the system (mass, angular momentum, charge, VEVs) that are determined only after we obtain the solution of the ODE system that obeys not only the UV but also the IR boundary conditions.

Let us now discuss the boundary conditions at the inner boundary. We are interested in black hole solutions, for which this inner boundary is the event horizon. We introduce the horizon in our system by requiring it to be the locus of $f(r_+) = 0$. Again, \`a priori, the number of free IR parameters is given by the order (i.e. 12) of the ODE system. However, some of these are fixed, imposing boundary conditions and requiring regularity of the solution at $r=r_+$. Our coupled ODE system can effectively be rewritten as a system of 6 second-order ODEs to find the regularity conditions. The horizon is a regular singular point with degeneracy 2 (i.e., the indicial root is 2), which means that the six functions have independent pair solutions where one of them is proportional to $\ln(r-r_+)$ and the other is a regular power law of $(r-r_+)$. Regularity at the horizon requires that we discard the logarithmic terms. We are left with six free IR parameters. Since we have a (non-extremal) horizon at $r=r_+$, $f$ and $g$ must vanish linearly and quadratically at the horizon. Moreover, we work in the $U(1)$ gauge where $\Phi$ is real and $A_t + w A_\psi$ vanishes linearly at $r=r_+$. After imposing these regularity conditions, we are left with 6 IR parameters plus the horizon parameter $r_+$.

Our field ansatz\"e~\eqref{3Q:ansatz} still has a residual coordinate freedom $\psi \to \psi+\a \, t$ under which
\begin{equation}\label{residualGaugeFreed}
\begin{split}
w(r) \to w(r) - \a , \qquad A_t(r) \to A_t(r) + \a \, A_\psi(r)  
\end{split}
\end{equation}
and the line element is left invariant.
We can use this residual gauge symmetry to fix the constant part of $w(r)$ to any value we wish. In the numerical analysis of Section~\ref{sec:numerics}, we will use this freedom to fix the asymptotic value of this function as $w_\infty=0$. However, for the analytical analysis of Sections~\ref{sec:hbh-pert} and~\ref{sec:susy-analysis}, it will be more convenient to use it to set the horizon value of this function to be $w|_{r=r_+} = 0$.

Altogether, we have eight free UV parameters, namely $\{ C_f, C_h, j, e, \b, \e, \mu, w_\infty \}$ and we have 6 IR parameters (the coefficients of the regular terms discussed above), i.e. a total of 8+6=14 free parameters for an ODE system of order 12 (2 first-order EoM plus 5 second order EoM). Therefore, black hole solutions depend on $2$ parameters plus the dimensionless horizon radius $y_+$ (which defines the location of the inner boundary of the problem), \emph{i.e.} a total of 3 parameters. These three parameters are related to the mass $E$, $U(1)$ charge $Q$ ($=Q_1=Q_2=Q_3$), and angular momentum $J$ ($=J_1=J_2$) of the black hole. These charges can be determined via holographic renormalisation~\cite{Bianchi:2001de, Bianchi:2001kw} which, for this particular system, is reviewed in Appendix A of~\cite{Dias:2022eyq} and was already employed to get the VEVs \eqref{VEVmaxwell}--\eqref{VEVscalar}. In terms of the expansion coefficients appearing in \eqref{asymp_exp}, these are\footnote{$M$ and $J$ are the conserved charges corresponding to $\p_{t'} = \p_t + \frac{1}{L} w_\infty \p_\psi$ and $\p_{\psi'} = \p_\psi$, respectively. $E$ is the mass of the system above that of empty AdS, $E = M-\frac{3N^2}{16L}$ (i.e., after subtracting the Casimir energy).}
\begin{equation}
\begin{split}
\label{ansatz_charges}
E = \frac{N^2}{4L} ( C_h - 3 C_f ) , \qquad J = N^2 j , \qquad Q = \frac{N^2}{L} \frac{e}{2}  . 
\end{split}
\end{equation}
The BPS bound \eqref{BPS_Bound} now takes the form
\begin{equation}
\begin{split}
\label{BPS_bound_1}
E \geq E_\BPS  \equiv 3 Q + 2 J / L . 
\end{split}
\end{equation}
The parameter $\mu$ appearing in \eqref{asymp_exp} is the chemical potential of the black hole. The Hawking temperature $T$, Bekenstein-Hawking entropy $S$, and angular velocity $\O_H$ of the horizon
\begin{equation}
\begin{split}
\label{ansatz_potentials}
T &= \frac{f'(r_+)}{4\pi\sqrt{g(r_+)}}  , \qquad S = \frac{N^2}{L^3} \pi r_+^3 \sqrt{h(r_+)} , \qquad  \O_H = \frac{1}{L} w_\infty - w(r_+) . 
\end{split}
\end{equation}
The charges \eqref{ansatz_charges} and thermodynamic quantities \eqref{ansatz_potentials} satisfy the first law of black hole mechanics
\begin{equation}
\begin{split}
\label{firstlaw}
\delta E = T \delta S + 3 \mu \delta Q + 2 \O_H \delta J. 
\end{split}
\end{equation}
The BPS limit is obtained by taking $E \to E_\BPS $. In this limit, regular solutions have $T \to 0^+$, $\mu \to 1^+$ and $L\O_H \to 1^-$.

%%%%%%%%%%%%%%%%%%%%%%%%%
\subsection{Supersymmetric static soliton}
\label{sec:soliton} 
%%%%%%%%%%%%%%%%%%%%%%%%%

The theory \eqref{action} admits a static, horizonless, regular supersymmetric solution, which we refer to as the \emph{soliton}. This solution was constructed in~\cite{Bhattacharyya:2010yg} and reproduced in~\cite{Markeviciute:2016ivy}.
For parametrically small values of $(E, Q)$, the solution is known analytically in a perturbative expansion in $\e$ (defined in \eqref{asymp_exp})~\cite{Bhattacharyya:2010yg}. The mass, charge, and angular momentum of such a `small' soliton is given by~\cite{Bhattacharyya:2010yg}
\begin{equation}
\begin{split}
\label{EQ_soliton}
\frac{E_sL}{N^2} = \frac{3Q_sL}{N^2} = \frac{3\e^2}{16} + \frac{\e^4}{256} + \frac{\e^6}{2560} + \frac{169\e^8}{2949120} + \frac{221\e^{10}}{22020096} + \CO(\e^{12}) , \qquad J_s = 0  \,.
\end{split}
\end{equation}
The soliton saturates the BPS bound \eqref{BPS_bound_1} and is, therefore, supersymmetric. It can also be checked that the chemical potential of the solution is $\mu_s = 1$, as expected. We refer the reader to~\cite{Bhattacharyya:2010yg} for the exact form of the metric, gauge, and scalar fields for the soliton and the numerical results (see also~\cite{Markeviciute:2016ivy}). The perturbative solution for the soliton to $\CO(\e^{15})$ can also be found in the accompanying {\tt Mathematica} file.

For large values of $(E_s, Q_s)$,~\cite{Bhattacharyya:2010yg, Markeviciute:2016ivy} constructed the soliton numerically. A rather surprising feature of this system is that the regular supersymmetric soliton does not exist for all values of $Q_s$. Rather, the one-parameter family of regular solitonic solutions, described perturbatively by \eqref{EQ_soliton}, ends at a critical value of the charge $Q_c/N^2 \approx 0.2613$, where the soliton becomes singular, and is of a special type as described in~\cite{Bhattacharyya:2010yg}.  Another one-parameter family of singular solitons branches off from this point and exists for all $Q_s \geq Q_c$. Interestingly, the regular and singular solitons that branch out of the special solution at $Q_s=Q_c$ exhibit an infinite set of damped self-similar oscillations around $Q_c$. Due to this spiral behaviour near $Q$, the regular supersymmetric solitons actually exist up to a maximum charge $Q_m/N^2 \approx 0.2643$.

%%%%%%%%%%%%%%%%%%%%%%%%%
\section{Cveti\v{c}-L\"u-Pope black hole}
\label{sec:CLP_bh}
%%%%%%%%%%%%%%%%%%%%%%%%%

In this section, we start by reviewing the known `bald' black hole solution of the theory \eqref{action}, which has $\Phi=0$ (subsection~\ref{sec:CLP_bh_subsec}). Then, we find a subset of bald black holes that are unstable to the condensation of the charged scalar field $\Phi$, and we find the onset curve of this instability (subsection~\ref{sec:Onset3Q}). Finally, we use a simple thermodynamic model~\cite{Bhattacharyya:2010yg} to find the leading order thermodynamics of the hairy black holes of the theory  \eqref{action} (subsection~\ref{sec:toymodel}). In layperson's terms, this model builds the hairy black hole, placing a small rotating bald black hole on top of the static supersymmetric soliton of Section~\ref{sec:soliton}.
%%%%%%%%%%%%%%%%%%%%%%%%%
\subsection{Cveti\v{c}-L\"u-Pope black hole}
\label{sec:CLP_bh_subsec}
%%%%%%%%%%%%%%%%%%%%%%%%%

The system \eqref{action} admits a black hole solution with vanishing scalar field $\Phi$ with $E$, $Q_{1,2,3}\equiv Q$, and $J_{1,2}\equiv J$ first constructed by Cveti\v{c}, L\"u and Pope in~\cite{CVETIC2004273}, which we refer to as the \emph{CLP black hole} or `bald' black hole (since $\Phi=0$). When $J=0$, this is the Behrndt-Cveti\v{c}-Sabra solution with $Q_{1,2,3}\equiv Q$ ~\cite{Behrndt_1999}; see also~\cite{Dias:2022eyq}) The CLP black hole is described by \eqref{3Q:ansatz} with
\begin{equation}
\begin{split}
\label{CLP_sol} 
&h_\CLP(r) = 1 +  \frac{\a^2 L^4}{r^4} \left( m   - \frac{e^2 L^2}{r^2} \right) , \qquad g_\CLP(r) = \frac{1}{h_\CLP(r)} , \\
&f_\CLP(r) =  \frac{r^2}{L^2} + 1 + \frac{1}{h_\CLP(r)} \left[ \frac{(2e - m) L^2}{r^2} +  \frac{e^2 L^4}{r^4} + \frac{ \a^2 e^2 L^6}{r^6}  \right]  , \\
&w_\CLP(r) = \frac{\a}{h_\CLP(r)} \left[ \frac{(e-m)L^3}{r^4} + \frac{e^2 L^5}{r^6}  \right]  -  \frac{\a}{h_\CLP(r_+)} \left[ \frac{(e-m)L^3}{r_+^4} + \frac{e^2 L^5}{r_+^6}  \right]  , \\
&A_t^\CLP(r) =  \frac{1}{h_\CLP(r_+)} \left( \frac{e L^2}{r_+^2} + \frac{\a^2 e^2 L^6}{r_+^6 }  \right) \left( 1  - \frac{r_+^2}{r^2} \right) , \\
&A_\psi^\CLP(r) = - \frac{\a eL^3}{r^2} , \qquad \Phi_\CLP(r) = 0 , \\
\end{split}
\end{equation}
The black hole horizon is at $r=r_+$, where $f(r_+) = 0$. We can use this latter condition to write the parameter $m$ in terms of $r_+$ as
\begin{equation}
\begin{split}
\label{m_def}
m = \frac{y_+^6  + ( y_+^2 + e )^2   - \a^2 e^2}{y_+^2 - \a^2 ( 1 + y_+^2 ) } , \qquad y_+ \equiv \frac{r_+}{L}  . 
\end{split}
\end{equation}
Using \eqref{ansatz_charges}, we find that the mass, charge, and angular momentum of the black hole are given by
\begin{equation}
\begin{split}
\label{BH_charges}
E_\CLP & =  \frac{N^2}{4L}  [ m ( \a^2 + 3  ) - 6 e  ]  , \qquad Q_\CLP = \frac{N^2 e}{2L} , \qquad J_\CLP = \frac{N^2}{2} \a ( m - e )  , 
\end{split}
\end{equation}
and using \eqref{ansatz_potentials}, we find the temperature, entropy, angular velocity, and chemical potential to be
\begin{equation}
\begin{split}\label{BH_thermo}
T_\CLP &= \frac{ 2 \a^2 [ e(e-1) - y_+^2 ( 1 + y_+^2 )^2 ] - ( 1 + \a^4 ) e^2 + y_+^4 + 2 y_+^6 }{2 \pi L ( e \a^2  + y_+^4 ) [ y_+^2 - \a^2 ( 1 + y_+^2 ) ]^{1/2}  } , \\
S_\CLP &= \frac{ N^2 \pi ( \a^2 e + y_+^4 ) }{  [ y_+^2 - \a^2 ( 1 + y_+^2 ) ]^{1/2} } , \\
\O_H^\CLP &= \frac{\a ( e + y_+^2 + y_+^4 ) }{ L ( e \a^2   + y_+^4 ) } , \\
\mu_\CLP &= \frac{e [ y_+^2 - \a^2 ( 1 + y_+^2 ) ] }{ e \a^2 + y_+^4 }  .
\end{split}
\end{equation}
It can be easily verified that the quantities in \eqref{BH_charges} and \eqref{BH_thermo} satisfy the first law \eqref{firstlaw} and the BPS bound \eqref{BPS_bound_1}. Requiring that $T_\CLP \geq 0$, we find that the black hole exists in the range
\begin{equation}
\begin{split}\label{blackholerange}
\left| e - \frac{\a^2}{(1-\a^2)^2} \right| \leq \frac{ [ y_+^2 - \a^2 ( 1 + y_+^2 ) ] [ 1 + 2 y_+^2 ( 1 - \a^2 ) ]^{1/2}  }{ ( 1 - \a^2 )^2 } , \qquad \a^2 < \frac{y_+^2}{1 + y_+^2 }  .
\end{split}
\end{equation}
The extremal black hole solution (i.e., with $T_\CLP=0$) saturates the first inequality at 
\begin{equation}
\begin{split}
\label{CLP_ext}
e_{ext} = \frac{ [ y_+^2 - \a^2 ( 1 + y_+^2 ) ] [ 1 + 2 y_+^2 ( 1 - \a^2 ) ]^{1/2}  - \a^2 }{ ( 1 - \a^2 )^2 } .
\end{split}
\end{equation}
The extremal CLP black hole is supersymmetric only in the limit
\begin{equation}
\begin{split}\label{extCLP}
e=e_{ext} \to e_\BPS = \frac{y_+^2}{2} ( 2 + y_+^2 )   , \qquad \a \to \a_\BPS = \frac{y_+^2}{2+y_+^2},
\end{split}
\end{equation}
in which case one recovers the familiar Gutowski-Reall (GR) black hole~\cite{Gutowski:2004ez}. Note that while the CLP black hole depends on three independent parameters $(y_+,\a,e)$, and extremal CLP depends on two parameters $(y_+,\a)$ with $e=e_{ext}$, the GR black hole only depends on one parameter $y_+$. As expected, the GR charges saturate the BPS bound \eqref{BPS_bound_1} and satisfy the charge constraint 
$\Delta_{\hbox{\tiny KLR}}=0$, with $\Delta_{\hbox{\tiny KLR}}$ defined in \eqref{charge_constraint} with $Q_{1,2,3}\equiv Q$ and $J_{1,2}\equiv J$.

One way to understand why the BPS limit of the non-extremal CLP black hole reduces the number of parameters by two instead of one is as follows. Using \eqref{m_def} and \eqref{BH_charges}, we evaluate
\begin{equation}
\begin{split}
\label{BPS_bound_calc}
E_\CLP - E_\BPS &= \frac{ N^2 ( 3 - \a ) ( 1 - \a )  }{4 L [ y_+^2 - \a^2 ( 1 + y_+^2 ) ] } \\
&\qquad \qquad \times \left\{ ( 1 - \a^2 ) e^2  + 2 \left[ \frac{2 \a^2}{1-\a} - ( 1 + 2 \a ) y_+^2 \right] e + y_+^4 ( 1 + y_+^2 ) \right\} . 
\end{split}
\end{equation}
where $E_\BPS$ is defined in \eqref{BPS_bound_1}. This is a quadratic polynomial in $e$, whose discriminant is
\begin{equation}
\begin{split}
\text{Disc} = - \frac{N^4(3-\a)^2 [ y_+^2 - \a ( 2  + y_+^2 ) ]^2}{ 4L^4 [ y_+^2  - \a^2 ( 1 + y_+^2 ) ] } .
\end{split}
\end{equation}
We immediately see that $\text{Disc} \leq 0$, and since the coefficient of $e^2$ in \eqref{BPS_bound_calc} is positive, this implies that $E_\CLP \geq E_\BPS$, which is, of course, the BPS bound \eqref{BPS_bound_1}. Furthermore, the BPS bound is saturated if, and only if, $\text{Disc} = 0$. Thus, to obtain the supersymmetric limit, we need to solve two equations, namely $E_\CLP = E_\BPS$ and $\text{Disc} = 0$, which reduces the number of parameters by two instead of one.

%%%%%%%%%%%%%%%%%%%%%%%%%
\subsection{Onset of scalar condensation instability}
\label{sec:Onset3Q}
%%%%%%%%%%%%%%%%%%%%%%%%%

In this subsection, we study the (static) onset of the dynamical stability of the CLP black hole to the condensation of a charged scalar field $\Phi$. At linear order, the scalar instability of the CLP black hole can be addressed by analysing linear perturbations of the charged scalar field $\Phi$ about the CLP black hole described by \eqref{3Q:ansatz} and \eqref{CLP_sol}. From \eqref{EoM}, we see that the linearised scalar equation is simply the charged Klein-Gordon equation,
\begin{equation}
\label{eq:linear3Q}
D^2 \Phi + \frac{4}{L^2} \Phi = 0. 
\end{equation}
To study the linearised stability of $\Phi$, we use the fact that $\partial_t$ and $\partial_\psi$ are Killing vector fields of the CLP black hole to Fourier decompose the perturbation as
\begin{equation}
\begin{split}
\Phi(t,r,\t,\phi,\psi) = \Phi_{\o,k,\ell,m} (r) e^{-i\o t}e^{i k \psi} Y_{\ell,m}(\theta,\phi) , 
\end{split}
\end{equation}
where $k \in \mzz$, $\ell \in \mzz_{\geq 0}$ and $m \in \mzz \cap [ -\ell,\ell]$. This introduces the frequency $\o$ of the mode, the azimuthal quantum number $k$ and the harmonic numbers $\ell$ and $m$ of $Y$. The case of interest for us is the simplest case where $k=0$ (no angular momentum along $\psi$) and $\ell=m=0$ (symmetry of $S^2$ is unbroken). This `$s$--wave' mode is where the instability appears first (the non-spherically symmetric `excited' unstable states are irrelevant here). Moreover, assuming that instability is present, instead of solving the eigenvalue problem for $\o$ we can immediately look for the onset of the instability with $\o=0$ which defines the marginal boundary in the parameter space of CLP black holes between stable (when ${\rm Im}\, \o<0$) and unstable (when ${\rm Im} \,\o>0$) CLP black holes (for some static CLP black holes, the reader can find the frequencies $\o$ in~\cite{Dias:2022eyq}, including the transitions from ${\rm Im}\, \o<0$ into ${\rm Im}\, \o>0$). Rotating CLP is a 3-parameter family of black holes and those for which $\o=0$ reduces to a 2-parameter family of black holes. This considerably reduces the computational cost if we want to identify the instability onset since instead of solving an eigenvalue problem for $\o$, we need to solve a simpler (nonlinear) eigenvalue problem e.g., for the charge parameter $e$. Ultimately, one obtains the onset surface $e_{onset}(y_+,\a)$.

In these conditions, we solve \eqref{eq:linear3Q} numerically, and to do this, we introduce an auxiliary field $p$ and a compact coordinate $y$,
\begin{equation}\label{eq:linearRedef}
\Phi(r)=\left(\frac{r_+}{r}\right)^2 p(r) , \qquad r=\frac{r_+}{\sqrt{1-y^2}} ,
\end{equation}
so that the conformal boundary is located at $y=1$ and the black hole event horizon at $y=0$. We will search for smooth solutions $p(y)$ which correspond to scalar fields $\Phi(y)$ that satisfy the boundary conditions \eqref{asymp_exp} at the conformal boundary and which are regular (in ingoing Eddinghton-Finkelstein coordinates) on the horizon. Under these conditions, \eqref{eq:linear3Q} reduces to an equation of the form
\begin{equation} \label{eq:quadratic_3Q}
L_2\big(y;\tilde{\l};\tilde{e},{\tilde\a}\big)\,p''(y)+L_1\big(y;\tilde{\l};\tilde{e},{\tilde\a}\big)\,p'(y)+L_0\big(y;\tilde{\l};\tilde{e},{\tilde\a}\big)\,p(y)=0,
\end{equation}
where we have defined $\tilde{\l}\equiv y_+^2$, $\tilde{e}\equiv e/y_+^2$ and ${\tilde\a} \equiv \a/y_+$. Here, $L_{2,1,0}$ are functions of the coordinate $y$ and of the parameters $\tilde{\lambda},\tilde{e},{\tilde\a}$ that are not enlightening to display. The boundary conditions for $p(y)$ follow directly from the equation of motion (note that the asymptotic decay for $\Phi$ required by \eqref{asymp_exp} is already incorporated in the field redefinition \eqref{eq:linearRedef}). These are:
\begin{equation}
\label{eq:quadraticBC_3Q}
p'(0)=0 \qquad \hbox{and} \qquad 
p'(1)-\frac{2 \,\tilde{e}^2 \big[1-\tilde{\a}^2 (1+\tilde{\l})\big]^2}{\tilde{\l} \big( 1+ \tilde{e}\,\tilde{\a}^2\big)^2}\,p(1)=0\,.
\end{equation}

Equation  \eqref{eq:quadratic_3Q} subject to the boundary conditions \eqref{eq:quadraticBC_3Q} 
constitutes a non-polynomial (nonlinear) double eigenvalue problem in $\tilde{e}$ and ${\tilde\a}$ for a given input value of $\tilde{\l}$. To find the eigenfunction $p(y)$ and eigenvalues $\{\tilde{e}, {\tilde\a} \}$ we use a Newton-Raphson root-finding algorithm tailored to solve nonlinear eigenvalue problem as introduced and explained in detail  in~\cite{Dias:2015nua}. More precisely, to solve such a double nonlinear eigenvalue problem, one needs to supplement the eigenvalue equation \eqref{eq:quadratic_3Q}--\eqref{eq:quadraticBC_3Q} with two additional conditions (one for each eigenvalue). One of them is a condition that chooses the normalisation of the (linear) eigenfunction $p(y)$, e.g. $p(0)=1$ at the horizon.  For the second supplementary condition, we choose to search for the values of the CLP parameters $\{ \tilde{\l}, \tilde{e}, {\tilde\a} \}$ that have a constant angular momentum $J/N^2$, as defined in \eqref{BH_charges}, which fixed as an input in our code.
To summarise, as input parameters, we give the value of $J/N^2$ of the black hole together with the normalisation condition $p(0)=1$, and our  Newton-Raphson code runs over the eigenvalue $\tilde{\l}$ to find $\tilde{e}$, ${\tilde\a}$ and $p(y)$ that solve not only the two supplementary conditions but also the eigenvalue problem \eqref{eq:quadratic_3Q}--\eqref{eq:quadraticBC_3Q}.

\begin{figure}
\centering
\includegraphics[width=0.5\textwidth]{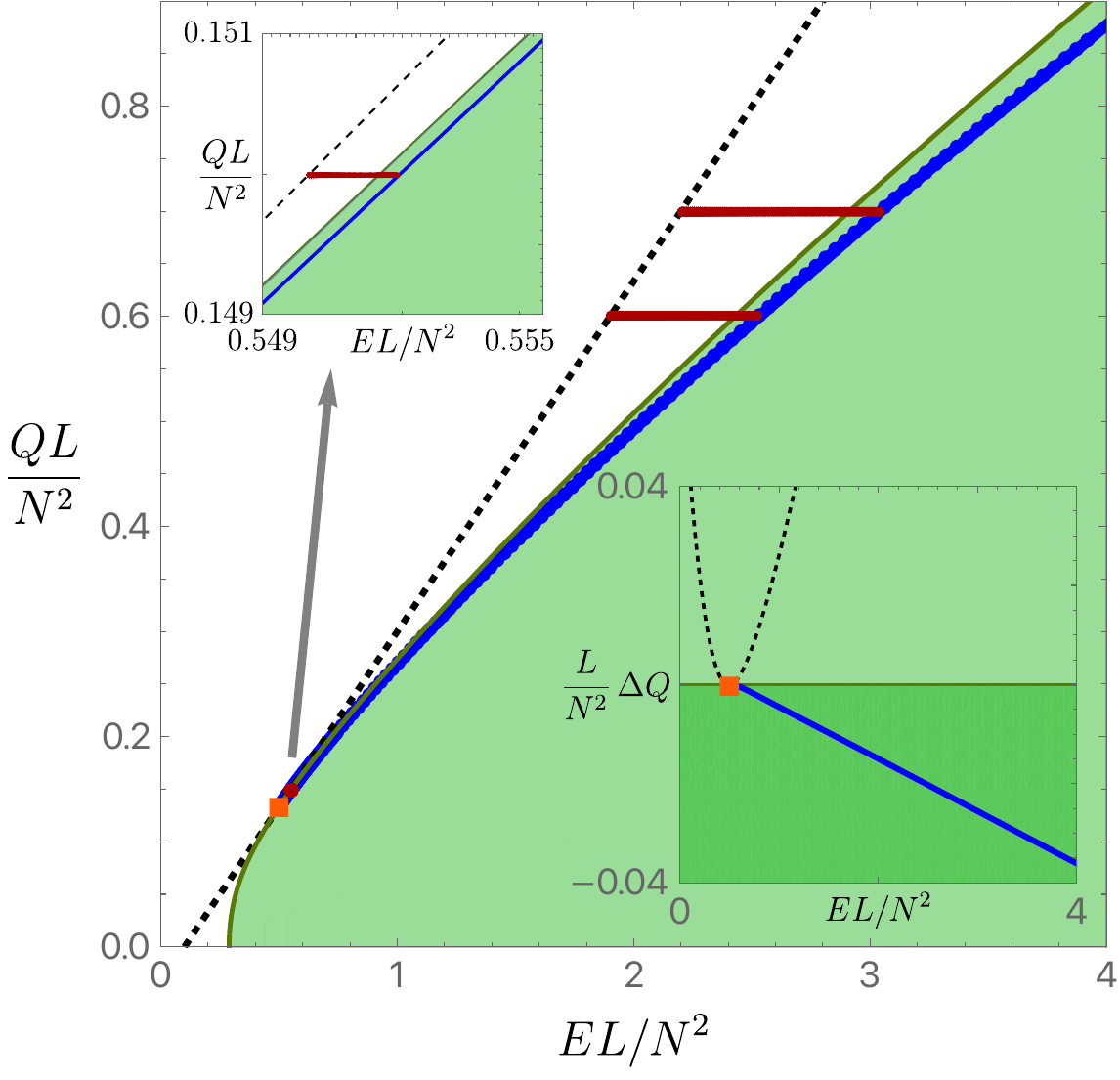}
\caption{Charge vs energy for $ J /N^2=0.05$ with the three red horizontal lines having $Q L/N^2=0.15, 0.6, 0.7$ (see zoom in the right-top inset plot for the $Q L/N^2=0.15$ case). The green area describes regular CLP black holes, with the upper dark-green boundary of this region being the extremal CLP with $T=0, S \neq 0$. The black dashed line describes the BPS line with $Q= \frac{1}{3}\left(E-2 J/L \right)$. The blue curve describes the instability onset of CLP black holes. It meets both the BPS and extremal CLP curves at the orange square with $(E, Q)\frac{L}{N^2}\simeq (0.502733, 0.134244)$ which describes the Gutowski-Reall supersymmetric black hole. The inset plot on the right-bottom represents the phase diagram of solutions $\Delta Q$ vs $E$ where $\Delta Q$ is the charge difference between a given solution and the extremal CLP black hole with the same $J/N^2$ and $E L/N^2$ (so $\Delta Q=0$ for the extremal CLP family).}
\label{figJ0p05:phasediagram}
\end{figure}

For a given angular momentum $J/N^2$, we can find the instability onset curve $Q_{\mathrm{onset}}(E)$. An example is the solid blue onset line displayed in Fig.~\ref{figJ0p05:phasediagram} for CLP with $ J /N^2=0.05$. We have checked that our numerical findings (obtained with an independent numerical code) exactly match those reported in the right panel of Fig.~5 of~\cite{Markeviciute:2018yal}.

%%%%%%%%%%%%%%%%%%%%%%%%%
\subsection{Hairy black hole: a non-interacting thermodynamic model}
\label{sec:toymodel}
%%%%%%%%%%%%%%%%%%%%%%%%%

Before embarking on the full construction of the hairy black hole solution in the following sections, we consider here a thermodynamic model for the hairy black hole, where we treat it as a {\it non}-interacting equilibrium mix of CLP black hole and static soliton described in the Sections~\ref{sec:CLP_bh} and~\ref{sec:soliton}, respectively. The analysis we perform is, in essence, the one completed in~\cite{Bhattacharyya:2010yg}. Of course, this model is a good approximation only for small thermodynamic charges since the equations of motion of the problem are highly non-linear. In Section~\ref{sec:hbh-pert}, we will perturbatively solve the equations of motion (using a matching asymptotic expansion) to find hairy black holes and confirm that the leading order thermodynamics of these solutions indeed reduces to the one predicted by the non-interacting thermodynamic model of this section. Remarkably, this simple thermodynamic model -- that does {\it not} use the equations of motion -- can produce the correct leading order thermodynamics for hairy black holes (as long as the temperature is not too close to zero, as we check later).

In this model, we take the rotating hairy black hole as a non-interacting mixture of the CLP black hole and the supersymmetric soliton. The absence of any interaction between the two phases means that the mass, charge, and angular momentum of the hairy black is simply the sum of the charges of its two constituents: 
\begin{equation}
\begin{split}
\label{mix-formulas}
E_\mix &= E_\CLP + 3 Q_s , \qquad Q_\mix = Q_\CLP + Q_s , \qquad J_\mix = J_\CLP + J_s = J_\CLP . 
\end{split}
\end{equation}
The subscripts CLP, $s$, and $mix$ denote the thermodynamic quantities of the soliton (of subsection~\ref{sec:soliton}), CLP black hole, and the non-interacting mix, respectively. The entropy of the non-interacting mix is 
\begin{equation}
\begin{split}
S_\mix = S_\CLP \big(E_\CLP, Q_\CLP , J_\CLP \big) = S_\CLP \big( E_\mix - 3 Q_s , Q_\mix - Q_s , J \big) .
\end{split}
\end{equation}
In the microcanonical ensemble where we fix $E$, $Q$, and $J$, the dominant phase is the one that maximises the entropy. It follows that the hairy black hole is dominant whenever the partition of charges among the two constituents is such that they maximise $S$ with respect to $Q_s$:
\begin{equation}
\begin{split}\label{entropy-extremization}
\pd{S_\mix}{Q_s} = 0 
\end{split}
\end{equation}
Using the first law \eqref{firstlaw}, we immediately find that the solution to this equation is
\begin{equation}
\label{e_hbh} 
\mu_\mix \equiv \mu_\CLP=\mu_s =1 \quad \implies \quad e_\mix = \frac{y_+^4}{y_+^2 - \a^2 ( 2 + y_+^2 )}, 
\end{equation}
where to obtain the second relation, we have used the CLP relations \eqref{BH_thermo}. Additionally, we can assign the soliton the same temperature and angular velocity as the CLP black hole because the former is a horizonless regular solution. In these conditions, the Killing vector field $\partial_t+\O_s \partial_\psi$ of the soliton follows the same orbits as the Killing horizon generator $\partial_t+\O_\CLP \partial_\psi$ as the black hole. In other words, the angular velocity and temperature of the non-interacting mix is 
\begin{equation}
\label{e_hbh2}
\O_\mix \equiv \O_\CLP = \O_s ,  \qquad T_\mix \equiv T_\CLP=T_s , 
\end{equation}
Altogether, the mass, charge, and angular momentum distribution among the two components that maximise the entropy is the one that also yields thermodynamic equilibrium between the two constituents of the hairy black hole. Substituting this into \eqref{blackholerange}, we find that the hairy black hole can exist as a non-interacting mix if and only if 
\begin{equation}
\begin{split}
\label{alpha_bound_hbh}
0 \leq \a \leq \frac{y_+^2}{2 + y_+^2} . 
\end{split}
\end{equation}
Motivated by this bound, we set
\begin{equation}
\begin{split}\label{j1q1non-int}
\a = \sqrt{1-\g}  \frac{y_+^2}{2+y_+^2} \qquad \implies \qquad  e_\mix = \frac{y_+^2(2+y_+^2)}{2+\g y_+^2} ,  
\end{split}
\end{equation}
where $0 \leq \g\leq 1$ guarantees that we are in the range \eqref{alpha_bound_hbh}. The thermodynamic quantities of the hairy black hole are then given by
\begin{equation}
\begin{split}\label{MQJ_mix}
\frac{E_{\mix} L}{N^2} &= \frac{N^2}{L} \frac{24 y_+^2  + 36 y_+^4 + 2 (\g + 8) y_+^6 + \g(4-\g) y_+^8}{4 (2 + \g y_+^2)^2} + 3 \frac{Q_s L}{N^2} , \\
\frac{Q_{\mix} L}{N^2} &= \frac{N^2}{L} \frac{y_+^2(2 + y_+^2)}{2(2 + \g y_+^2)} + \frac{Q_sL}{N^2} , \qquad \frac{J_{\mix} }{N^2} = \sqrt{1-\g} \frac{y_+^4 ( 6 + (4+\g) y_+^2 + \g y_+^4 ) }{ 2 ( 2 + \g y_+^2 )^2 } , \\
\mu_{\mix} &= 1 , \qquad \O_{\mix} L = \sqrt{1-\g} , \qquad T_{\mix}  L  = \frac{\g y_+[4 +(3+\g)y_+^2 + \g y_+^4]^{1/2}}{\pi(2+\g y_+^2)} , \\
\frac{S_{\mix}}{N^2} &= \pi y_+^3 \frac{[\g  y_+^4+(3+\g)y_+^2+4]^{1/2}}{2+\g y_+^2} .
\end{split}
\end{equation}
\eqref{MQJ_mix} describes the thermodynamics of the hairy black hole modelled as a non-interacting mix of the CLP black hole and soliton. We expect this model to yield a good approximation to leading order in $y_+$ and $Q_s$, and we will show that this is true in Section~\ref{sec:hbh-pert}.

\begin{figure}[H]
\centering
\includegraphics[width=0.5\textwidth]{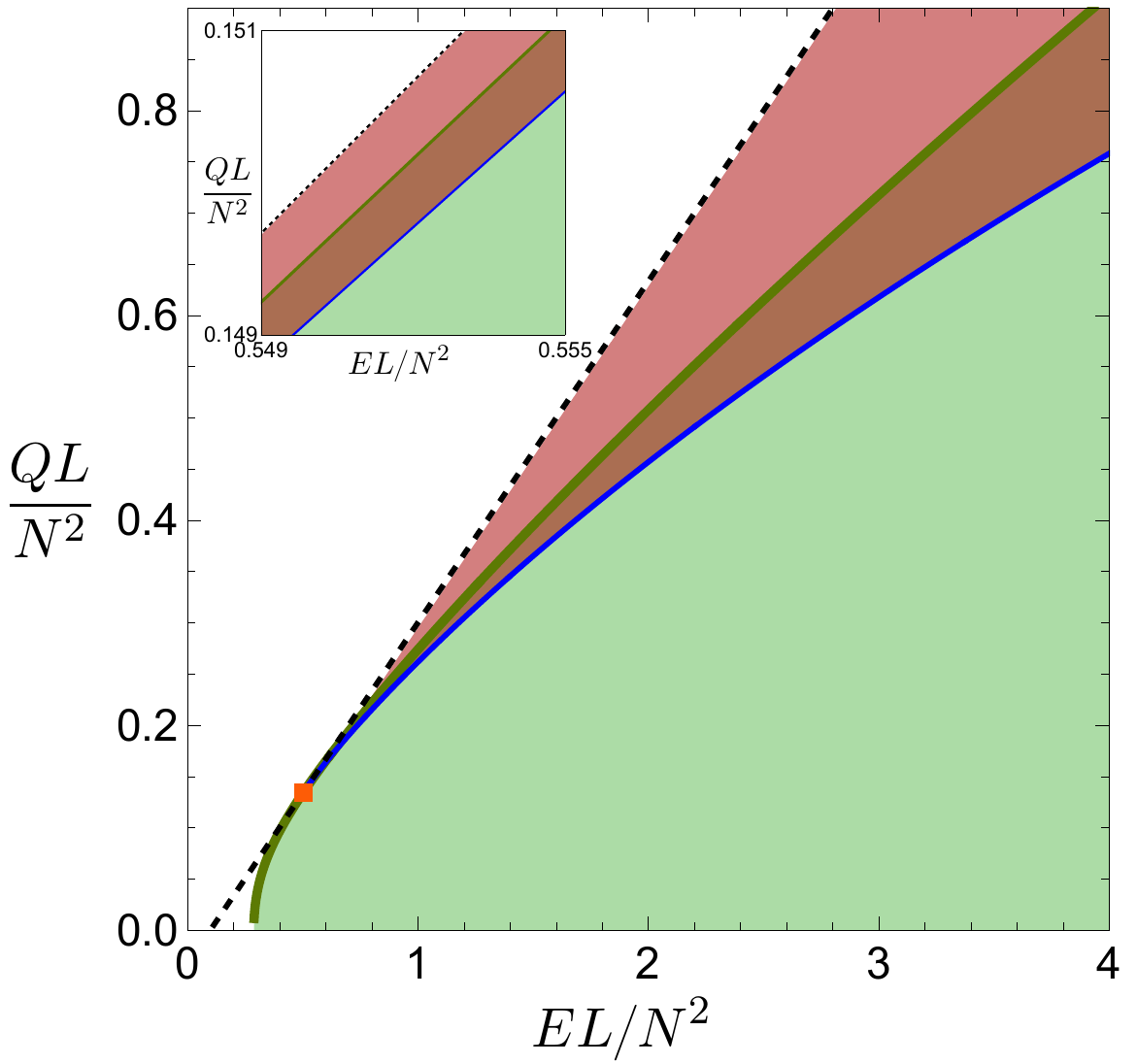}
\caption{Charge vs. energy phase diagram for $J/N^2 = 0.05$. The rotating hairy black hole is a non-interacting equilibrium mix in the red-shaded region. For small $E$, $Q$, and $J$ (which is the regime in which this thermodynamic model is valid), the blue curve coincides precisely with the instability onset curve in Fig.~\ref{figJ0p05:phasediagram}. The hairy black holes start existing precisely where the CLP black holes become unstable to scalar condensation (blue curve) and continue until the BPS bound (black-dashed curve). In particular, hairy black holes also exist in the region above the green curve (which describes the extremal boundary of the existence of CLP black holes) in a region where CLP black holes are no longer present (CLP exist in the region below the extremal black curve in the green-shaded region). As in Fig.~\ref{figJ0p05:phasediagram}, the red square describes the BPS Gutowski-Reall black hole. The extremal CLP, onset, and BPS curves meet at this single GR point.}
\label{fig:HBH_toy}
\end{figure}

The hairy black hole exists as a non-interacting mix whenever
\begin{equation}
\begin{split}
y_+ \geq 0 , \qquad Q_s \geq 0 , \qquad 0 \leq \g \leq 1 . 
\end{split}
\end{equation}
In Fig.~\ref{fig:HBH_toy}, we plot this region for fixed $J/N^2=0.05$. From the above analysis, it is clear that the CLP component of the mixture provides all the angular momentum of the hairy black hole, and the soliton's component contribution to the mixture is to provide the scalar condensate, which contributes to the mass and charge of the hairy black hole through $Q_s$. Thus, in Fig.~\ref{fig:HBH_toy}, $Q_s=0$ (no scalar condensate) gives the merger blue curve while  $\g=0$ yields the BPS dashed black line where $E\to E_\BPS=3Q+2J/L$ in \eqref{MQJ_mix}. Hairy black holes exist between these two curves. The thermodynamic model predictions are at most valid only when $(E, Q, J)$ are all parametrically small.

From \eqref{MQJ_mix}, one sees that the BPS limit is attained when  $\g\to \g_\BPS=0$ since $E\to  E_\BPS$, $T \to 0$, $\mu\to 1$ and  $\O_H L\to 1$.\footnote{\label{foot:g} The thermodynamic model predicts $\g_\BPS = 0$. In the exact perturbative solution (see Section~\ref{sec:hbh-pert}), this receives corrections proportional to the horizon radius and scalar condensate.} Furthermore, in this limit, the entropy $S$ goes to a {\it finite non-zero} value, which suggests that the hairy black hole reduces to a regular supersymmetric hairy black hole in the BPS limit. However, as we will show in Section~\ref{sec:MainResults}, this prediction fails to hold. This will be the main result of our paper. The thermodynamic model and perturbative analyses do yield very good approximations for small $(E, Q, J)$ as long as the temperature of the system is not to close to $T=0$ (and this is one of the reasons why they can still be very useful).

There is another feature of the thermodynamic model that will guide our numerical search of hairy black holes in Section~\ref{sec:numerics}. Recall from the discussion of Section~\ref{sec:soliton} that the one-parameter family of regular (static) solitons ends at some critical charge $Q_c$, and a new one-parameter family of singular solitons starts. In other words, there is a phase transition in the phase space of solitons at $Q=Q_c$. To the extent that the thermodynamic model is still a reasonably valid description of the hairy black hole for larger charges, one expects to see some structural transition at some critical charge $Q_c(J)$ (probably dependent on $J$) in the phase space of hairy black holes as we move from small to larger charges: the qualitative behaviour of hairy black holes might change when we cross this transition boundary.\footnote{\label{foot:Qc} Strictly speaking, this critical charge should be described by a surface $Q=Q_c(J,E)$ in the phase space $(E, Q, J)$ which reduces to a curve $Q=Q_c(J,E_\BPS)\equiv Q_c(J)$ only in the BPS limit. From here on, we will loosely refer to this surface as simply $Q_c(J)$ since there is a clear distinction between black hole families only when we analyse how they approach the BPS limit.}   We will see that this is indeed the case (see later Figs.~\ref{fig:epsilonE}--\ref{fig:entropyT} and associated discussions).

%%%%%%%%%%%%%%%%%%%
\section{Hairy black holes: perturbative solution}
\label{sec:hbh-pert}
%%%%%%%%%%%%%%%%%%%

The outcome of Section~\ref{sec:Onset3Q} shows that for certain values of $E$, $Q$, and $J$, the CLP black hole is unstable to the condensation of the charged scalar field of \eqref{EoM}, and the considerations of Section~\ref{sec:toymodel} suggest that the hairy black hole should emerge as a new dominant phase in the microcanonical ensemble whenever the CLP black hole is unstable. This further hints that the hairy black hole might be the endpoint, or at least a metastable configuration, of the dynamical time evolution of a CLP black hole that is unstable to scalar condensation. Of course, these arguments rely partially on the crude thermodynamic model of Section~\ref{sec:toymodel}, which is expected only to be a first-order approximation to the actual hairy black hole solution (and, as we find later, only when the temperature is not too small). Therefore, we must make these ideas more solid to find the phase diagram of hairy black holes.

To do this, we solve the coupled system of non-linear ODEs \eqref{EoM} for the field ansatz\"e  \eqref{3Q:ansatz} subject to the boundary conditions discussed in Section~\ref{sec:ansatz} to find the hairy black hole solutions of \eqref{action}. In Section~\ref{sec:numerics}, we will solve the nonlinear boundary value problem exactly using numerical methods. In this section, we find the hairy black holes in perturbation theory using techniques that were introduced and used in~\cite{Bhattacharyya:2010yg} (see also~\cite{Dias:2022eyq}). Here, we present a summary of the perturbative procedure.

In our perturbative construction, we treat the scalar field as a small perturbation on the CLP black hole. Our perturbative parameter is the VEV $\e$ of the operator dual to $\Phi$ which appears in the asymptotic expansion of this field in \eqref{asymp_exp}. We expand the functions appearing in our ansatz\"e \eqref{3Q:ansatz} -- collectively denoted as ${\mathfrak F}^I =\{ f,g,h,w,A_t,A_\psi\}$ and $\Phi$ -- as\footnote{We perturb the CLP black hole with $\Phi$ at $\CO(\e)$ which then backreacts on the black hole at $\CO(\e^2)$, which then perturbs $\Phi$ again at $\CO(\e^3)$, and so on. Consequently, the scalar field gets excited at odd orders in $\e$ whereas the functions ${\mathfrak F}^I$ are excited at even orders in $\e$ in \eqref{main_exp}.}
\begin{equation}
\begin{split}
\label{main_exp}
{\mathfrak F}^I(r) = \sum_{n=0}^\infty \e^{2n} {\mathfrak F}^I_{(2n)}(r), \qquad  \Phi(r) = \sum_{n=0}^\infty \e^{2n+1} \Phi_{(2n+1)}(r) , 
\end{split}
\end{equation}
where the base solution ($n=0$) is the CLP black hole \eqref{CLP_sol}. Plugging in \eqref{main_exp} into the EoM \eqref{EoM}, we find coupled ordinary linear differential equations at each order in $\e$. However, since the CLP solution is so complicated, these differential equations are impossible to solve analytically. To simplify the problem, we perform a second expansion in the size of the hairy black hole $y_+ \equiv \frac{r_+}{L}$. This is more subtle compared to the expansion \eqref{main_exp} in $\e$ as we need to decide how the radial coordinate $r$ scales with $y_+$ when we perform the expansion. First keeping $r$ fixed, we expand all functions as\footnote{In our expansion \eqref{far_exp}, we allow for terms proportional to $\ln y_+$ as well (and indeed, such terms do appear in the final solution). In \eqref{far_exp} all terms of the form $y_+^{2k} \ln^m(y_+)$ are categorized as $\CO(y_+^{2k})$.}
\begin{equation}
\begin{split}
\label{far_exp}
{\mathfrak F}^I_{(2n)}(r) = \sum_{k=0}^\infty y_+^{2k} {\mathfrak F}^{I,\far}_{(2n,2k)}(r)  , \qquad \Phi_{(2n+1)}(r) = \sum_{k=0}^\infty y_+^{2k} \Phi^\far_{(2n+1,2k)}(r) . 
\end{split}
\end{equation}
The parameters $e$ and $\a$ appearing in the base CLP solution are similarly expanded in $\e$ and $y_+$. Motivated by the thermodynamic model of Section~\ref{sec:toymodel} -- specifically, by \eqref{j1q1non-int} -- we first rewrite
\begin{equation}\label{introduceGamma}
\begin{split}
e = y_+^2 + y_+^4 a  , \qquad \a = \frac{1}{2} y_+^2 \sqrt{1-\g} \,. 
\end{split}
\end{equation}
The thermodynamic model calculation of Section~\ref{sec:toymodel} tells us that $a$ is $\CO(1)$ as $\e,y_+ \to 0$. We therefore expand $a$ in $\e$ and $y_+$ as
\begin{equation}
\begin{split}
\label{a_exp}
a = \sum_{n=0}^\infty \e^{2n} a_{(2n)} , \qquad a_{(2n)} = \sum_{k=0}^\infty y_+^{2k} a_{(2n,2k)} . 
\end{split}
\end{equation}
We plug in the expansions \eqref{main_exp}, \eqref{far_exp} and \eqref{a_exp} into the EoM \eqref{EoM}. The differential equations we now get at each order in $\e$ and $y_+^2$ are much simpler, and we can solve them analytically (see Appendix~\ref{app:far_field_eq} for the explicit form of the equations).

\begin{figure}[H]
\centering
\includegraphics[width=0.7\textwidth]{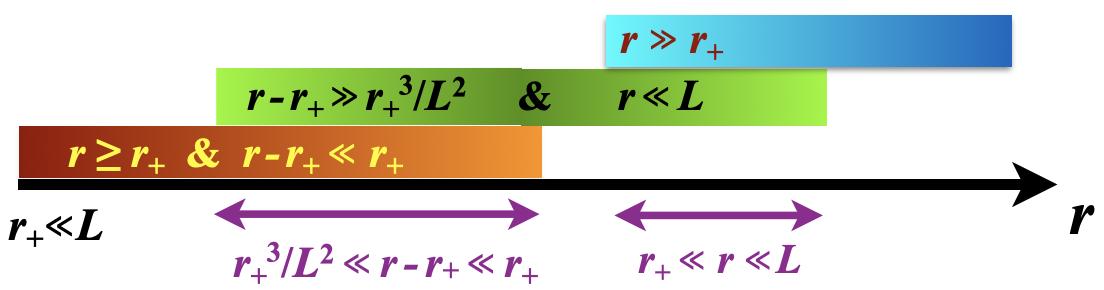}
\caption{The three regions of the matching asymptotic expansion and their overlapping regions.}
\label{fig:matching}
\end{figure}

The perturbative expansion described by \eqref{far_exp} is only valid when $r \gg r_+$. We can see why this is so by expanding the base CLP solution in $y_+$. For example, at small $y_+$, $f_\CLP(r)$ has the expansion
\begin{equation}
\begin{split}
\label{fCLPfar_exp}
f_\CLP(r) &= \left( \frac{r^2}{L^2}  + 1 \right) + \left( - \frac{2L^2}{r^2} \right) y_+^2 + \CO(y_+^4) .
\end{split}
\end{equation}
We see that the subleading term in \eqref{fCLPfar_exp} is small compared to the leading term only  when $r \gg r_+$. This holds for all the other functions in the CLP solution as well. It follows that the solution constructed via the perturbative expansion \eqref{far_exp} is valid only in the \emph{far-field region}, $r \gg r_+$ and breaks down when $r\sim r_+$. This region is shown in blue in Fig.~\ref{fig:matching}.

To construct the solution in the region $r\sim r_+$, we define new coordinates
\begin{equation}
\begin{split}
\label{ytau_def}
y = \frac{r}{r_+} , \qquad \tau = \frac{t}{r_+} . 
\end{split}
\end{equation}
We now expand all the functions in $y_+$, keeping $y$ fixed as
\begin{equation}
\begin{split}
\label{int_field_exp}
{\mathfrak F}^I_{(2n)}(r_+ y) = \sum_{k=0}^\infty y_+^{2k+\b_I} {\mathfrak F}^{I,\intt}_{(2n,2k)}(y)  , \qquad \Phi_{(2n+1)}(r_+ y) = \sum_{k=0}^\infty y_+^{2k} \Phi^\intt_{(2n+1,2k)}(y), \\
\end{split}
\end{equation}
The leading coefficients $\b_I$ are fixed by requiring that the metric and gauge field scale appropriately as $y_+ \to 0$ (see~\cite{Bhattacharyya:2010yg} for details). Substituting \eqref{int_field_exp} and \eqref{a_exp} into EoM \eqref{EoM}, we find a set of differential equations (summarized in Appendix~\ref{app:int_field_eq}) at each order in $\e$ and $y_+$ that can be solved analytically.

As with the far-field solution, the intermediate-field expansion has a limited regime of validity. To find the region where it is valid, we turn to the expansion of the base CLP solution, which now has an expansion
\begin{equation}
\begin{split}
\label{fCLPint_exp}
\! f_\CLP(r_+ y) = \frac{(y^2-1)^2}{y^4} \left[ 1 + \left( \frac{y^2+y^4}{y^2-1}  - (1-\g) \frac{1-5y^2+4y^4+8y^6}{y^6(y^2-1)} \right) y_+^2 + \CO(y_+^4) \right]  . 
\end{split}
\end{equation}
where we used the fact that $a_{(0,0)} = \frac{1-\g}{2}$ which is predicted by the thermodynamic model of Section~\ref{sec:toymodel} (see equation \eqref{j1q1non-int}) and confirmed by the explicit construction done in this section. From \eqref{fCLPint_exp}, we see that the subleading term is small compared to the leading term only when $y \ll \frac{1}{y_+}$ (or $r \ll L$) and when $y - 1 \gg y_+^2$ (or $r - r_+  \gg \frac{r_+^3}{L^2}$). It follows that the solution constructed via the expansion \eqref{fCLPint_exp} is valid only in the \emph{intermediate-field region}, $r \ll L$ and $r - r_+ \gg \frac{r_+^3}{L^2}$ and breaks down when either $r \sim L$ or $r-r_+ \sim \frac{r_+^3}{L^2}$. This region is shown in green in Fig.~\ref{fig:matching}.

The far-field solution above already covers the region $r \sim L$ (when $r_+ \ll L$). To construct the solution in the region $r - r_+  \sim \frac{r_+^3}{L^2}$, we define new coordinates
\begin{equation}
\begin{split}
\label{zT_def}
z = \frac{L^2}{r_+^3} ( r - r_+ ) , \qquad T = \frac{r_+}{L^2} t .
\end{split}
\end{equation}
We now expand all fields in $y_+$ keeping $z$ fixed as
\begin{equation}
\begin{split}
\label{near_field_exp}
{\mathfrak F}^I_{(2n)} \bigg( r_+ + \frac{z r_+^3}{L^2} \bigg) &= \sum_{k=0}^\infty y_+^{2k+\g_I} {\mathfrak F}^{I,\near}_{(2n,2k)}(z) , \\
\Phi_{(2n+1)} \bigg( r_+ +  \frac{z r_+^3}{L^2} \bigg) &= \sum_{k=0}^\infty y_+^{2k} \Phi^\near_{(2n+1,2k)}(z). 
\end{split}
\end{equation}
The leading coefficients $\g_I$ are fixed by requiring that the metric and gauge field scale appropriately as $y_+ \to 0$~\cite{Bhattacharyya:2010yg}. Substituting \eqref{near_field_exp} and \eqref{a_exp} into EoM \eqref{EoM}, we find a set of differential equations (summarized in Appendix~\ref{app:near_field_eq}) at each order in $\e$ and $y_+$ that can be solved analytically. This expansion is valid down to the horizon $z=0$ \emph{if and only if} $\g$ is $\CO(1)$ as $y_+ \to 0$. To see this, we turn to the expansion of the base CLP solution, which has the expansion
\begin{equation} \label{orderGamma1}
\begin{split}
& f_\CLP\bigg( r_+ + \frac{z r_+^3}{L^2} \bigg) = 4 z ( z + \g ) y_+^4 \left[ 1 + \left( - \frac{1 + 4 \g}{2(z+\g)} + \frac{9+\g}{4}  - 3 z  \right) y_+^2  + \CO(y_+^4) \right]  . 
\end{split}
\end{equation}
As long as $\g = \CO(1)$, we see that the subleading term is small compared to the first as long as $z \ll y_+^{-2}$ (or $r - r_+ \ll r_+$). It follows that the solution constructed via the expansion \eqref{near_field_exp} is valid in the \emph{near-field region}, $r - r_+ \ll r_+$ and $r \geq r_+$ and breaks down when $r-r_+ \sim r_+$. This region is shown in orange in Fig.~\ref{fig:matching}. The intermediate-field solution covers the region $r-r_+ \sim r_+$, and we are only interested in constructing the black hole solution up to the horizon. It follows that we do not need to introduce any new region, and the far-field, intermediate-field, and near-field solutions completely describe the hairy black hole solution everywhere in spacetime.

We end this discussion with two comments. The first is regarding the boundary conditions required to solve the differential equations. For the far-field solution, we impose asymptotically AdS boundary conditions \eqref{asymp_exp}. For the near-field solution, we impose regularity on the horizon. The remaining integration constants are fixed by matching the far-field, intermediate-field, and near-field solutions in the regions where their domain of validity overlap. These overlap regions are shown in Fig.~\ref{fig:matching} (purple arrows). The second comment concerns what happens if $\g \neq \CO(1)$. To study this case, we take $\g = y_+^2 \g'$ and keep $\g'$ fixed as $y_+ \to 0$. In this case, the base CLP black hole has the expansion
\begin{equation} \label{orderGamma2}
\begin{split}
f_\CLP\bigg( r_+ + \frac{z r_+^3}{L^2} \bigg) &= 4 z^2 y_+^4 \left[ 1 + \left( \frac{9}{4} - 3 z + \frac{\g'-1/2}{z} \right) y_+^2  + \CO(y_+^4) \right]  . 
\end{split}
\end{equation}
From this, the subleading term is small compared to the first only if $z \gg y_+^2$. Consequently, in this case, to construct the full solution, we would need to introduce another region of spacetime, namely the ``near-near-field region'' where $z \sim y_+^2$.\footnote{If $\g'=1/2$, then the subleading term is always small compared to the leading one, but the subsubleading $\CO(y_+^4)$ term is large compared to the subleading $\CO(y_+^2)$ term when $z \sim y_+^2$. Either way, the perturbative expansion is still valid only when $z \gg y_+^2$.} In the same way, when $\g=\CO(y_+^4)$, we would need to introduce \emph{two} new regions to cover the entire solution. If $\g=\CO(y_+^{2k})$, we must construct the perturbative expansion in $k+3$ regions to cover the entire spacetime outside the horizon. For simplicity, we restrict ourselves to the $k=0$ case. It follows from this that our perturbative construction is valid as long as $\g \gg y_+^2$.

\subsection*{Thermodynamics in the Microcanonical Ensemble}

\begin{figure}[!h]
\centering
\includegraphics[width=0.5\textwidth]{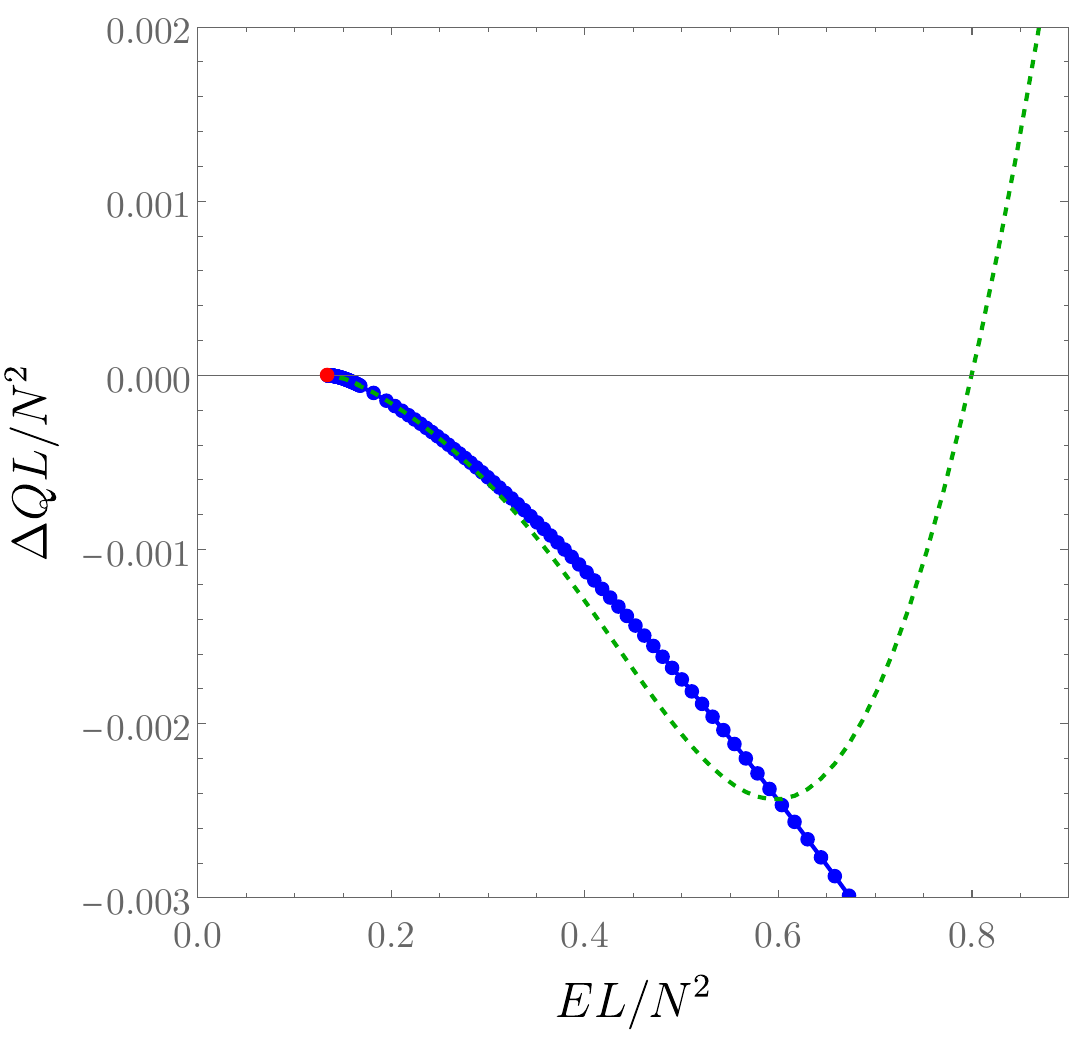}
\caption{Merger curve of hairy and CLP black holes for $J/N^2=0.05$. The exact numerical curve is the blue curve while the green dashed curve is the perturbative merger curve given by \eqref{hbh_pert_E}-\eqref{hbh_pert_Q} with $\epsilon=0$. As expected, the perturbative analysis is a good approximation for small $E$. $\Delta Q$ is the charge difference between a given solution and the extremal CLP black hole with the same $J/N^2$ and $E L/N^2$ (so $\Delta Q=0$ for the extremal CLP family) as defined by \eqref{extCLP}. As seen in Fig.~\ref{figJ0p05:phasediagram} (see its inset plot), the extremal CLP and merger curves meet at the BPS orange square that describes the Gutowski-Reall supersymmetric black hole.
}
\label{fig:onsetNumPert}
\end{figure}

Once we have the explicit solutions for $\{ f,g,h,w, A_t, A_\psi,\Phi\}$ in the three regions that satisfy the boundary and matching conditions of the problem (see accompanying {\tt Mathematica} file), we can use \eqref{ansatz_charges} and \eqref{ansatz_potentials} to find the thermodynamic quantities of the hairy black hole. Below and in the accompanying {\tt Mathematica} file, we present the results for hairy black hole thermodynamics to the order that we have evaluated them: 
\begin{subequations}
\label{hbh_pert_thermo}
\begin{equation}
\begin{split}
\label{hbh_pert_E}
\frac{EL}{N^2} &= \bigg( \frac{3 y_+^2}{2}  + \frac{3y_+^4}{4} ( 3 - 2 \g )   + \frac{y_+^6}{8}  (9 \g ^2-23 \g +20 ) + \frac{y_+^8}{16} [ -12 \g ^3+27 \g ^2-127 \g \\
&\qquad \quad  + 28 - 48 \g  \ln (2 \g  y_+^4) ]  + \frac{y_+^{10} }{32} [ 15 \g ^4-\g ^3+606 \g ^2+719 \g -192 \g  \zeta (3) \\
&\qquad \quad + 269  + ( 360 \g ^2 +648 \g + 96 ) \ln(2 \g  y_+^4 ) +192 \g  \ln^2(2   \g  y_+^4) ] + \CO(y_+^{12})  \bigg) \\
&\quad + \e^2 \bigg( \frac{3}{16} + \frac{3 y_+^2}{16} - \frac{3y_+^4}{32} \left[ 7 \g + 5  + 4 \ln (2 \g  y_+^4 ) \right]  + \frac{y_+^6}{64 \g } [ 63 \g ^3+8 \pi ^2 \g ^2 + 265 \g ^2 \\
&\qquad \quad +336 \g  - 48 \g  \zeta (3) + 24 + 24 \g ( 10 \g + 9 )  \ln  (2 \g  y_+^4 )  + 48 \g  \ln ^2 (2 \g  y_+^4 ) ]  \\
&\qquad \quad + \CO(y_+^8)  \bigg) + \e^4 \bigg( \frac{1}{256} + \frac{29 y_+^2}{768} + \frac{y_+^4}{4608} [ -72 \pi ^2 \g - 399 \g - 1147  \\
&\qquad \quad - 648 \ln (2 \g  y_+^4 ) ] + \CO(y_+^6)  \bigg) + \e^6 \bigg( \frac{1}{2560} + \CO(y_+^2) \bigg) + \CO(\e^8) , \\
\end{split}
\end{equation}
\begin{equation}
\begin{split}
\label{hbh_pert_Q}
\!\! \frac{QL}{N^2} &= \bigg( \frac{y_+^2}{2} +  \frac{y_+^4}{4} (1-\g)  + \frac{y_+^6}{8} ( \g ^2 - \g + 2 )  + \frac{y_+^8}{16} [-\g ^3-2 \g ^2-24 \g - 1 \\
&\qquad \quad - 16 \g \ln (2 \g  y_+^4) ] + \frac{y_+^{10} }{32} [ \g ^4+9 \g ^3+130 \g ^2 + 317 \g  - 64 \g  \zeta (3) + 77 \\
&\qquad \quad + ( 88 \g^2 + 248 \g   + 32 )  \ln (2 \g  y_+^4) + 64 \g  \ln ^2(2  \g  y_+^4)  ]  + \CO(y_+^{12})  \bigg) \\
&\quad + \e^2 \bigg( \frac{1}{16} + \frac{y_+^2}{16} + \frac{y_+^4}{32}  [ -6 \g - 7 - 4 \ln (2 \g  y_+^4 )  ] + \frac{y_+^6}{192 \gamma } [ 39 \g^3+8 \pi ^2 \g ^2+235 \g ^2 \\
&\qquad \quad + 426 \g -48 \g  \zeta (3)+24  + ( 216 \g^2  + 264 \g ) \ln(2 \g  y_+^4 ) +48 \g  \ln^2 (2 \g y_+^4 ) ] \\
&\qquad \quad + \CO(y_+^8) \bigg) + \e^4 \bigg( \frac{1}{768} + \frac{29 y_+^2}{2304} + \frac{y_+^4 }{13824} [ -72 \pi ^2 \g -318 \g -1309 \\
&\qquad \quad - 648 \log  (2 \g  y_+^4 ) ]  + \CO(y_+^6)  \bigg)  + \e^6 \bigg( \frac{1}{7680} + \CO(y_+^2) \bigg) + \CO(\e^8) , \\
\end{split}
\end{equation}
\begin{equation}
\begin{split}
\label{hbh_pert_J}
\!\! \frac{J}{N^2} &= \sqrt{1-\g} \bigg[  \bigg( \frac{3 y_+^4}{4} + \frac{y_+^6}{8} (7-5 \g ) + \frac{y_+^8}{16} (7 \g ^2-15 \g +14) + \frac{y_+^{10} }{32} [ -9 \g ^3+14 \g ^2 \\
&\qquad \quad -104 \g +15 - 48 \g \ln (2 \g  y_+^4 ) ]  + \frac{y_+^{12} }{64} [ 11 \g ^4+11 \g ^3+526 \g ^2+807 \g \\
&\qquad \quad -192 \g \zeta(3)+251 + ( 328 \g^2  + 680 \g + 96 )  \ln  (2 \g  y_+^4 ) +192 \g \ln^2 (2 \g  y_+^4 )  ] \\
&\qquad \quad + \CO(y_+^{14}) \bigg) + \e^2 \bigg( \frac{3 y_+^4}{32}-\frac{3 y_+^6}{64} [ 6 \g +15 + 8 \ln (2 \g  y_+^4) ] + \CO(y_+^8) \bigg) \\
&\quad + \e^4  \bigg( \frac{9y_+^4}{512}  + \CO(y_+^6) \bigg)  +  \e^6 \bigg( \CO(y_+^4) \bigg)  + \CO (\e^8) \bigg],
\end{split}
\end{equation}
\begin{equation}
\begin{split}
\label{hbh_pert_mu}
\!\! \mu &= \bigg( 1 + \frac{\g  y_+^4}{2} + \frac{y_+^6}{2} [  - \g^2 - 5 \g - 1 - 4 \g  \ln (2 \g  y_+^4) ]  + \frac{y_+^8 }{8} [ 3 \g ^3+50 \g ^2+175 \g + 39 \\
&\qquad \quad - 32 \g  \zeta (3) + ( 32 \g^2  + 136 \g + 16 )  \ln  (2 \g  y_+^4 )  + 32 \g  \ln^2 (2 \g y_+^4 ) ] + \CO(y_+^{10}) \bigg) \\
&\quad + \e ^2 \bigg( \frac{\g  y_+^4}{6} + \frac{y_+^6}{144} [ - 51 \g ^2-389 \g -24  - 204 \g  \ln (2 \g  y_+^4) ]  + \CO(y_+^{8}) \bigg) \\
&\quad + \e^4 \bigg( \frac{287 \g  y_+^4}{5760} + \CO(y_+^6) \bigg) + \e^6 \bigg( \CO(y_+^4) \bigg) + \CO(\e^8) , \\
\end{split}
\end{equation}
\begin{equation}
\begin{split}
\label{hbh_pert_Omega}
\!\! \O_H L &= \sqrt{1-\g} \bigg[ \bigg(1 + \frac{y_+^2}{2} + \frac{\g  y_+^4}{4} + \frac{y_+^6}{4} [ -\g ^2-5 \g - 1  - 4 \g  \ln (2 \g  y_+^4) ] + \frac{y_+^8}{16} [ 3 \g^3  \\
&\quad + 50 \g^2+175 \g - 32 \g  \zeta (3) + 39  + ( 32 \g ^2 +136 \g +16 )  \ln (2 \g  y_+^4) \\
&\quad + 32 \g  \ln^2(2 \g y_+^4) ]  + \CO(y_+^{10}) \bigg) + \e^2 \bigg( - \frac{\g  y_+^2}{16}  + \frac{ \g y_+^4 }{192} [ 18 \g +73  + 72 \ln (2 \g  y_+^4) ]  \\
&\quad + \CO(y_+^6) \bigg) + \e^4 \bigg( - \frac{\g y_+^2 }{256} + \CO(y_+^4) \bigg) + \e^6 \bigg( \CO(y_+^2) \bigg) + \CO (\e^8) \bigg],
\end{split}
\end{equation}
\begin{equation}
\begin{split}
\label{hbh_pert_T}
T L &= \bigg( \frac{\g y_+}{\pi}  + \frac{y_+^3}{8\pi} (-3 \g ^2+7 \g - 8 ) + \frac{y_+^5}{128\pi} [ 23 \g ^3+42 \g ^2+399 \g  -16 \\
&\quad + 256 \g  \ln (2 \g  y_+^4 ) ]  + \frac{y_+^7}{1024\pi} [ -91 \g ^4-899 \g^3 - 9665 \g ^2 + 4096 \g  \zeta(3) - 18953\g \\
&\quad  - 4696 -256 (27 \g ^2+57 \g +8) \ln (2 \g  y_+^4)-4096 \g  \ln^2 (2 \g  y_+^4 )  ] + \CO(y_+^9) \bigg) \\
& +\e^2 \bigg( \frac{\g  y_+}{8\pi} + \frac{y_+^3}{192\pi} [ -51 \g ^2-281 \g - 24 - 144 \g  \ln (2 \g y_+^4 ) ]  + \CO(y_+^5) \bigg) \\
&+  \e ^4 \bigg( \frac{3 \g  y_+}{128\pi} + \CO(y_+^3)  \bigg) + \e^6 \bigg( \CO(y_+) \bigg) + \CO(\e^8) , \\
\end{split}
\end{equation}
\begin{equation}
\begin{split}
\label{hbh_pert_S}
\frac{S}{N^2} &= \pi y_+^3 + ( 1 - \g ) \bigg[ \bigg( \frac{3 \pi y_+^5}{8} + \frac{\pi y_+^7}{128}  (39-23\g)  + \frac{\pi y_+^9}{1024}  ( 91 \g^2-198 \g +235 ) \\
&\quad + \frac{\pi y_+^{11} }{32768} [ -1451 \g ^3 + 353 \g^2 - 29377 \g  + 1803 - 16384 \g  \ln(2 \g  y_+^4) ]  \\
&\quad + \frac{\pi y_+^{13} }{262144} [ 5797 \g ^4+29900 \g ^3+579006 \g^2  - 4 \g (65536 \zeta (3)-311011)\\
&\quad + 328517  +16384 (23 \g ^2+61 \g +8 ) \ln  (2 \g  y_+^4 ) + 262144 \g  \ln ^2 (2 \g  y_+^4 ) ] \\
&\quad + \CO(y_+^{15}) \bigg) + \e^2 \bigg( \frac{3\pi }{32} y_+^5 + \frac{\pi y_+^7}{256} [ - 35 \g - 177 - 96 \ln (2 \g  y_+^4 ) ]  + \CO(y_+^9) \bigg) \\
&+ \e^4 \bigg( \frac{3\pi }{128} ( 1 - \g ) y_+^5 + \CO(y_+^7)  \bigg)  + \e^6 \bigg( \CO(y_+^5 ) \bigg) + \CO(\e^8) \bigg] . 
\end{split}
\end{equation}
\end{subequations}
We can check that the quantities \eqref{hbh_pert_E}--\eqref{hbh_pert_S} satisfy the first law \eqref{firstlaw} up to the following order in perturbation theory
\begin{equation}
\begin{split}
\delta E - ( T \delta S + 3 \mu \delta Q + 2 \O_H \delta J ) &= \d \g [ \CO(\e^0 y_+^{12} , \e^2 y_+^8, \e^4 y_+^6 , \e^6 y_+^4 )  ] \\
& \qquad \qquad + \d y_+ [ \CO(\e^0 y_+^{11} , \e^2 y_+^7 , \e^4 y_+^5 , \e^6 y_+^3) ] \\
&\qquad \qquad + \d \e [ \CO(\e^1 y_+^8  , \e^3  y_+^6 , \e^5 y_+^4) ] . 
\end{split}
\end{equation}
It can also be checked that in the static limit, where $\g \to 1$, the results here reduce to those of~\cite{Bhattacharyya:2010yg}. When $y_+=0$, we reproduce the thermodynamics of the soliton \eqref{EQ_soliton} (truncated at $\CO{(\e^6)}$). Finally, when $\e=0$, we recover the CLP thermodynamics \eqref{BH_charges}--\eqref{BH_thermo} (truncated at $\CO(y_+^6)$), for the 2-parameter CLP sub-family of black holes that merge (along a 2-dimensional surface) with the hairy black hole family: this is shown in Fig.~\ref{fig:onsetNumPert}.

At leading order \eqref{hbh_pert_E}-\eqref{hbh_pert_S} further agrees with the merger curve predicted by the thermodynamic model. To see this, we expand \eqref{MQJ_mix} in powers of $y_+$ to find
\begin{equation}
\begin{split}
\label{MQJ_mixSer}
\frac{E_\mix L}{N^2} &= \bigg( \frac{3 y_+^2}{2}  + \frac{3y_+^4}{4} ( 3 - 2 \g )   + \frac{y_+^6}{8}  (9 \g ^2-17 \g +8 )  + \CO(y_+^8)  \bigg) + \frac{3 Q_sL}{N^2}  ,  \\
\frac{Q_\mix L}{N^2} &= \bigg( \frac{y_+^2}{2} +  \frac{y_+^4}{4} (1-\g )  - \frac{y_+^6}{8} \g(1-\g)  + \CO(y_+^8)  \bigg) +\frac{Q_sL}{N^2}  , \\
\frac{J_\mix }{N^2} &= \sqrt{1-\g } \left( \frac{3 y_+^4}{4} + \frac{y_+^6}{8} (4-5 \g ) + \frac{y_+^8}{16} \g (7 \g  - 6 ) + \CO(y_+^{10}) \right) , \\
\mu_\mix  &= 1 , \qquad \O_\mix  L = \sqrt{1-\g} , \qquad T_\mix L = \frac{\g y_+}{\pi} + \frac{3}{8\pi} \g ( 1 - \g ) y_+^3 + \CO(y_+^5) , \\
\frac{S_\mix}{N^2} &= \pi y_+^3 + ( 1 - \g ) \left[ \frac{3\pi}{8} y_+^5 - \frac{\pi y_+^7}{128}  ( 23 \g + 9 ) + \CO(y_+^9) \right]  .
\end{split}
\end{equation}
Using \eqref{EQ_soliton}, we find that the prediction of the thermodynamic model matches the exact calculation \eqref{hbh_pert_thermo} to leading order in $y_+$.\footnote{The non-interacting model is expected to reproduce the leading $\g$ dependence of all thermodynamic quantities. This appears first at subleading order in $E_\mix$, $Q_\mix$, $\mu_\mix$, $S_\mix$, so these quantities match the exact results \eqref{hbh_pert_E}, \eqref{hbh_pert_Q}, \eqref{hbh_pert_mu}, and \eqref{hbh_pert_S} to subleading order in $y_+$.}

%%%%%%%%%%%%%%%%%%%%%%%%%%%%%%%%%%
\subsection*{BPS Limit?}

In this paper, we are particularly interested in the BPS limit
\begin{equation}
\begin{split}
\label{BPS_limit}
E \to E_\BPS = 3 Q + 2 J/L
\end{split}
\end{equation}
of the hairy black hole. If this limit exists and the limiting solution has a smooth horizon, we expect $\O_H L\to 1^-$, $\mu \to 1^+$, and $T \to 0^+$. To low orders in the perturbative expansion, this limit is equivalent to $\g \to \g_\BPS(\e,y_+)$ where
\begin{equation}
\begin{split}
\label{gamma_BPS}
\g_\BPS(\e,y_+) = y_+^2 - \frac{3}{4} y_+^4 + \frac{1}{2} y_+^6 + \CO(y_+^8) + \e^2 \left( - \frac{y_+^4}{8}  + \CO(y_+^6) \right) + \CO(\e^4 y_+^4, \e^6).
\end{split}
\end{equation}
To leading order in $y_+$, this is consistent with the thermodynamic model, which predicts that $\g_\BPS = 0$. Replacing $\g$ by $\g_\BPS$ as given in \eqref{gamma_BPS} into \eqref{hbh_pert_thermo}, we find
\begin{equation}
\begin{split}
\label{BPS_limit_pert}
E - E_\BPS &= \CO(y_+^{12},\e^2 y_+^8, \e^4 y_+^6 , \e^6 y_+^4 , \e^6 y_+^4,\e^8 )  , \\
\mu - 1 &= \CO(y_+^8,\e^2 y_+^8, \e^4 y_+^6,\e^8) , \\
\O_H L - 1 &=  \CO(y_+^8 , \e^2 y_+^6 , \e^4 y_+^4 , \e^6 y_+^2 ,\e^8) , \\
TL &= \CO(y_+^7 , \e^2 y_+^5 , \e^4 y_+^3 , \e^6 y_+ ,\e^8 ) . \\
\end{split}
\end{equation}
This suggests that the BPS limit of the hairy black hole we constructed in perturbation theory is a regular solution. Moreover, the entropy of this limiting solution is
\begin{equation}
\begin{split}
\label{BPS_limit_pert_S}
\frac{S}{N^2} &= \bigg( \pi y_+^3 + \frac{3\pi}{8} y_+^5 - \frac{9\pi y_+^7}{128} + \frac{27\pi y_+^9}{1024} + \CO(y_+^{11} )  \bigg)  + \e^2 \bigg( \frac{3\pi y_+^5}{32} \\
&\quad - \frac{3\pi y_+^7}{256} [ 67 + 32 \ln (2y_+^6) ]  + \CO(y_+^9) \bigg) + \e^4 \left( \frac{3\pi y_+^5}{128} + \CO(y_+^7) \right) + \CO ( \e^6 y_+^5 , \e^8 ). 
\end{split}
\end{equation}
Since the entropy is finite in the BPS limit, the perturbative construction predicts that the BPS limit is, in fact, a supersymmetric hairy black hole! Our perturbative analysis, therefore, seems to \emph{confirm} the prediction of~\cite{Bhattacharyya:2010yg} and the numerical analysis of~\cite{Markeviciute:2018yal, Markeviciute:2019exa}.

There is, however, an issue with this analysis. As mentioned in the discussions of  \eqref{orderGamma1}--\eqref{orderGamma2}, our perturbative construction is only valid when $\g = \CO(1)$ when $y_+$ is small. However, since from \eqref{gamma_BPS} one has $\g_\BPS = \CO(y_+^2)$, it is not clear that we can actually trust \eqref{BPS_limit_pert} and \eqref{BPS_limit_pert_S}. Furthermore, even if the limit $\g \to \g_\BPS$ is valid, it is not clear that \eqref{BPS_limit_pert} will continue to hold to higher orders in the perturbative expansion. Indeed, we will perform a more careful analysis in Section~\ref{sec:MainResults} and, to much surprise, we will conclude that the abovementioned prediction does \emph{not} hold! To be more precise, the careful analysis of Section~\ref{sec:MainResults} will show that as we take $\g \to \g_\BPS$ keeping $\e$ and $y_+$ fixed, the hairy black hole reaches the extremality bound \emph{before} the BPS bound. This finding is further supported in section~\ref{sec:numerics}, where we gather strong numerical evidence that in the BPS limit (i.e., $E \to E_\BPS$ keeping $Q$ and $J$ fixed), our numerical solutions approach $E\to  E_\BPS$, $T \to 0$, $\mu\to 1$, and $\O_H L\to 1$ but their entropy approaches {\it zero}.\footnote{The BPS limit considered for the analytical solutions ($\g \to \g_\BPS$ keeping $\e$ and $y_+$ fixed) is different from the one considered for the numerical solutions ($E \to E_\BPS$ keeping $Q$ and $J$ fixed). Consequently, the limiting solutions are different.}

The analytic and numerical analyses discussed above attempt to find supersymmetric hairy black holes by taking limits of non-supersymmetric solutions. In Section~\ref{sec:susy-analysis}, we will attempt to find supersymmetric hairy black holes directly by solving the BPS equations. In agreement with our numerical findings of Sections~\ref{sec:numerics}--\ref{sec:MainResults}, we will again find no evidence that such a regular hairy supersymmetric black hole exists. This raises the question: why is the perturbative analysis of this Section~\ref{sec:hbh-pert} (and associated non-interacting thermodynamic model) failing to deliver the correct result? This fundamental question will be fully addressed in Section~\ref{sec:PerturbativeFailing}, and we postpone further discussion on this issue till then. The reader should, however, be already aware that at the end of the day, the non-interacting thermodynamic model and the perturbative results will prove to be a rather good approximation whenever $(E, Q, J)$ are all parametrically small as long as the temperature of the system is not too close to $T=0$ (in Section~\ref{sec:conclusion} we will discuss how we can eventually fix the model to provide a good description also for arbitrarily small $T$).

%%%%%%%%%%%%%%%%%%%%%%%%%
\section{Hairy black holes: numerical solutions}
\label{sec:numerics}
%%%%%%%%%%%%%%%%%%%%%%%%%

In Section~\ref{sec:hbh-pert}, we found the thermodynamic properties of hairy black holes with small $(E, Q, J)$ using perturbation theory with a matching asymptotic expansion. We concluded that the leading order thermodynamics of the perturbative analysis agrees with the one obtained from the non-interacting thermodynamic model of Section~\ref{sec:toymodel} and~\cite{Bhattacharyya:2010yg}. Both analyses predict that the BPS limit of hairy black holes is a two-parameter family of regular supersymmetric hairy black holes first conjectured to exist in~\cite{Bhattacharyya:2010yg}. The numerical analysis performed in~\cite{Markeviciute:2018yal, Markeviciute:2018cqs} found evidence of this conjecture of~\cite{Bhattacharyya:2010yg}. In this section, we solve the coupled system of non-linear ODEs \eqref{EoM} for the field ansatz\"e  \eqref{3Q:ansatz} subject to the boundary conditions discussed in Section~\ref{sec:ansatz} to find exactly (to numerical accuracy) the hairy black hole solutions of \eqref{action}. The analysis of this section captures hairy black holes with any $(E, Q, J)$. For small $(E, Q, J)$, the numerical solutions will test the validity of the perturbative construction of Section~\ref{sec:hbh-pert}.

As briefly stated above, the hairy black holes of the theory \eqref{action} were already studied with some detail in~\cite{Markeviciute:2018yal, Markeviciute:2018cqs}. In particular, the transitions in the qualitative behaviour of the solutions crudely conjectured from the thermodynamic model of Section~\ref{sec:toymodel} were identified and well documented in~\cite{Markeviciute:2018yal, Markeviciute:2018cqs}. In this sense, our work will simply reinforce and complement the findings and discussions of~\cite{Markeviciute:2018yal, Markeviciute:2018cqs}. The true novelty of our work concerns the exploration of the $T\to 0$ limit of the hairy black holes (both numerically and analytically; the latter done in Section~\ref{sec:hbh-pert}).~\cite{Markeviciute:2018yal, Markeviciute:2018cqs} managed to find hairy black holes with temperature as low as $TL = 5 \times 10^{-3}$ and the low-temperature data gathered in these references strongly suggested that the entropy of the hairy black holes would reach a {\it finite} value at $T=0$ while also having $\mu=1$, $\O_H L =1$ and $E=E_\BPS$. That is to say, ~\cite{Markeviciute:2018yal, Markeviciute:2018cqs} found evidence that, as originally conjectured in the non-interacting thermodynamic model of~\cite{Bhattacharyya:2010yg},  the zero temperature limit of the hairy black holes of \eqref{action} should be a two-parameter family of (regular) supersymmetric hairy black holes. When the amplitude of the scalar condensate vanishes, such hairy supersymmetric solutions reduce to the `bald' Gutowski-Reall black hole. The latter is the red square in Fig.~\ref{figJ0p05:phasediagram} and, if the BPS solutions of~\cite{Bhattacharyya:2010yg, Markeviciute:2018yal, Markeviciute:2018cqs} are regular, then supersymmetric hairy black holes exist along the BPS curve of Fig.~\ref{figJ0p05:phasediagram} (at least) above the GR red square, and this would solve a long-standing puzzle regarding the existence of supersymmetric black hole solutions with $\Delta_{\hbox{\tiny KLR}}\neq 0$ (defined in \eqref{charge_constraint}).

It is in this context that the main motivation of our work emerges. In the previous section, we started by finding perturbative solutions with $T\neq 0$ and then studied their $T\to 0$ limit (which turns to agree with the conclusions of~\cite{Bhattacharyya:2010yg, Markeviciute:2018yal, Markeviciute:2018cqs}). To have bullet-proof evidence for the expectations of~\cite{Bhattacharyya:2010yg, Markeviciute:2018yal, Markeviciute:2018cqs}, in the present section, we aim to use enhanced numerical methods to find hairy black holes with even colder horizons than those of~\cite{Markeviciute:2018yal, Markeviciute:2018cqs}, and thus test, to our best limit, the conjecture of originally proposed in~\cite{Bhattacharyya:2010yg}.\footnote{In Section~\ref{sec:susy-analysis}, we will add a final test to the expectations of~\cite{Bhattacharyya:2010yg, Markeviciute:2018yal, Markeviciute:2018cqs} by analysing the BPS equations of the system.} We will be able to reach temperatures as low as $TL \sim 10^{-7}$ (i.e. 4 orders of magnitude lower than~\cite{Markeviciute:2018yal, Markeviciute:2018cqs}). We will gather strong evidence that below $T L\sim 10^{-3}$, the behaviour of the system changes. The entropy acquires a temperature dependence such that as $T\to 0$, the system approaches a {\it zero entropy} solution (with $\mu \to 1$ and $\O_H L\to 1$ and $E=E_\BPS$). Therefore, at the end of the day, against the expectations of~\cite{Bhattacharyya:2010yg, Markeviciute:2018yal, Markeviciute:2018cqs}, the BPS limit is {\it singular} and there should {\it not} exist a regular BPS hairy black hole solution in the sector of the theory we study.   

When solving the EoM \eqref{EoM} numerically, the boundary conditions discussed in Section~\ref{sec:ansatz} can be imposed naturally if we use the residual gauge freedom \eqref{residualGaugeFreed} to work in the gauge where $w|_{r\to\infty}=0$ (i.e. $w_\infty=0$), redefine the fields in the ansatz\"e \eqref{3Q:ansatz} as\footnote{Here, to rewrite $A_t$ in terms of $q_3$, $q_6$, and $q_7$, we have used the fact that when $\Phi \neq 0$, the EoM \eqref{EoM} implies that $A_t(r) + w(r) A_\psi(r)$ vanishes on the horizon.}
\begin{equation} \label{3Q:FieldsRedef}
\begin{split}
& f= \frac{r^2}{L^2}\left( 1-\frac{r_+^2}{r^2} \right) q_1 , \qquad 
 g= q_1\,q_2 , \qquad h= q_4 , \qquad 
w = - \frac{1}{L}\,\frac{r_+^4}{r^4}\, q_6 , \\
& A_t= \left( 1-\frac{r_+^2}{r^2} \right) q_3 + \frac{r_+^6}{r^6} q_6\,q_7 ,  \qquad   
A_\psi = L\,\frac{r_+^2}{r^2}\, q_7,  \qquad   
\Phi= 2 \, \frac{r_+^2}{r^2} \,q_5 \sqrt{2+\frac{r_+^4}{r^4}\,q_5^2} ,
\end{split} 
\end{equation}
and look for solutions $q_j$, ($j=1,2,\cdots 7$) that are everywhere smooth. Note that $g= q_1 \,q_2$ accounts for the fact that $g_{rr}=g/f\sim q_2$ in \eqref{3Q:ansatz} and the peculiar redefinition of $\Phi$ in terms of $q_5$ was introduced to avoid square root terms of the form $\sqrt{4+\Phi^2}$ in the EoM. For the numerical search of the hairy solutions, it is also convenient to introduce the new time and radial coordinates  and dimensionless horizon radius,
\begin{equation}\label{3Q:defY}
t= L\, T, \qquad r=\frac{r_+}{1-y^2}\,;\qquad y_+=\frac{r_+}{L},
\end{equation}
where the compact radial coordinate ranges  between $y=0$ (i.e., $r=r_+$) and $y=1$ (i.e., $r\to \infty$). The coordinates $T$ and $y$ introduced here should not be confused with similar coordinates introduced in Section~\ref{sec:hbh-pert}.

We can now specify the boundary conditions for the auxiliary fields $q_j$ at the asymptotic boundary. Demanding that our solutions are asymptotically AdS$_5$ at $y=1$ (see \eqref{asymp_exp} with $w_\infty=0$) requires that $q_{1}(1)=1=q_{2}(1)=1$. The EoM then require that $q_{4}(1)=1$.  Later, in \eqref{3Q:EQJ}, we will find that $Q$ is a function of $q_{3}(1)$ and $q_{3}'(1)$, $Q=\frac{N^2}{L}\frac{1}{4}y_+^2(q_{3}'+2 q_{3})|_{y=1}$. To introduce $Q$ in our numerical code as an input parameter (that will allow us to run lines of constant $Q$), we thus use this condition to give a mixed boundary condition for $q_3$. Finally, the EoM require that $q_5$ and $q_6$ also satisfy mixed boundary conditions, while $q_7$ must obey a Neumann boundary condition. Altogether, we impose the following boundary conditions at the asymptotically AdS$_5$ boundary ($y=1$):
\begin{equation} \label{3Q:BCqI}
\begin{split}
&q_1|_{y=1}=1,\qquad q_2 |_{y=1}=1   ,\qquad   
q_3'|_{y=1} =-2 q_3 |_{y=1}+\frac{4 Q}{y_+^2}\,\frac{L}{N^2},
\qquad q_4 |_{y=1}=1, \\ 
&  q_5'|_{y=1}=\frac{2}{y_+^2} q_3^2 \, q_5  |_{y=1} ,\qquad  q_6'  |_{y=1} = \frac{1}{y_+^2} q_7  (q_3'+2 q_3 )  |_{y=1} ,\qquad  q_7' |_{y=1}=0\,. 
\end{split} 
\end{equation}
At the horizon ($y=0$), the boundary conditions derived from the EoM are that all $q_j$'s ($j=1,\cdots,7$) obey Neumann boundary conditions (thus, all $q_j|_{y=0}$ are free parameters to be determined). However, we can introduce the angular velocity at the horizon, $\O_H$, in our numerical code as a boundary condition (so that we can directly look for constant $\O_H$ solutions should we wish to do so). Therefore, we impose $q_6|_{y=0}=\O_H$ as a boundary condition to explicitly introduce the input parameter $\O_H$ in the problem. Altogether, at  the horizon, we impose the boundary conditions
\begin{equation} \label{3Q:BCqH}
q'_j|_{y=0}=0 \:\: \hbox{for}\:\: j=1,2,3,4,5,7\,;\qquad q_6|_{y=0}=\O_H L \,.
\end{equation}
We now discuss our numerical strategy to find the nonlinear solutions of our boundary-value problem. At the asymptotic boundary, we have 8 free UV parameters. Indeed, after introducing the field redefinitions and imposing the boundary conditions above, a Taylor expansion of the EoM about $y=1$ reveals that $\{ q_1'', q_3, q_3', q_4'', q_5, q_6, q_7\}|_{y=1}$ are 7 unknown constants \`a priori. On the other hand, at the horizon, we have 7 free IR parameters. Indeed, a Taylor expansion of the EoM about $y=0$ finds that $q_j |_{y=0}$ are unknown before solving the boundary value problem. We have $7+7=14$ free parameters for an ODE system of order 12 (2 first-order EoM plus 5 second-order EoM).  It follows that our black hole solutions depend on $2$ parameters plus the dimensionless horizon radius $y_+$ (which defines the inner boundary of the problem), \emph{i.e.} a total of 3 parameters. We can take these parameters to be, e.g. the dimensionless electric charge $Q L/N^2$, dimensionless angular momentum $J/N^2$ and the dimensionless radius $y_+=r_+/L$ (the latter is related to the temperature and entropy of the solutions; see \eqref{3Q:TOS}). 

 In practice, we do the following to look for solutions with a given $Q$ and $J$. 
 Recall that the charge $Q L/N^2$ is introduced in the problem in the boundary conditions \eqref{3Q:BCqI}. 
 Hence, we can fix $Q$ by simply giving it as an input parameter. Solving the nonlinear boundary-value problem in these conditions necessarily returns solutions with the given $Q$.  We have yet to introduce the angular momentum. 
 Later, in \eqref{3Q:EQJ}, we will find that  $J=N^2\,\frac{1}{2}y_+^4 q_{6}|_{y=1}$. If we also give $J$ as an input parameter in our code, we can view this as a normalisation condition for the value of the function $q_6$ at $y=1$. That is to say, if for a given $Q$ and $J$ (input parameters), we solve the seven coupled ODEs of our boundary-value problem {\it simultaneously} together with the additional normalisation condition $q_{6}|_{y=1}=\frac{2}{y_+^4}\frac{J}{N^2}$, we can find the seven unknown functions $q_j(y)$ {\it including} the value of the angular velocity $\O_H L =q_6|_{y=0}$ that ultimately allows that $q_6$ obeys the normalisation condition that defines the angular momentum of the solution.\footnote{We could have also followed a different strategy. Indeed, much like we did with $Q$, we could have introduced $J$ in the problem by imposing the boundary condition  $q_{6}|_{y=1}=\frac{2}{y_+^4}\frac{J}{N^2}$ in \eqref{3Q:BCqI}, i.e.  instead of the mixed boundary condition given for $q_6$ in  \eqref{3Q:BCqI}. If we were to follow this route, to have a well-posed problem, we would then have to replace the IR Dirichlet boundary condition for $q_6$ in \eqref{3Q:BCqH} -- which introduces the horizon angular velocity $\O_H$ -- by the Neumann boundary condition $q'_6|_{y=0}=0$. In this case, we would introduce both $Q$ and $J$ as input UV parameters through the boundary conditions at $y=1$ (and, in the end, we would obtain $\O_H$ by simply reading the value of $q_6|_{y=0}$).} 
 
We solve our nonlinear boundary-value problem with the additional normalisation condition using a Newton-Raphson algorithm. For the numerical grid discretisation, we use a pseudospectral collocation with a Chebyshev-Lobatto grid, and the Newton-Raphson linear equations are solved by LU decomposition.  
These methods are reviewed and explained in detail in the review~\cite{Dias:2015nua} and used in a similar context e.g. in~\cite{Dias:2015pda, Dias:2016eto, Dias:2017uyv, Dias:2017opt, Bena:2018vtu, Bea:2020ees,Dias:2024vsc}. Our solutions have analytical polynomial expansions at all the boundaries of the integration domain. Thus, the pseudospectral collocation guarantees that the numerical results have exponential convergence with the number of grid points. We further use the first law to check our numerics. In the worst cases, our solutions satisfy these relations with an error smaller than $0.1\%$. As a final check of our full nonlinear numerical results, we will compare them against the perturbative expansion results of Section~\ref{sec:hbh-pert}. 

As usual, to initiate the Newton-Raphson algorithm, one needs an educated seed. The hairy black holes merge with the CLP (i.e. the CLP) black holes when the condensate $\Phi$ (i.e. $q_5$) vanishes. Therefore, it is natural to expect that the CLP solution with a small $q_5$ perturbation can be used as a seed for the solution near the merger. 

To scan the 3-dimensional parameter space of hairy black holes, we can fix $J/N^2$ and $Q L/N^2$ and run the numerical code for several values of $y_+$. Keeping the same  $J /N^2$, we can repeat the exercise for other values of  $Q L/N^2$. Of course, we can also complete similar runs for other values of  $J /N^2$. Once we have the numerical solutions $q_j(y)$, the thermodynamic quantities are read straightforwardly from the holographic renormalization expressions \eqref{ansatz_charges} and \eqref{ansatz_potentials} (see Appendix of~\cite{Dias:2022eyq} for a derivation of these expressions). In terms of the functions $q_j$, these are given by
\begin{subequations}
\begin{equation}
\begin{split}
\label{3Q:EQJ}
E &=\frac{N^2}{L}\,\frac{y_+^2 }{16}\left[ 
 9 + y_+^2 \left( 9-\frac{3}{2}\,q_1''(1)+\frac{1}{2}\,q_4''(1) \right) \right],
\\
Q &=  \frac{N^2}{L}\,\frac{y_+^2}{4} \Big[q_3'(1)+2q_3(1)\Big],
 \qquad J = N^2\, \frac{y_+^4}{2} q_6(1) , 
\end{split}
\end{equation}
and
\begin{equation}
\begin{split}
\label{3Q:TOS}
T=\frac{1}{L}\, \frac{y_+}{2\pi} \frac{\sqrt{q_1(0)}} {\sqrt{q_2(0)}}, \qquad  \O_H=\frac{1}{L} \,q_6(0), \qquad S= N^2 \pi y_+^3\sqrt{q_4(0)}\,.
\end{split}
\end{equation}
\end{subequations}

\begin{figure}[H]
\centering
\includegraphics[width=0.325\textwidth]{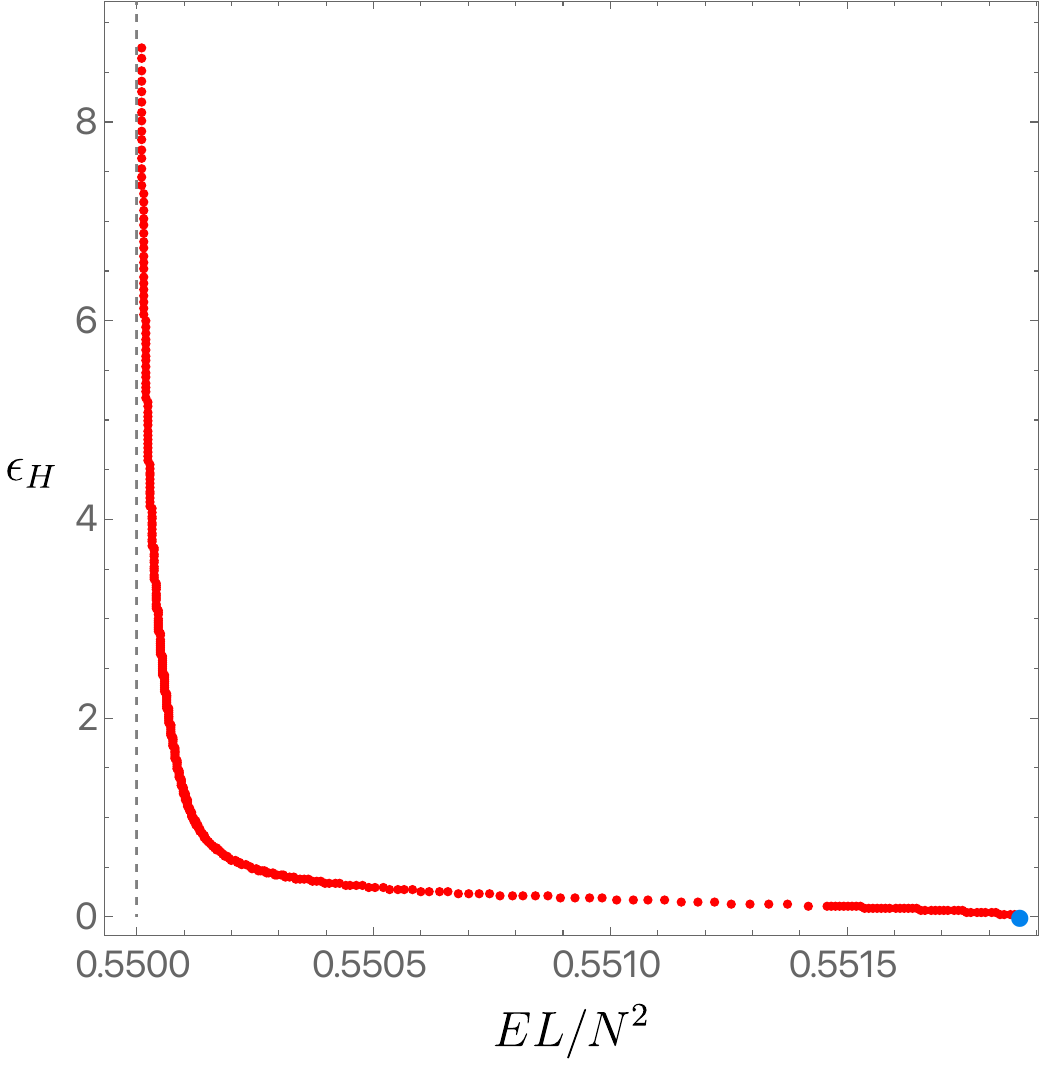}
\includegraphics[width=0.325\textwidth]{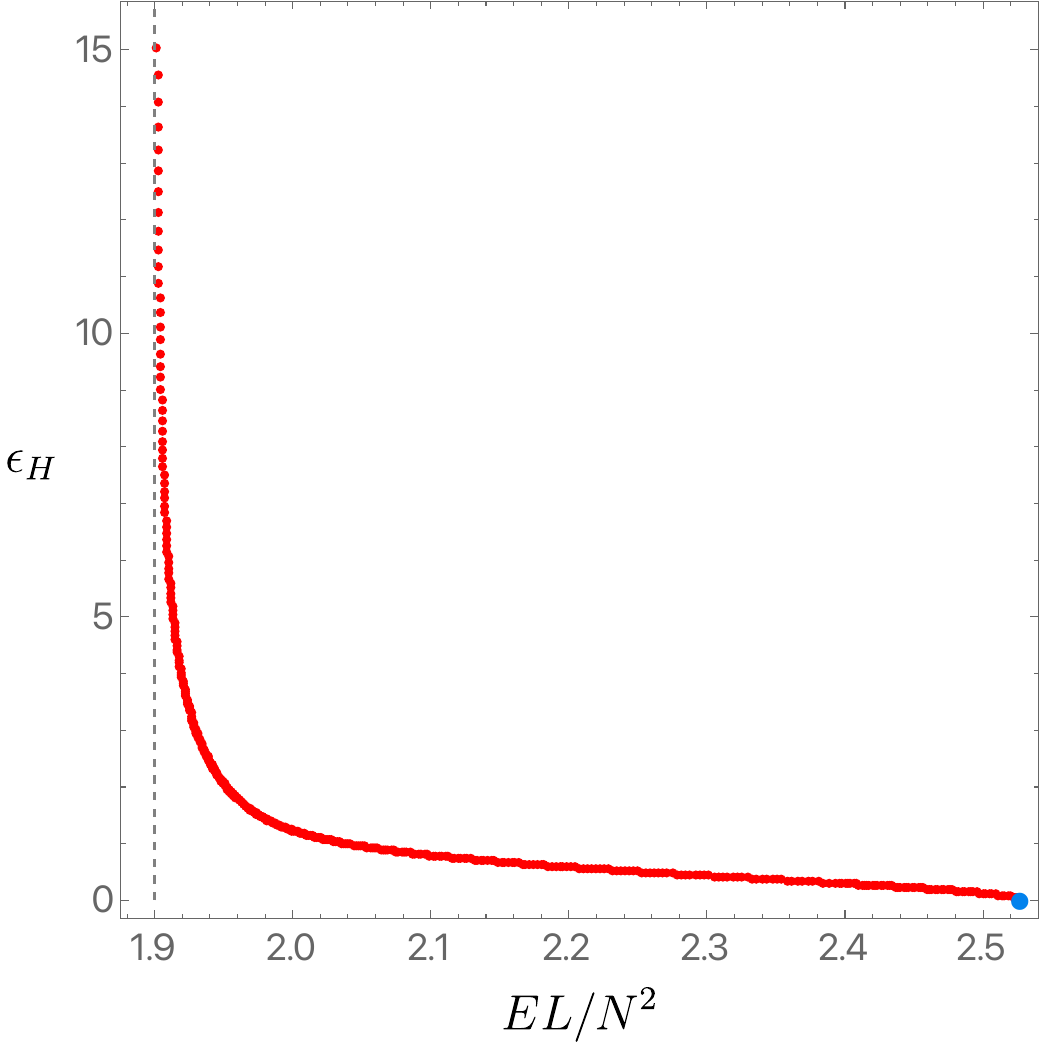}
\includegraphics[width=0.325\textwidth]{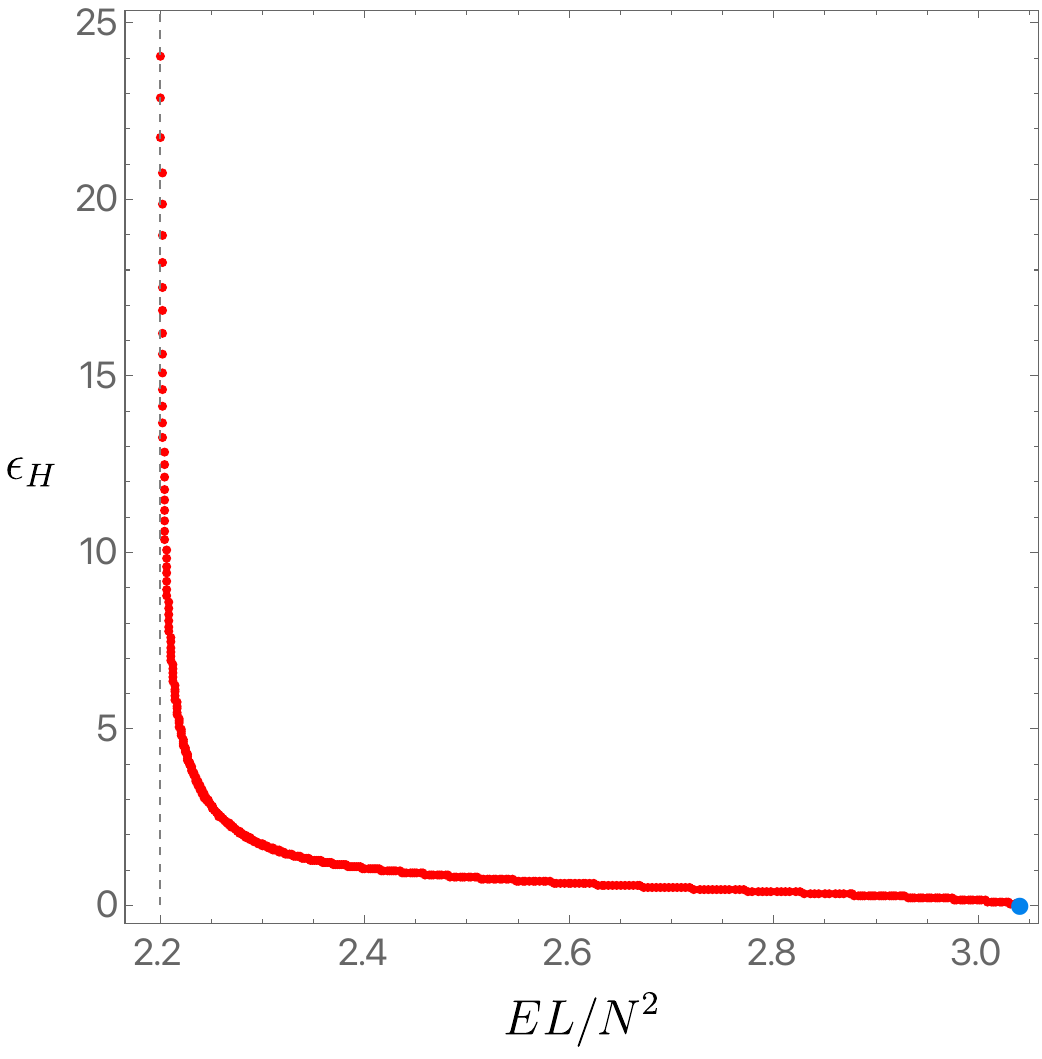}
\vspace{-0.25cm}
\caption{Value of the charged scalar field at the horizon as a function of energy. Here and in Figs.~\ref{fig:epsilonT}--\ref{fig:entropyT}, these plots describe black holes with $ J /N^2=0.05$ and $Q L/N^2=0.15<Q_c(J)$ (left), $Q L/N^2=0.6\gtrsim Q_c(J)$ (middle), $Q L/N^2=0.7 > Q_c(J)$ (right). The blue disk is the merger point between the hairy and the CLP families and the dashed vertical line (when present) signals $E=E_\BPS$. }
\label{fig:epsilonE}
\end{figure}

\begin{figure}[H]
\centering
\includegraphics[width=0.325\textwidth]{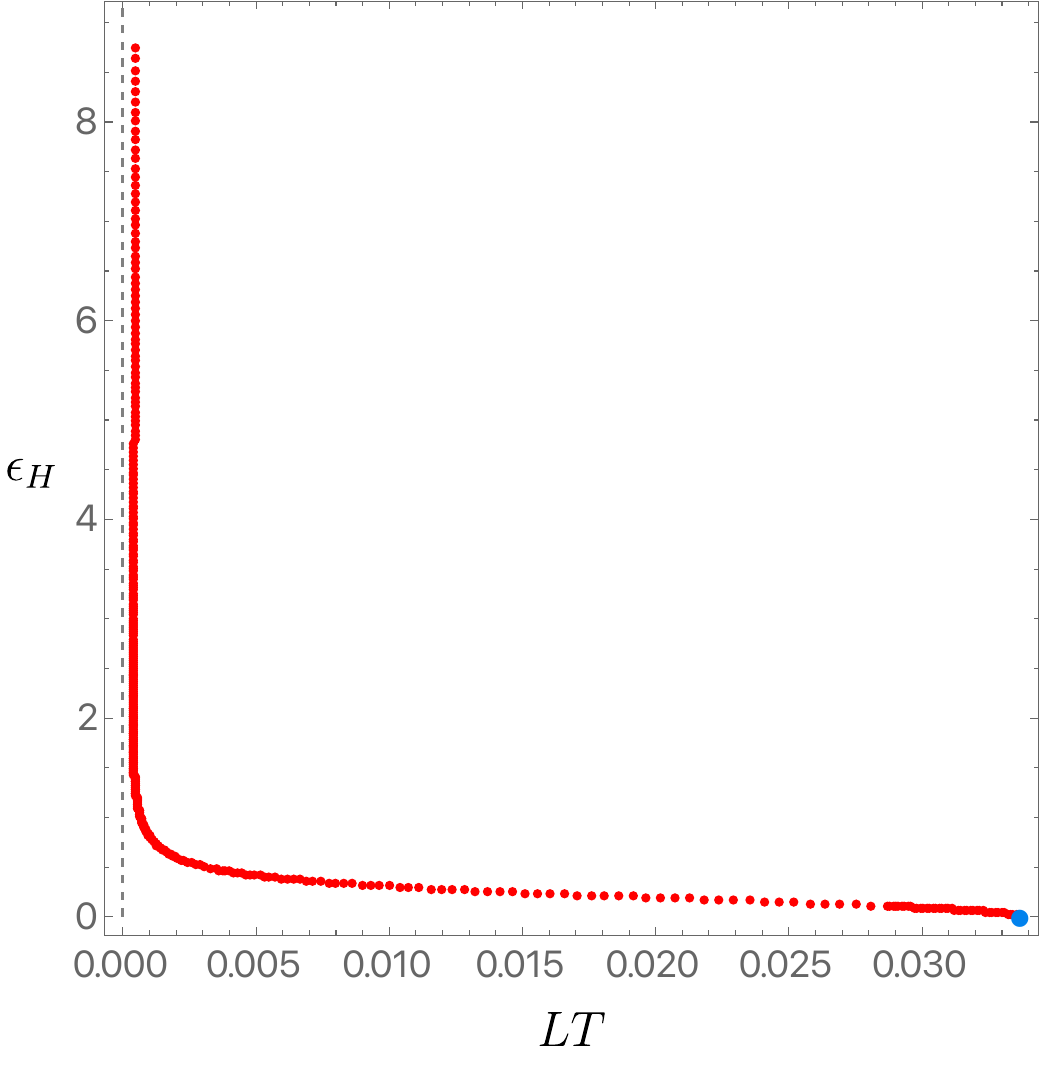}
\includegraphics[width=0.325\textwidth]{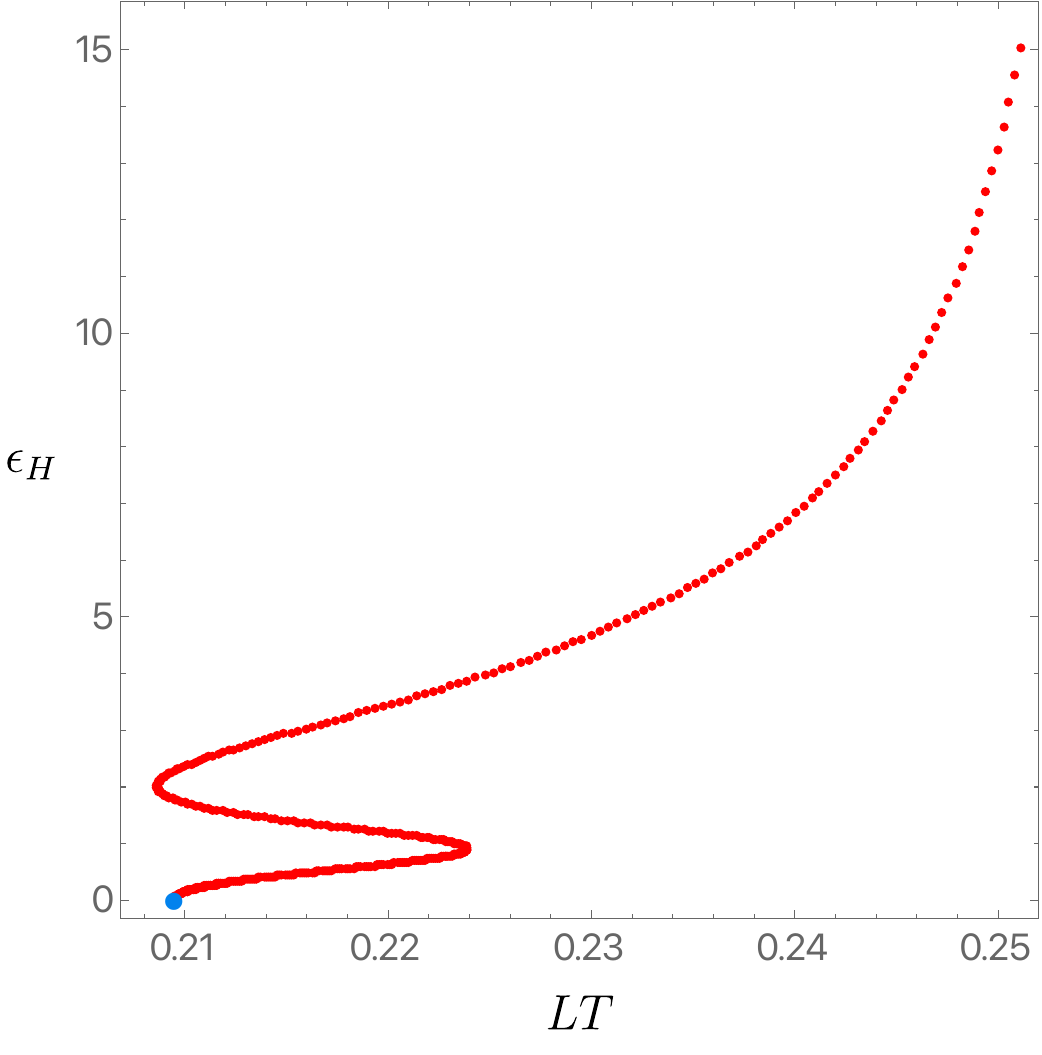}
\includegraphics[width=0.325\textwidth]{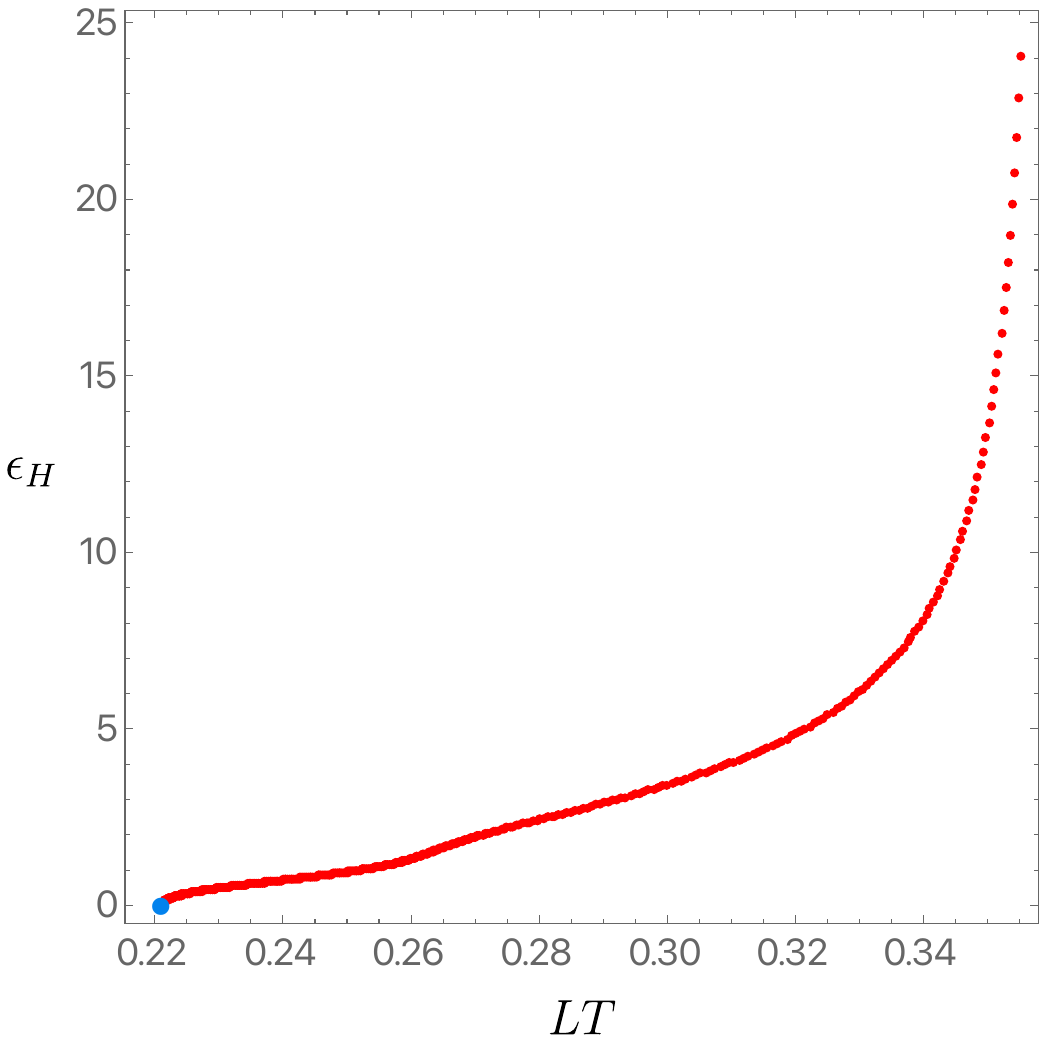}
\vspace{-0.25cm}
\caption{Value of the charged scalar field at the horizon as a function of temperature.}  
\label{fig:epsilonT}
\end{figure}

\begin{figure}[H]
\centering
\includegraphics[width=0.325\textwidth]{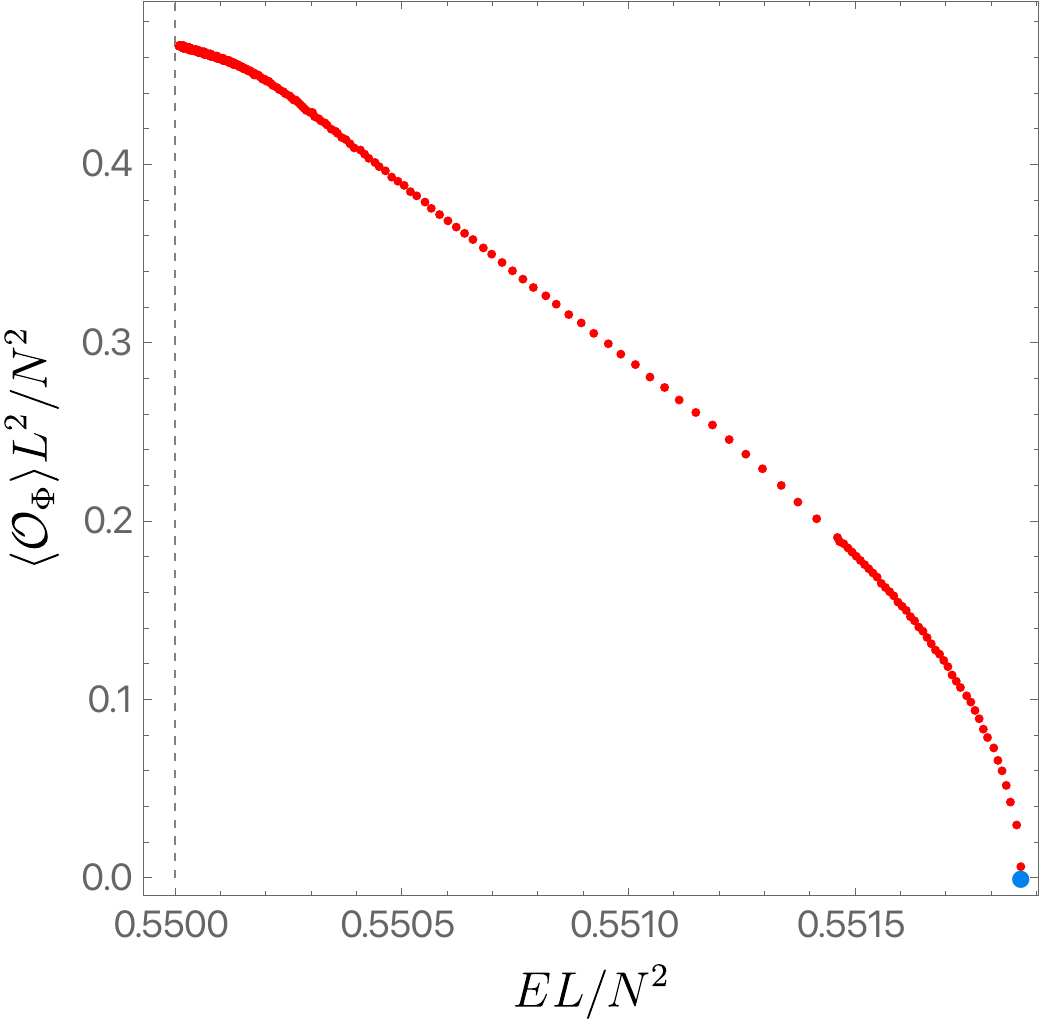}
\includegraphics[width=0.325\textwidth]{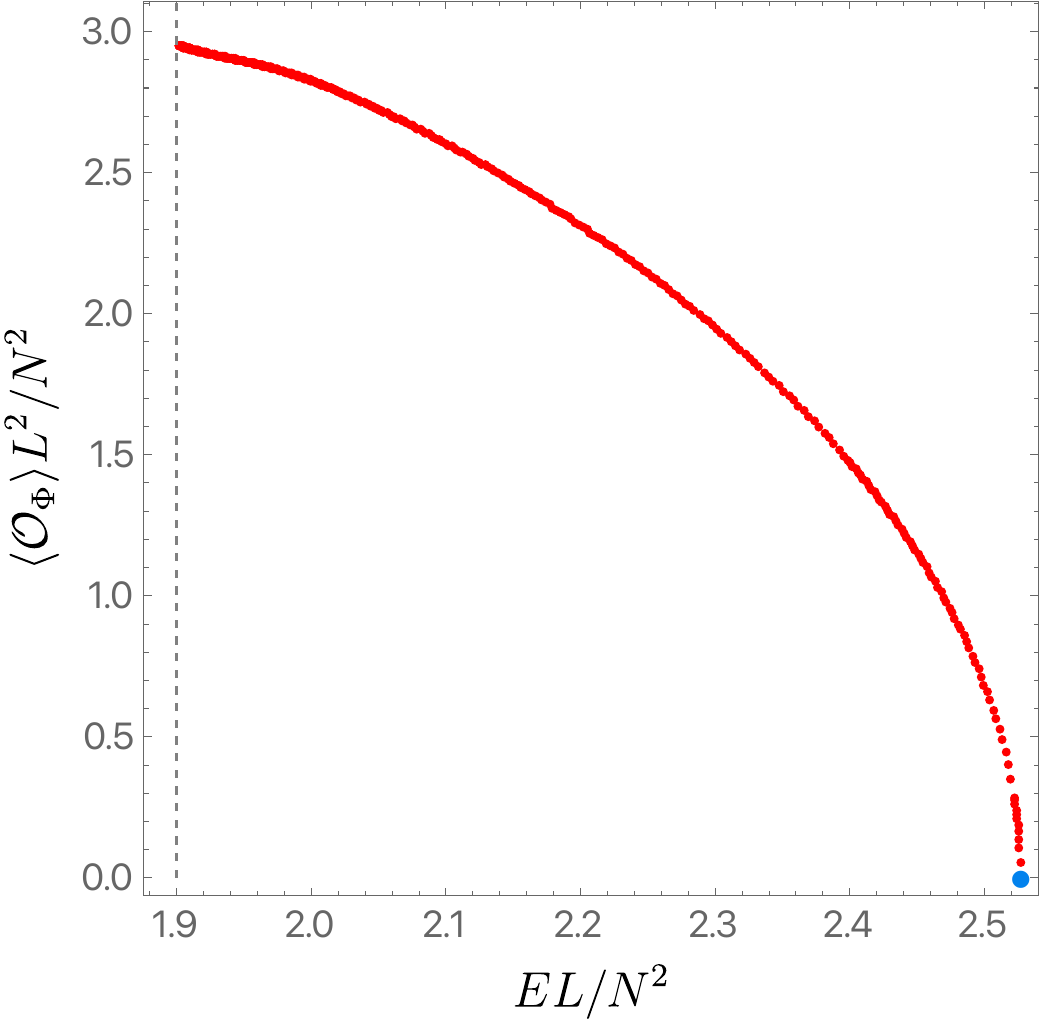}
\includegraphics[width=0.325\textwidth]{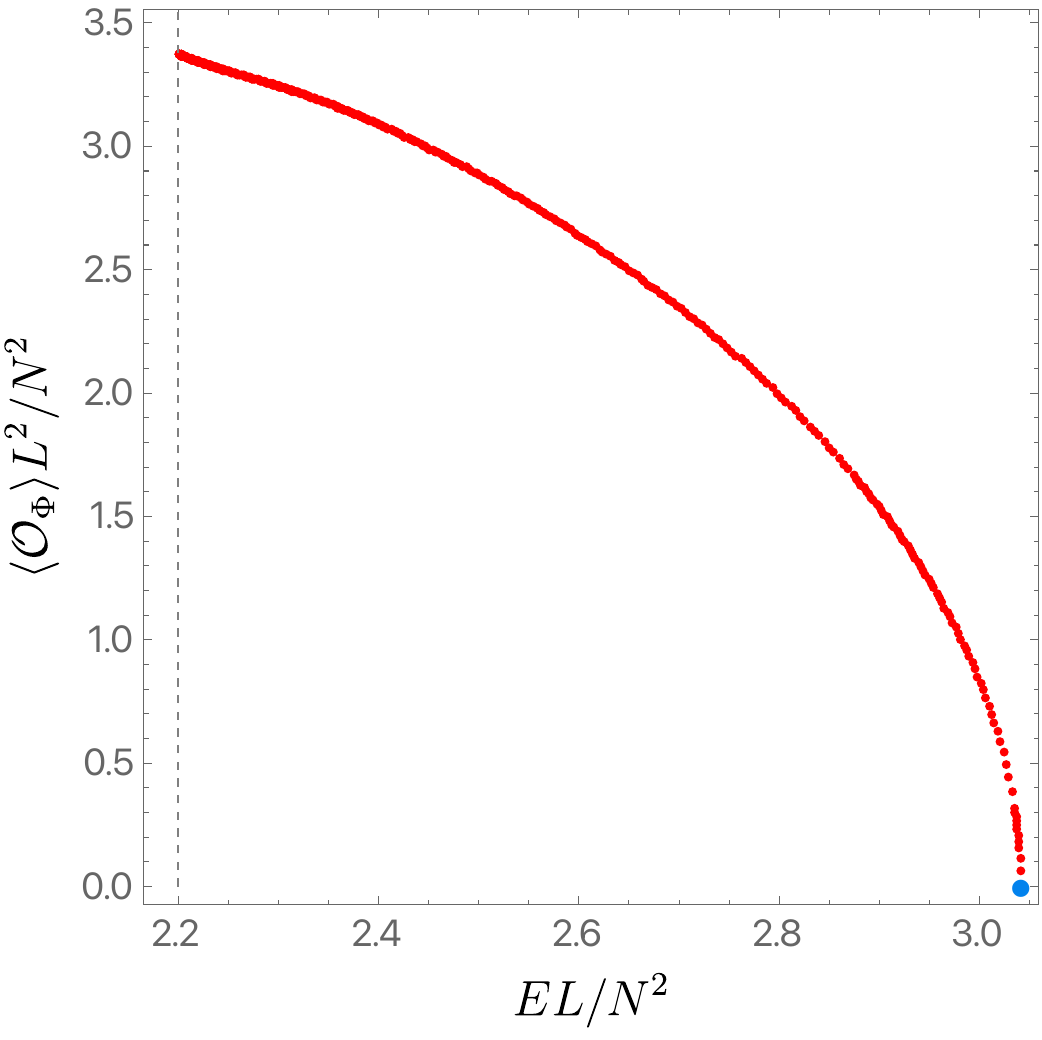}
\vspace{-0.25cm}
\caption{VEV of the operator dual to charged scalar field as a function of energy.}
\label{fig:vevE}
\end{figure}

\begin{figure}[H]
\centering
\includegraphics[width=0.325\textwidth]{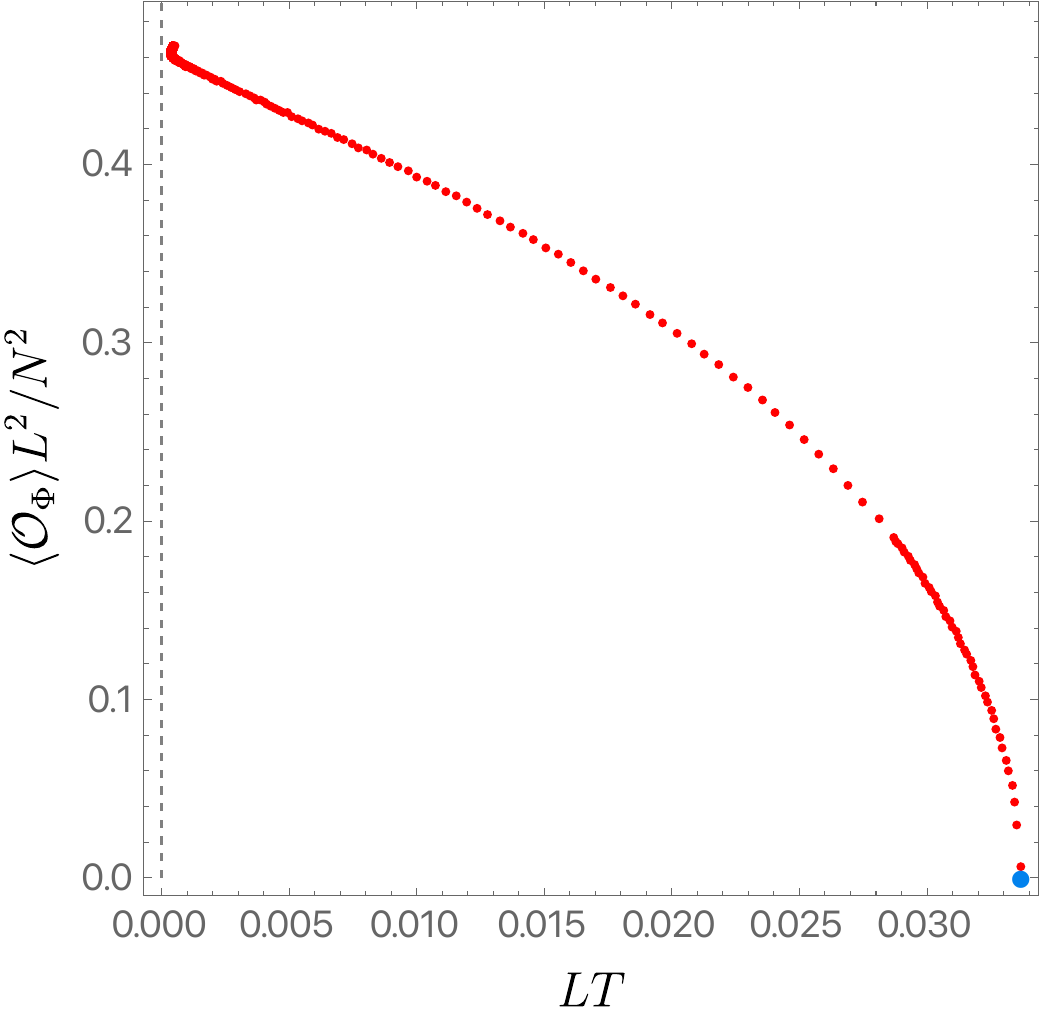}
\includegraphics[width=0.325\textwidth]{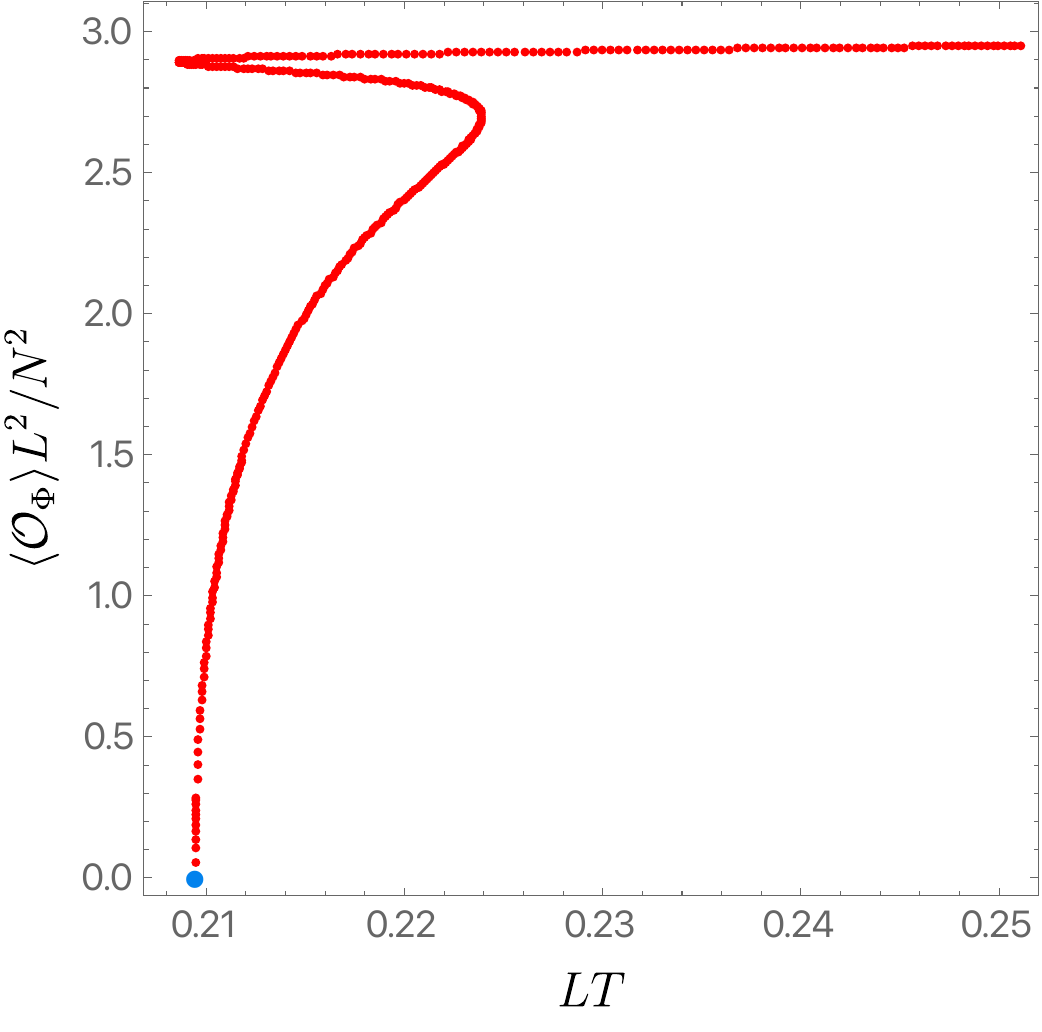}
\includegraphics[width=0.325\textwidth]{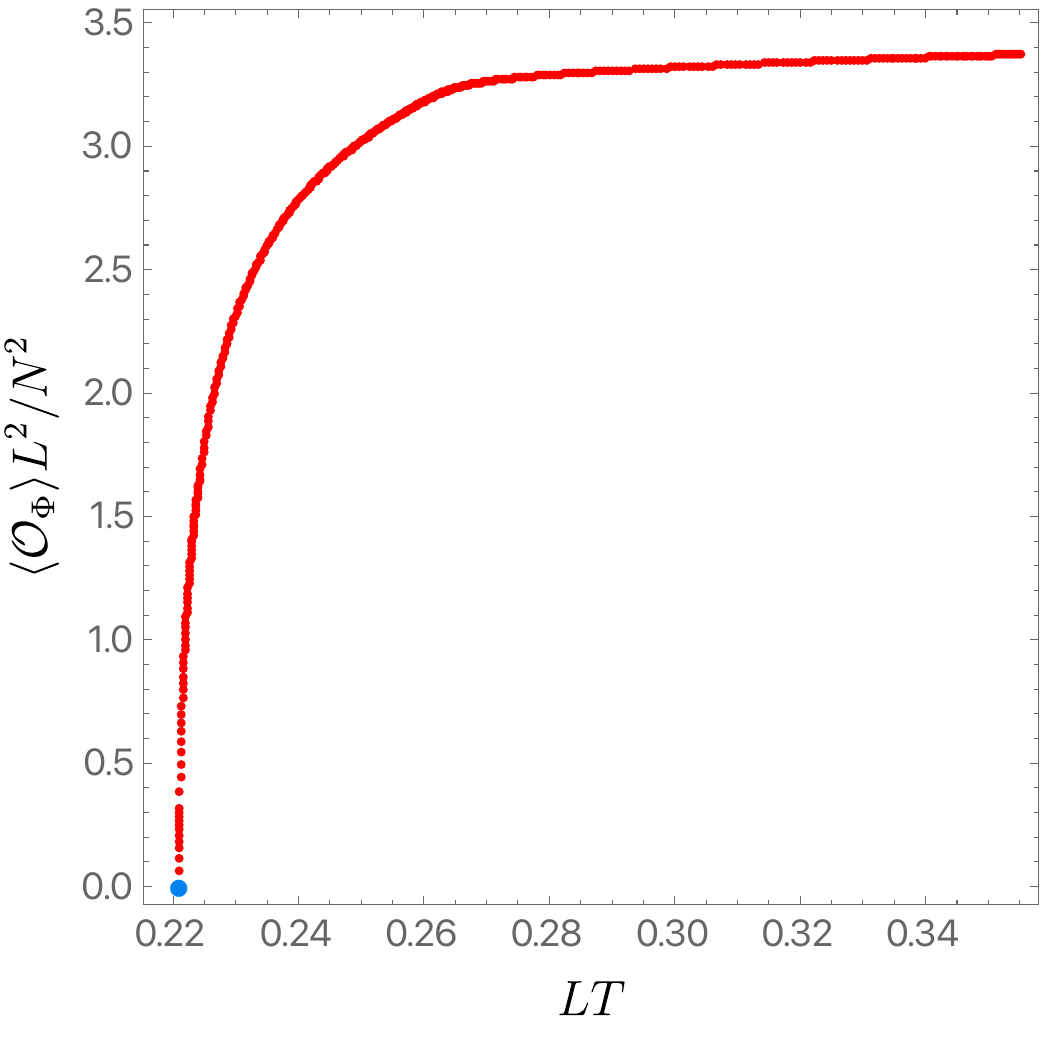}
\vspace{-0.25cm}
\caption{VEV of the operator dual to charged scalar field as a function of temperature.}  
\label{fig:vevT}
\end{figure}

\begin{figure}[H]
\centering
\includegraphics[width=0.325\textwidth]{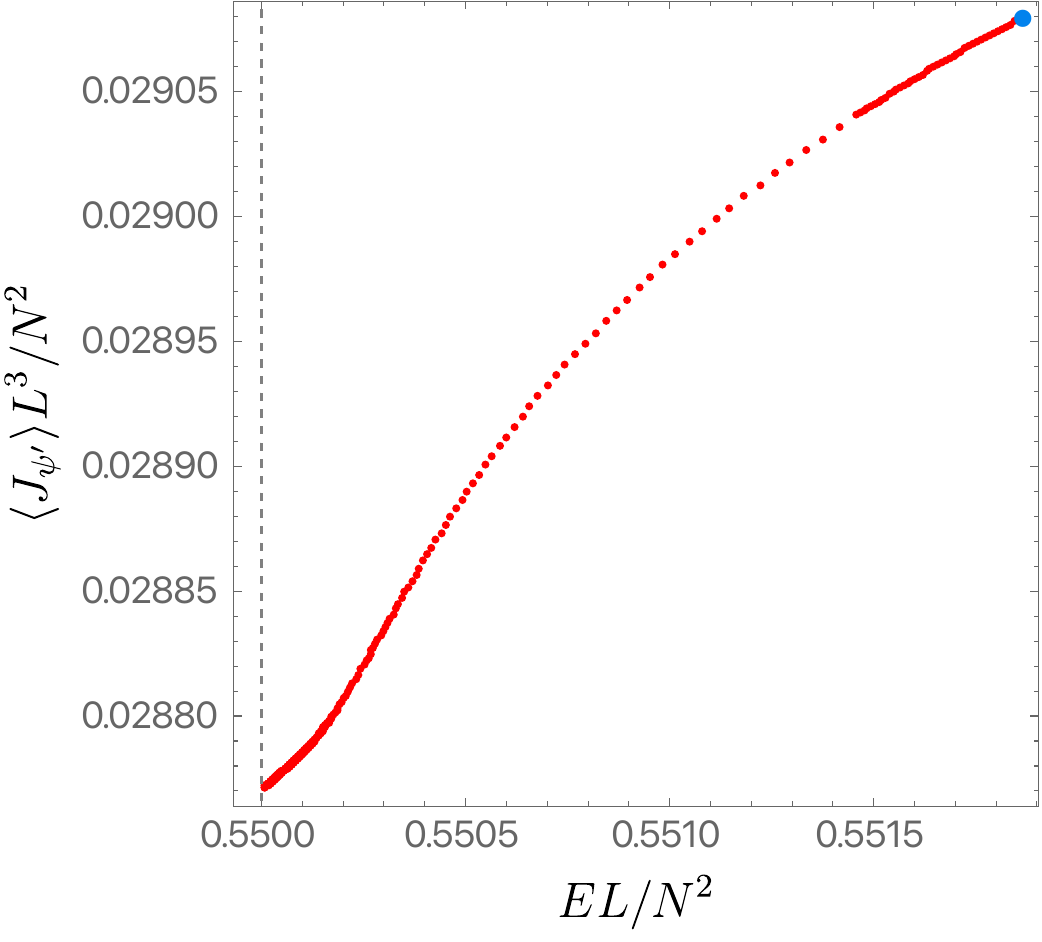}
\includegraphics[width=0.325\textwidth]{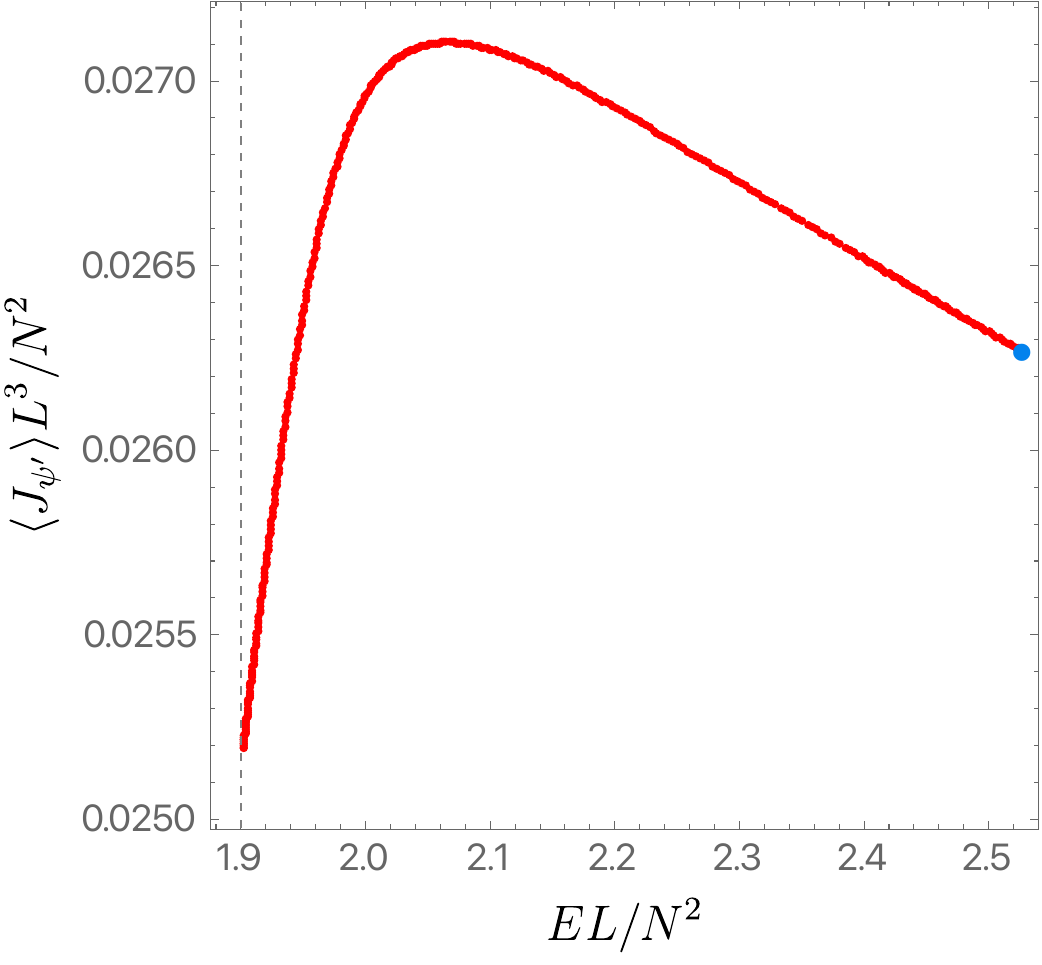}
\includegraphics[width=0.325\textwidth]{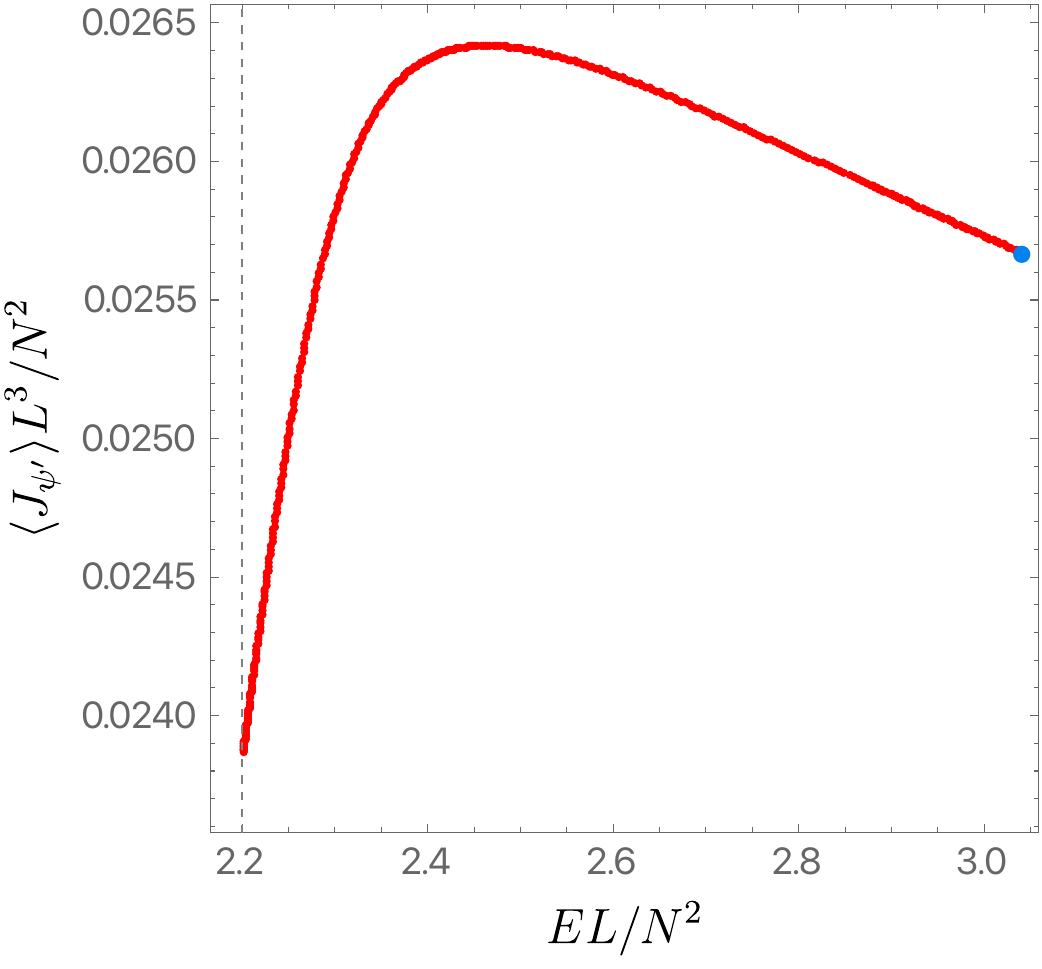}
\vspace{-0.25cm}
\caption{VEV of the operator dual to $A_\psi'$ as a function of energy.}
\label{fig:VevApsiE}
\end{figure}

We now turn to the outcome of our numerical construction, which is summarised in the plots below (Figs.~\ref{fig:epsilonE}--\ref{fig:entropyT}). Each of these plots shows some physical observables of our hairy rotating black holes (HBH) with fixed $J/N^2 = 0.05$ and three values of the charge: $QL/N^2 = 0.15$ (left panels), $QL/N^2 = 0.6$ (middle panels) and $QL/N^2 = 0.7$ (right panels). These plots represent, for specific values of $J$ and $Q$, the otherwise generic qualitative behaviour of HBH solutions with charge $Q<Q_c(J)$ (left panels), $Q$ slightly above $Q_c(J)$ (middle panels) and $Q>Q_c(J)$ (right panels), where $Q_c(J)$ (see footnote~\ref{foot:Qc}) is a critical charge expected to be present in the system from the thermodynamic model analysis of Section~\ref{sec:toymodel} and whose existence was also confirmed in~\cite{Markeviciute:2018yal, Markeviciute:2018cqs}. 
As argued using the thermodynamic model of Section~\ref{sec:toymodel} and demonstrated in ~\cite{Markeviciute:2018yal, Markeviciute:2018cqs}, one expects to find substantial differences in the structure of the rotating HBHs when we compare solutions with  $Q<Q_c(J)$ and  $Q>Q_c(J)$, i.e. the surface $Q=Q_c(J)$ should mark a sharp transition boundary in the qualitative behaviour of some physical properties (see footnote~\ref{foot:Qc}). For the value  $J/N^2 = 0.05$ displayed in Figs.~\ref{fig:epsilonE}--\ref{fig:entropyT}, one finds that $Q_c(J)L/N^2 \lesssim 0.6$. Pinpointing this critical value is computationally very costly, and (at least in the present study) there is no strong motivation to do so. What is important for us is to study the properties of solutions with $Q<Q_c(J)$ and $Q>Q_c(J)$ and identify significant physical differences between the two families. In each of the plots of Figs.~\ref{fig:epsilonE}--\ref{fig:entropyT}, the blue disk denotes the merger point between the HBH and the CLP BH. For reference, this blue point is the point with the given charge $Q$ on the blue onset curve of Fig.~\ref{figJ0p05:phasediagram}. In some of these plots, we will also find a dashed grey vertical line that identifies the BPS energy $E_\BPS \equiv 3 Q + 2 J/L$. We will find strong numerical evidence that HBHs `start' at the merger blue surface (a point in our plots at fixed $J$ and $Q$) and terminate at the BPS surface $E=E_\BPS$ (a point in our plots at fixed $J$ and $Q$).
%%%%%%%%%%%%%%%%%%
%%%%%%%%%%%%%%%%%%
%%%%%%%%%%%%%%%%%%
In Figs.~\ref{fig:epsilonE}--\ref{fig:vevT}, we display properties of the scalar field, be it the value at the horizon or its VEV. This will be important for the discussions of Section~\ref{sec:conclusion}. We start by displaying the value of the charged scalar field at the horizon $\e_H$ as a function of the dimensionless energy $E N^2/L$ (Fig.~\ref{fig:epsilonE}) and as a function of the hairy temperature $L\,T$ (Fig.~\ref{fig:epsilonT}) of the HBH.
Both figures show that the scalar field vanishes (blue disk) when the HBH merges with the CLP BH (which certainly has $\Phi=0$). Fig.~\ref{fig:epsilonE} illustrates that at fixed $J$ and $Q$, the energy of the HBH always decreases as it moves away from this merger, and $\e_H$ starts increasing significantly. This figure does not show a difference between BHs with $Q$ bigger or smaller than $Q_c(J)$.
Such a difference can be found in Fig.~\ref{fig:epsilonT} where we see that for $Q>Q_c(J)$ (right panel), the temperature of HBH increases (for fixed $J$ and $Q$) when it moves away from the merger point. For $Q\gtrsim Q_c(J)$ (middle panel), the temperature of the HBH ultimately still ends up increasing sufficiently far from the merger blue point. Still, it displays an oscillating behaviour for $T$ close to the one of the merger point. The number of oscillations may increase without bound as $Q\to Q_c(J)^+$ (but also as $Q\to Q_c(J)^-$; not shown). On the other hand, HBHs with $Q<Q_c(J)$ (left panel) have a completely distinct behaviour: their temperature decreases down to zero as they move away from the merger blue point. Our numerics suggests that $\e_H$ can grow arbitrarily large (possibly without bound) as $T\to 0$ (as illustrated in the left panel of Fig.~\ref{fig:epsilonT}). This feature will be very relevant in later discussions.  
%%%%%%%%%%%%%%%%%%
%%%%%%%%%%%%%%%%%%
%%%%%%%%%%%%%%%%%% 
In Figs.~\ref{fig:vevE}--\ref{fig:vevT}, we plot the dimensionless VEV $\, \langle \CO_{\Phi}\rangle L^2/N^2 \,$  of the operator dual to the charged scalar field  as function of the dimensionless energy $E L/N^2$  (Fig.~\ref{fig:vevE}) or as a function of the dimensionless temperature $L\,T$ (Fig.~\ref{fig:vevT}) of the HBH at fixed $J$ and $Q$. The plots of the VEV as a function of $E$ (Fig.~\ref{fig:vevE}) show similar monotonic behaviour for all values of fixed $Q$ -- the VEV increases as the energy decreases and as the HBHs move away from the merger blue point (where the VEV vanishes). A significant distinct qualitative behaviour between HBHs with $Q<Q_c(J)$ and $Q>Q_c(J)$ occurs when we look into the plot of VEV vs temperature (Fig.~\ref{fig:vevT}). We see that for $Q>Q_c(J)$ (right panel), the temperature and the VEV increase monotonically as we move away from the merger blue point. When $Q\gtrsim Q_c(J)$ (middle panel), the temperature of the HBH ultimately still ends up increasing sufficiently far from the merger blue point but displays an oscillating behaviour for $T$ close to the one of the merger point. The number of oscillations may increase without bound as $Q\to Q_c^+$ (but also as $Q\to Q_c^-$; not shown). We have a radically different behaviour for $Q<Q_c(J)$ (left panel). The temperature decreases down to zero, and the HBHs move away from the merger blue point, although the VEV still increases along this path. As $T\to 0$ one starts observing an oscillatory behaviour, we cannot exclude the possibility that there are many turning points in these oscillations before $T=0$ is reached.
%%%%%%%%%%%%%%%%%%
For completeness, in Fig.~\ref{fig:VevApsiE}, we plot the dimensionless VEV $\langle J_{\psi'} \rangle L^3/N^2$ of operator dual to the azimuthal component $A_{\psi'}$ of the gauge potential as a function of the dimensionless energy. It always starts at a finite value at the merger blue point (as expected since rotating CLP BHs must have $A_{\psi}$). For $Q<Q_c(J)$ (left panel), it decreases monotonically as the energy decreases. At the same time, for $Q>Q_c(J)$ (middle and right panels), it increases until it reaches a maximum and then starts decreasing as the energy decreases further. In all cases, this approaches a finite value as $E\to E_\BPS$ (see dashed grey vertical line).
%%%%%%%%%%%%%%%%%%

The results for the chemical potentials $\mu$, angular velocity, $\O_H$, and entropy $S$ as a function of energy $E$ or temperature $T$ of the HBHs are postponed to Section~\ref{sec:MainResults}.

%%%%%%%%%%%%%%%%%%%
\section{Supersymmetric hairy black hole?}
\label{sec:MainResults}
%%%%%%%%%%%%%%%%%%%

In the previous sections, we used analytical and numerical techniques to construct non-supersymmetric hairy black holes parameterised by three independent parameters: $E$, $Q$, and $J$. This section considers the supersymmetric limit $E \to E_\BPS = 3 Q + 2 J/L$ of our HBH solution. For some of our numerical plots, we will instead take the extremal limit $T \to 0$ and verify that $E \to E_\BPS$. We study the behaviour of various HBH thermodynamic quantities as we approach this limit.
The fundamental question we want to address is whether we can use our $T\neq 0$ HBHs in the limit where $T\to 0$ to find evidence for the existence, or not, of regular supersymmetric HBHs (which would have $\Delta_{\hbox{\tiny KLR}}\neq 0$, with $\Delta_{\hbox{\tiny KLR}}$ defined in \eqref{charge_constraint}).
%%%%%%%%%%%%%%%%%%%%%%%%%%%%%%%%%%%%%%%%%%%%%%%%
\subsection{Hairy black holes: numerical results \label{sec:NumericalResults}} 
%%%%%%%%%%%%%%%%%%%%%%%%%%%%%%%%%%%%%%%%%%%%%%%%

\begin{figure}[H]
\centering
\includegraphics[width=0.325\textwidth]{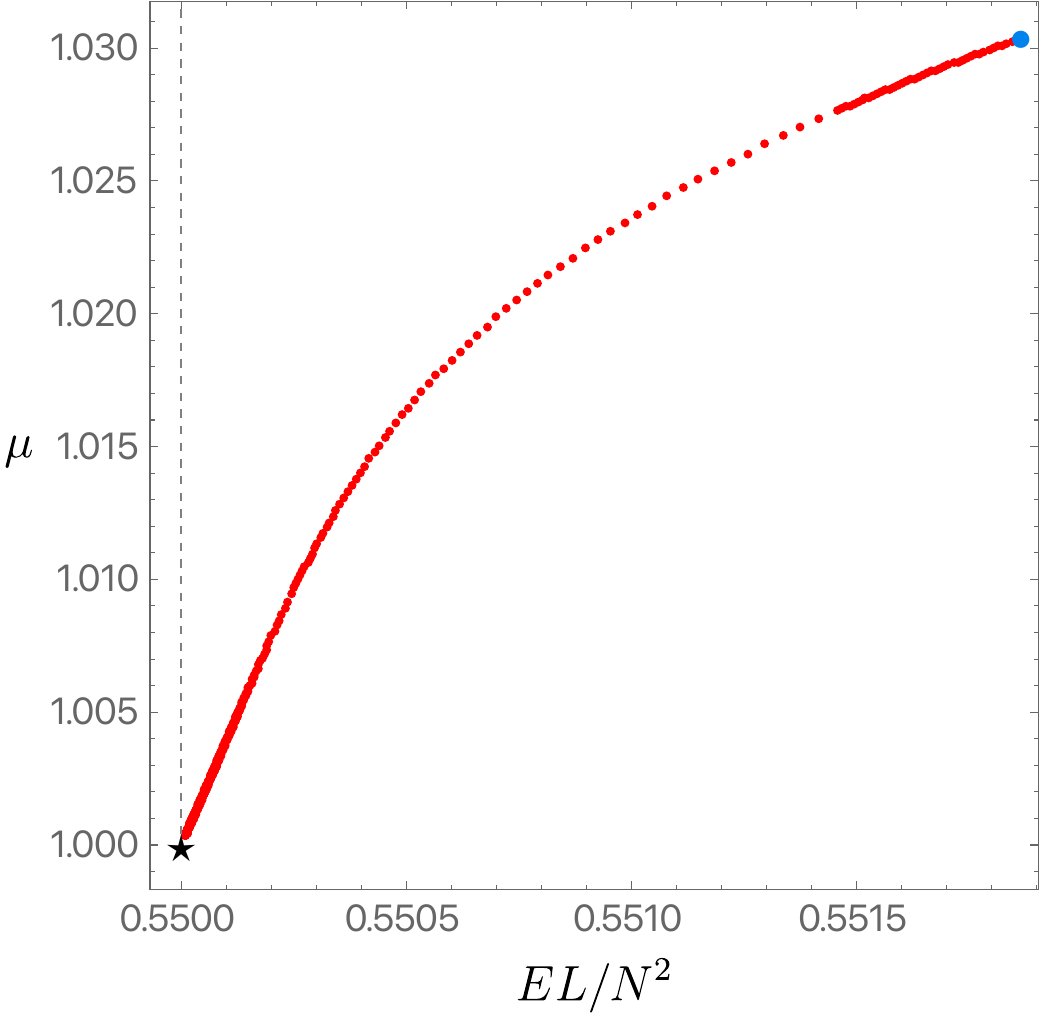}
\,\includegraphics[width=0.325\textwidth]{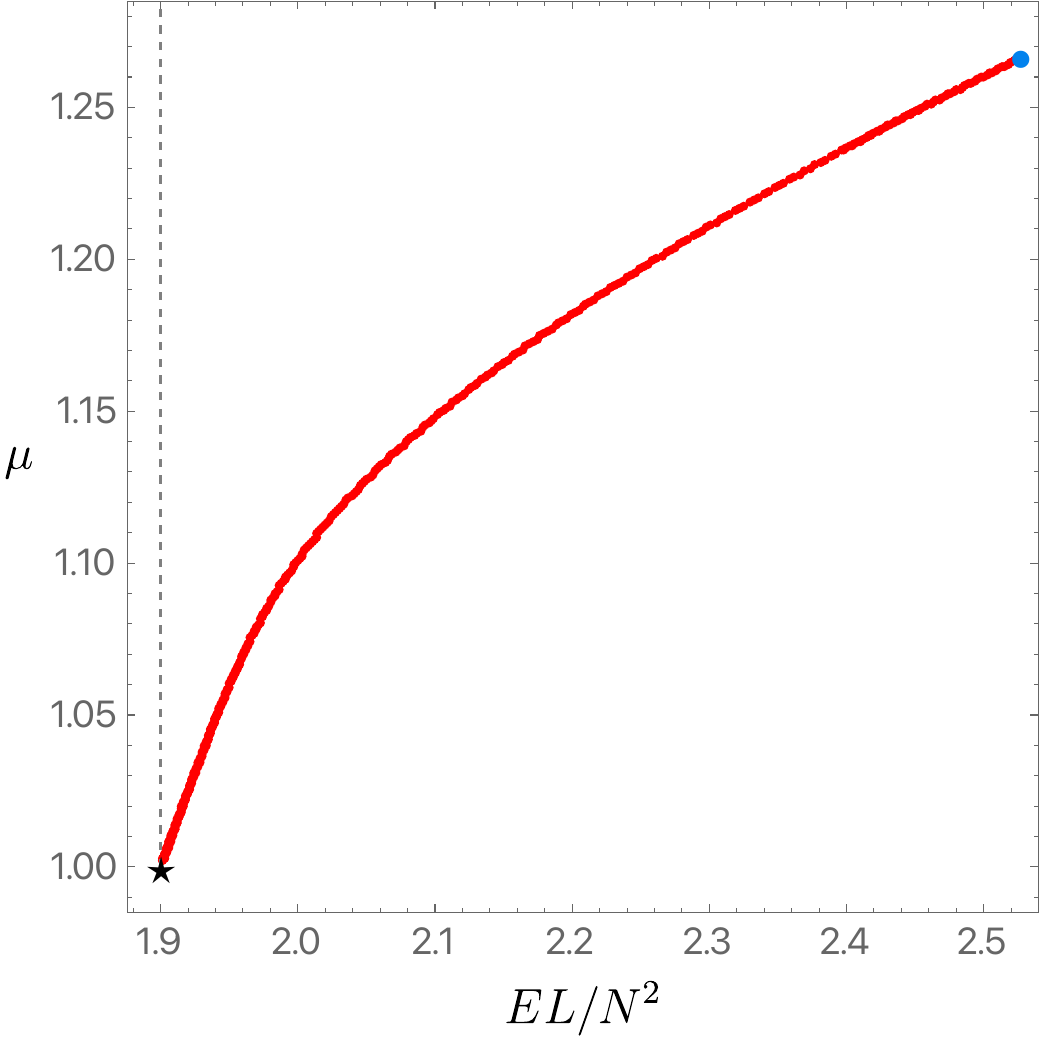}
\includegraphics[width=0.325\textwidth]{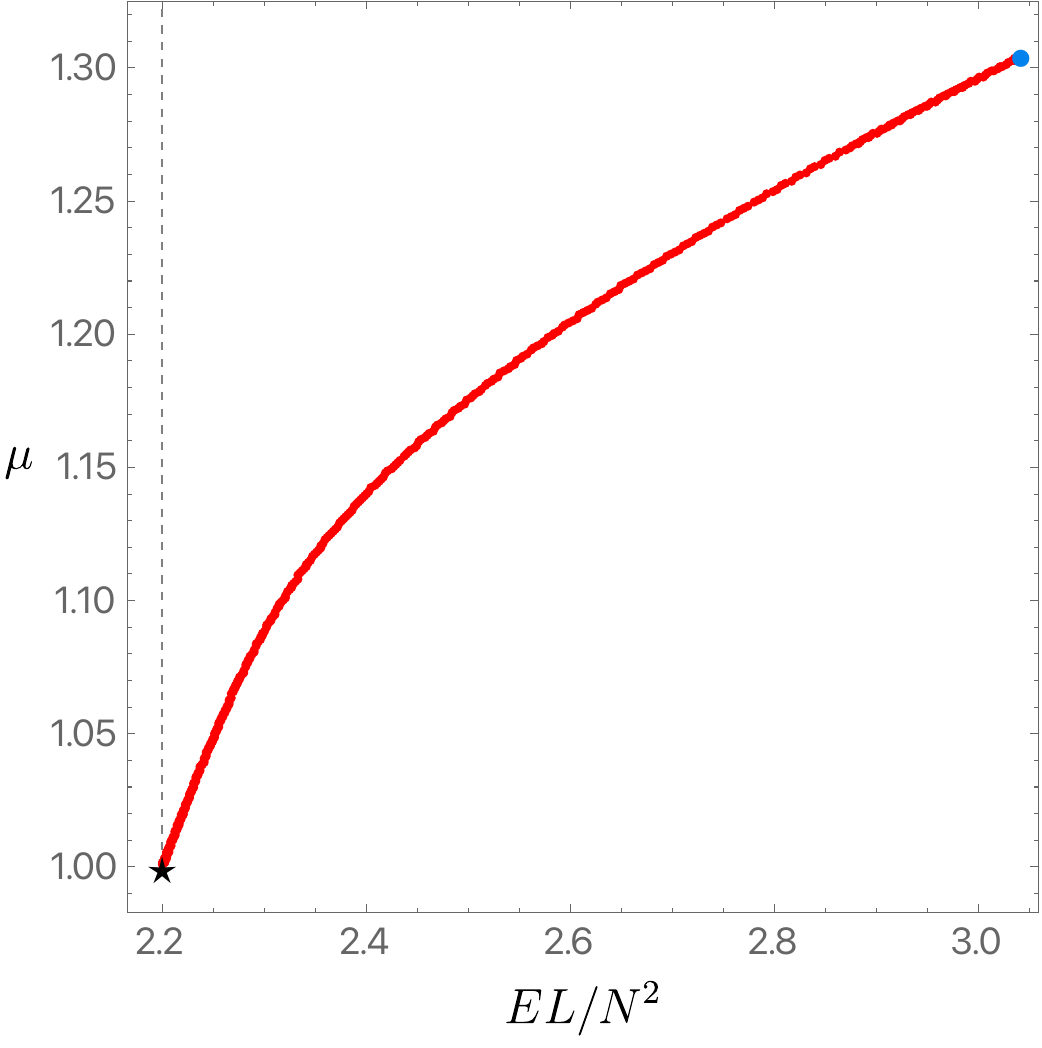}
\vspace{-1cm}
\caption{Chemical potential as a function of energy.}
\label{fig:MuE}
\end{figure}

\begin{figure}[H]
\centering
\includegraphics[width=0.325\textwidth]{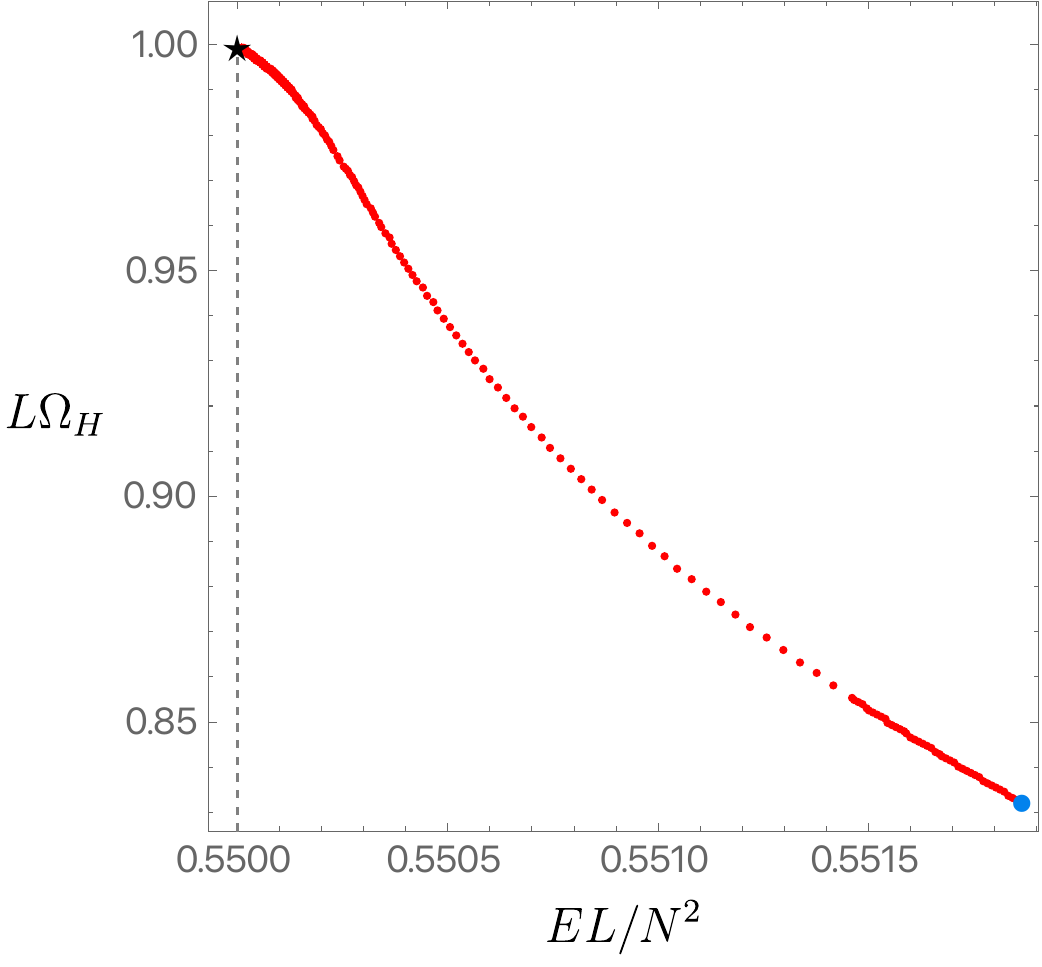}\includegraphics[width=0.325\textwidth]{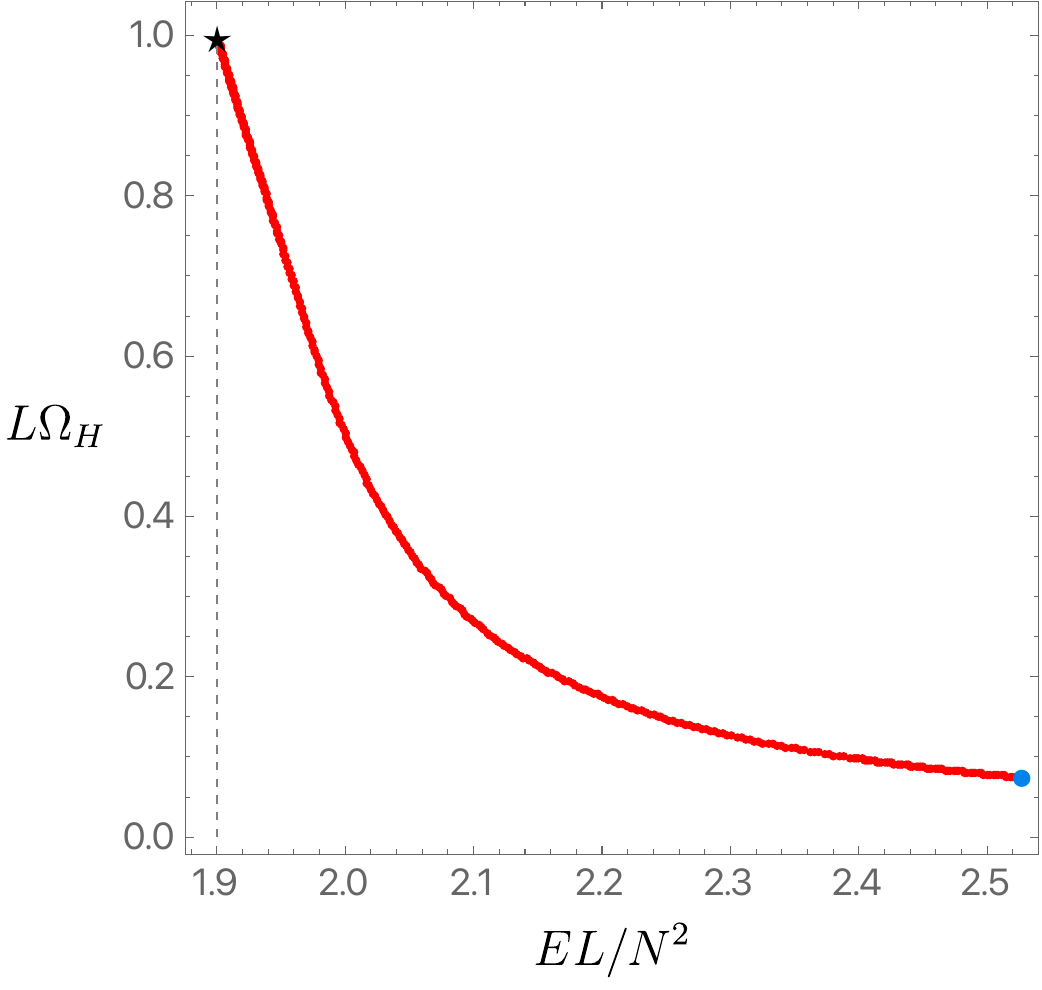}\includegraphics[width=0.325\textwidth]{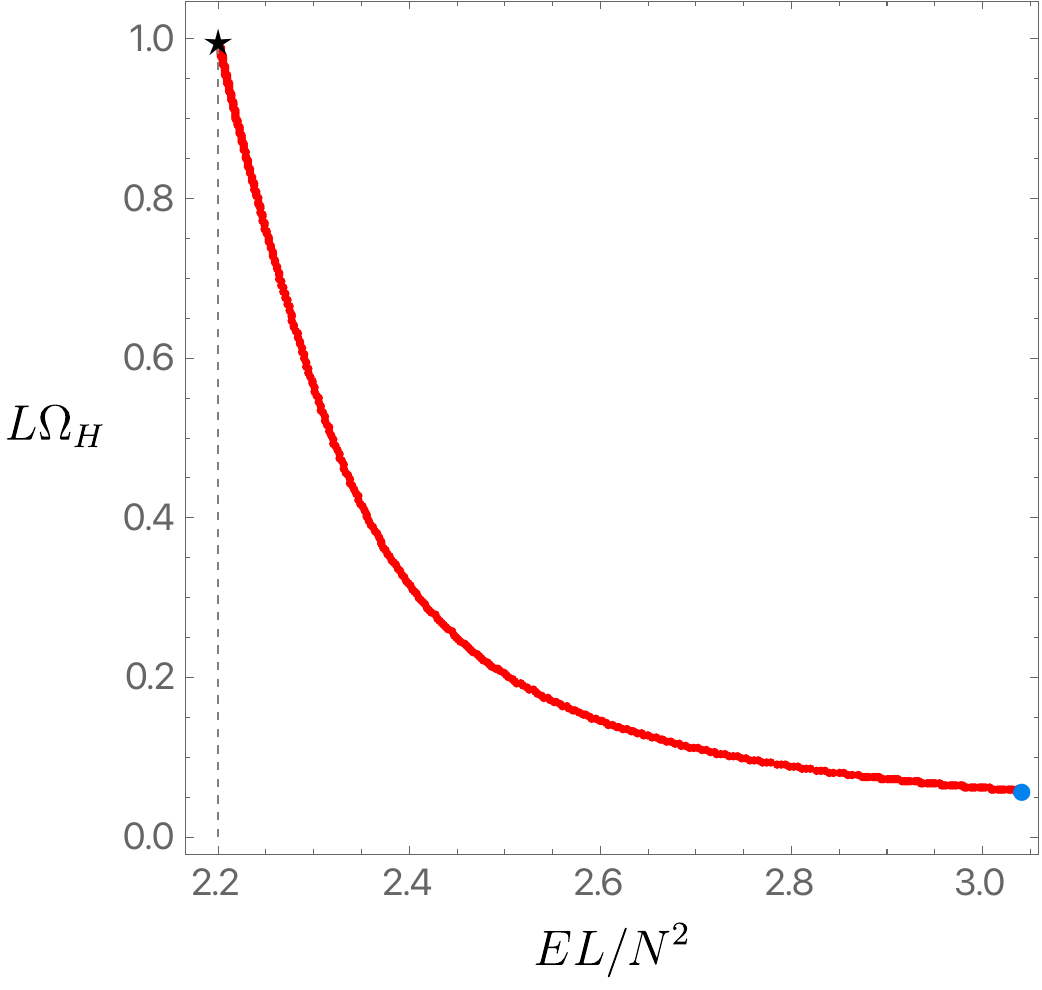}
\vspace{-0.5cm}
\caption{Angular velocity at the horizon as a function of energy.}
\label{fig:OmegaE}
\end{figure}

\begin{figure}[H]
\centering
\includegraphics[width=0.325\textwidth]{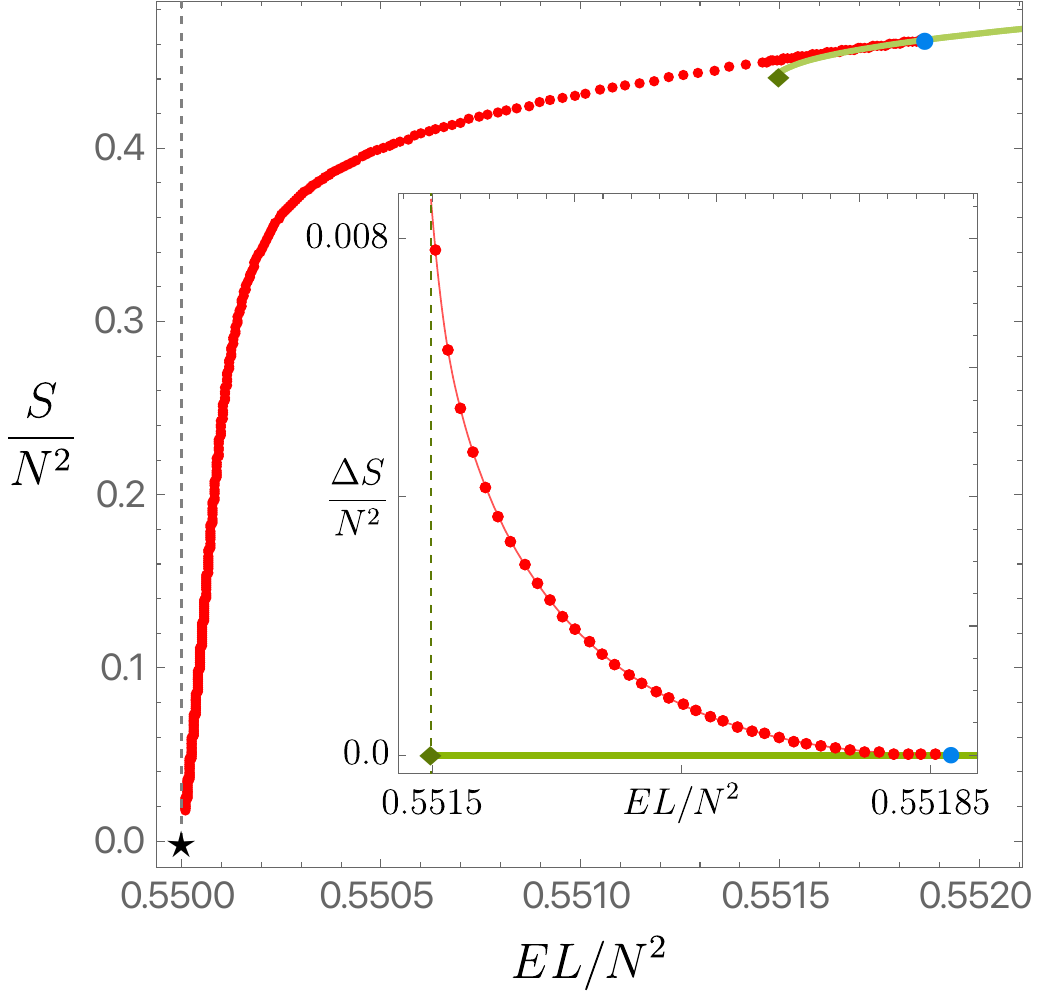}
\includegraphics[width=0.325\textwidth]{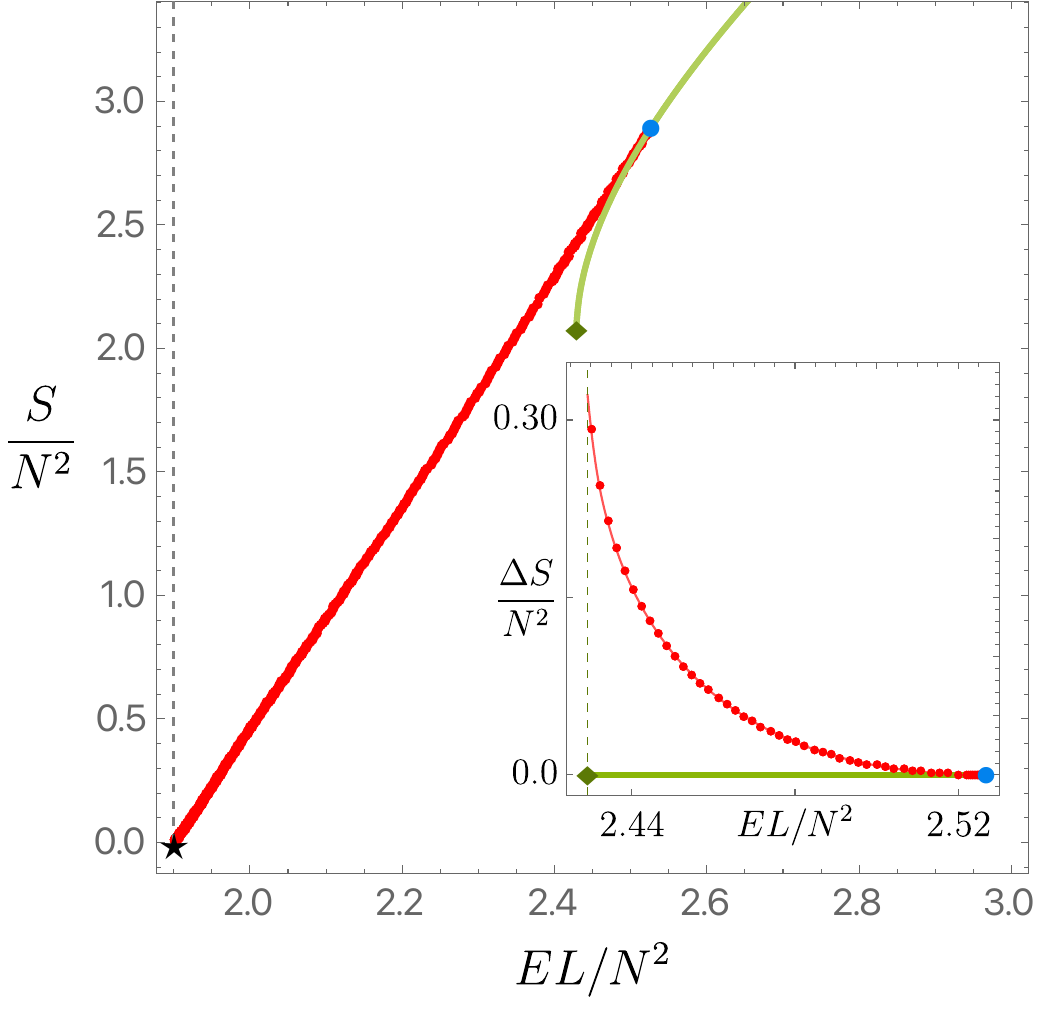}
\includegraphics[width=0.325\textwidth]{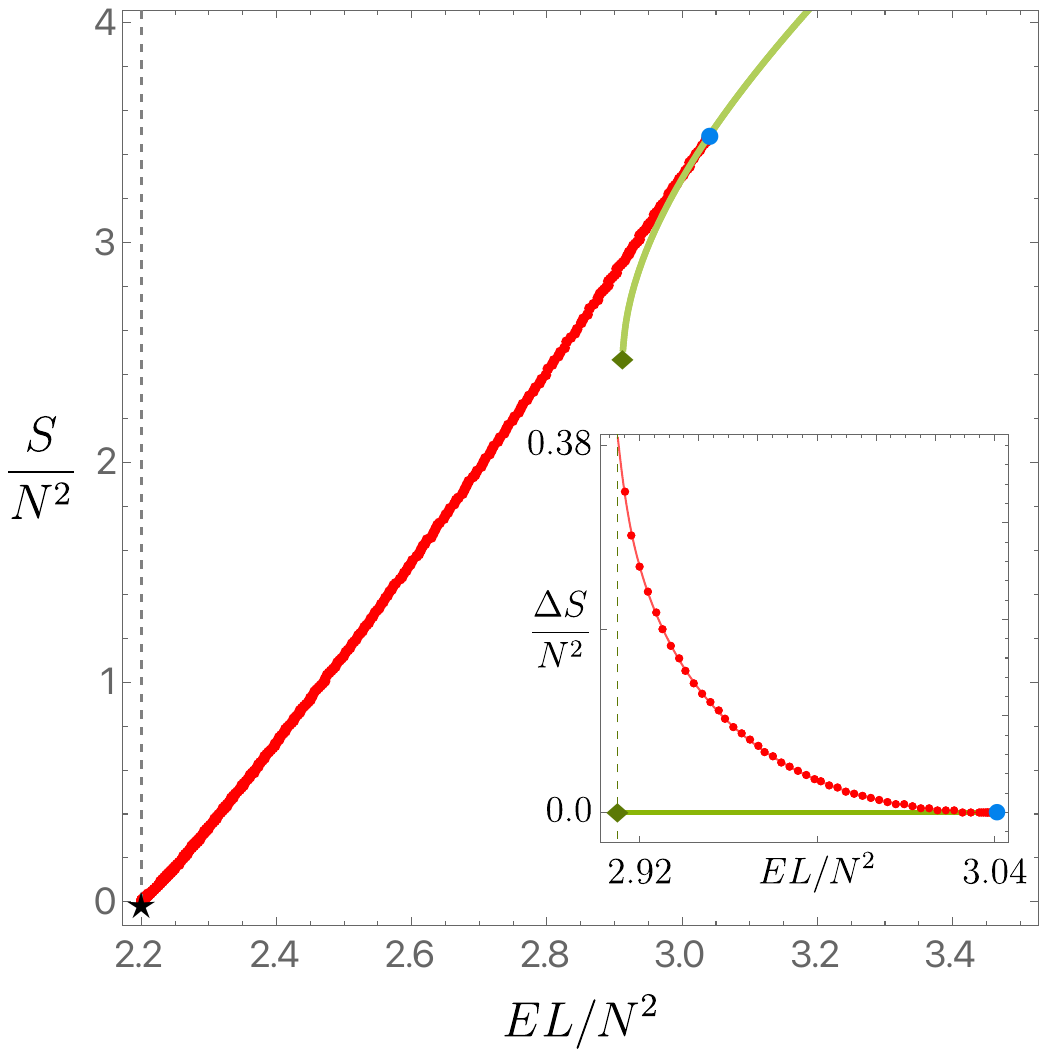}
\vspace{-0.5cm}
\caption{Entropy of the hairy (red disks) and CLP (green curve) black holes as a function of the energy. Green diamond pinpoints the extremal ($T=0$) CLP black hole.}
\label{fig:entropyE}
\end{figure}

\begin{figure}[H]
\centering
\includegraphics[width=0.325\textwidth]{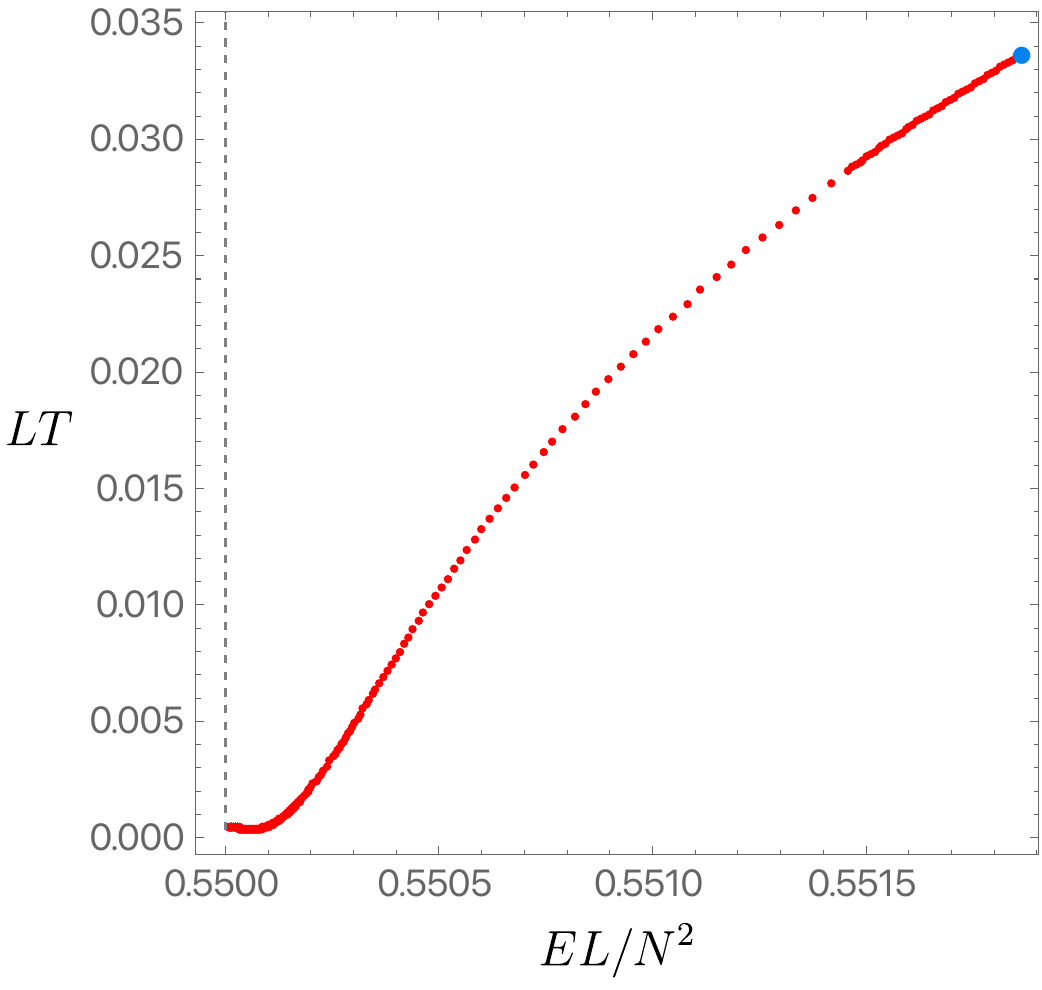}
\includegraphics[width=0.325\textwidth]{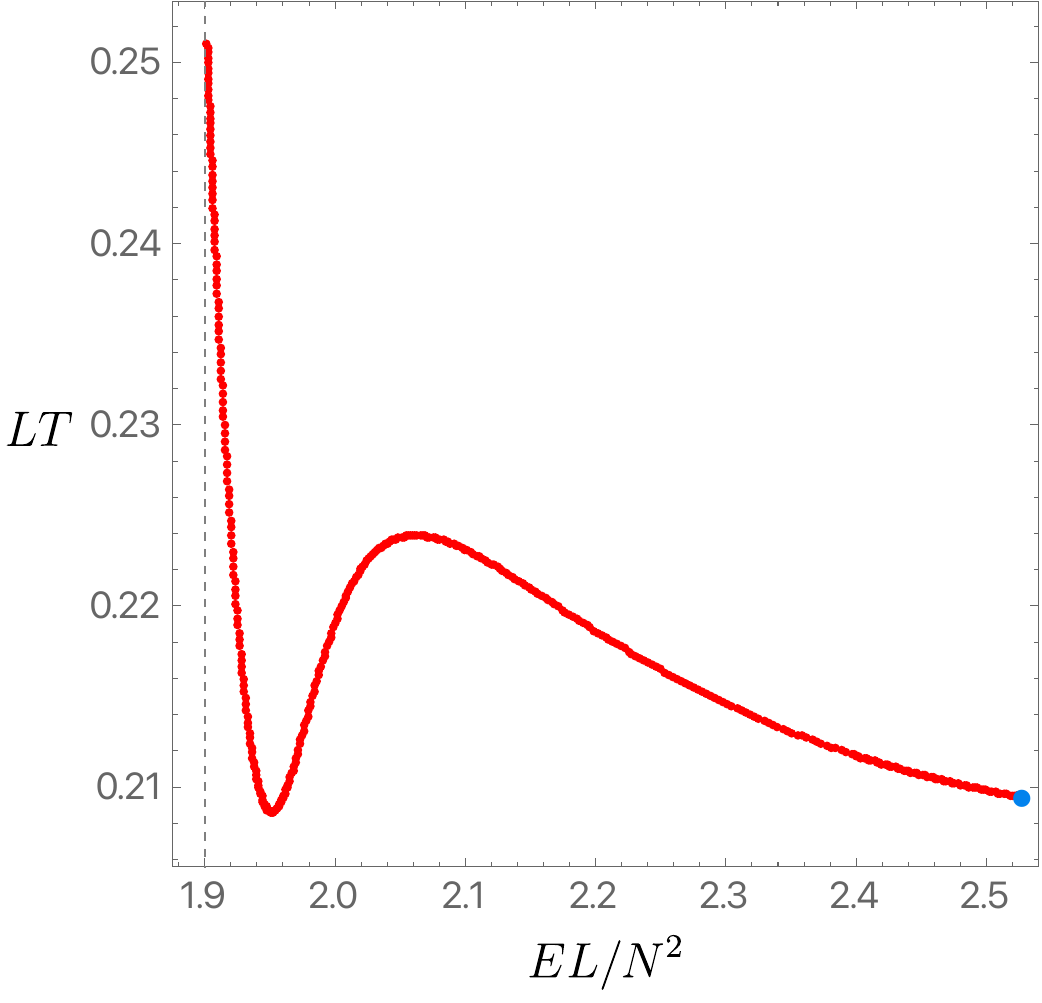}
\includegraphics[width=0.325\textwidth]{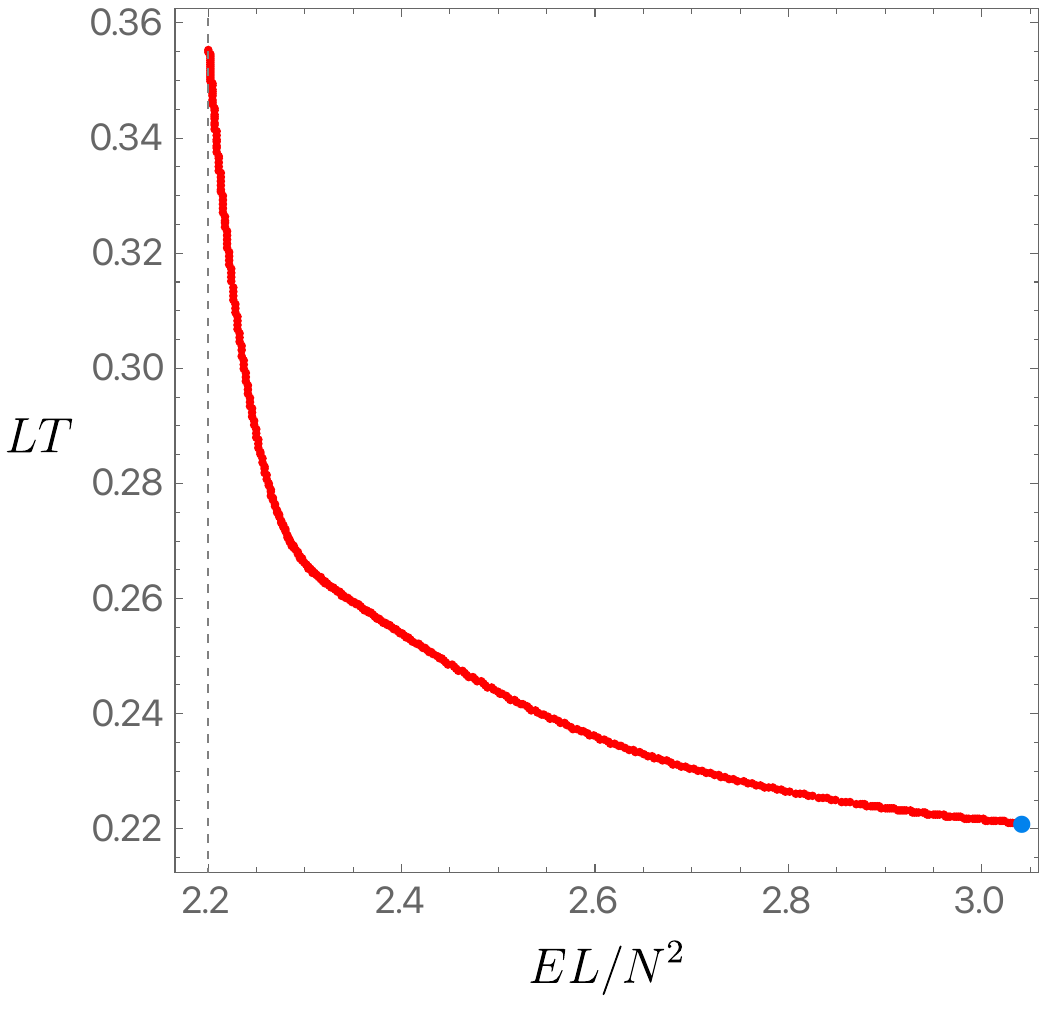}
\vspace{-0.5cm}
\caption{Temperature as a function of the energy.}
\label{fig:temperatureE}
\end{figure}

\begin{figure}[H]
\centering
\includegraphics[width=0.325\textwidth]{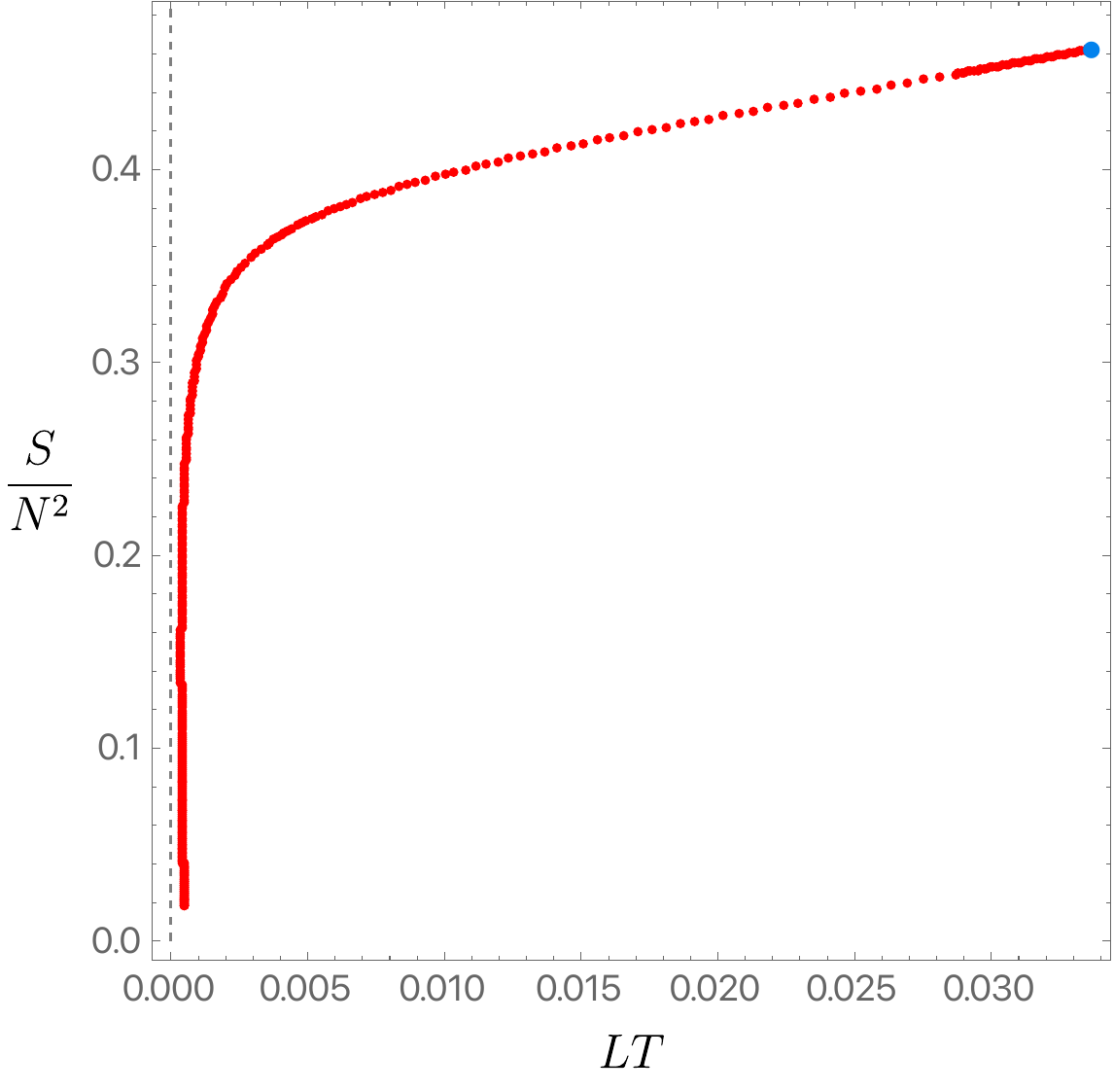}~\includegraphics[width=0.325\textwidth]{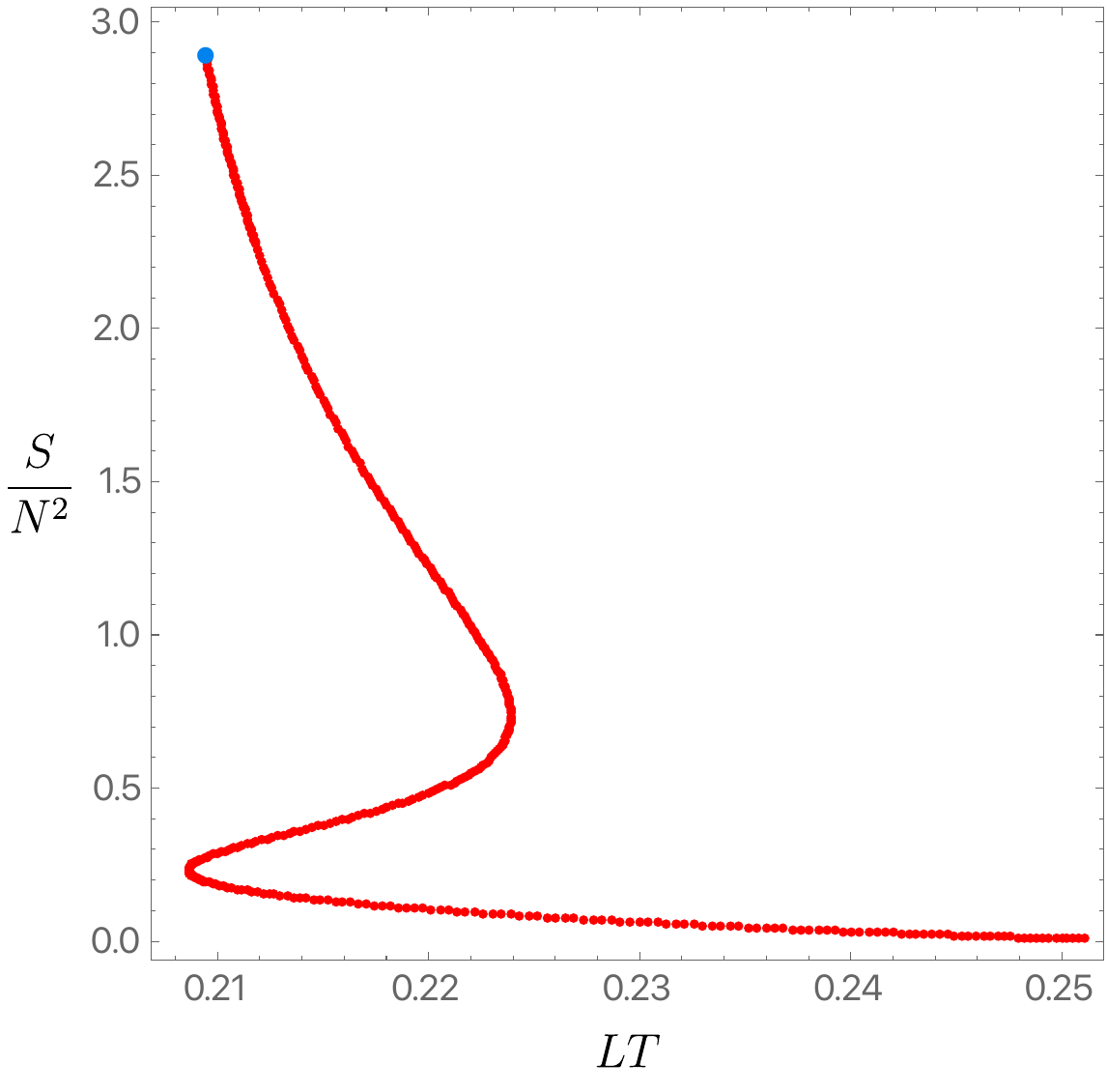}
~\includegraphics[width=0.325\textwidth]{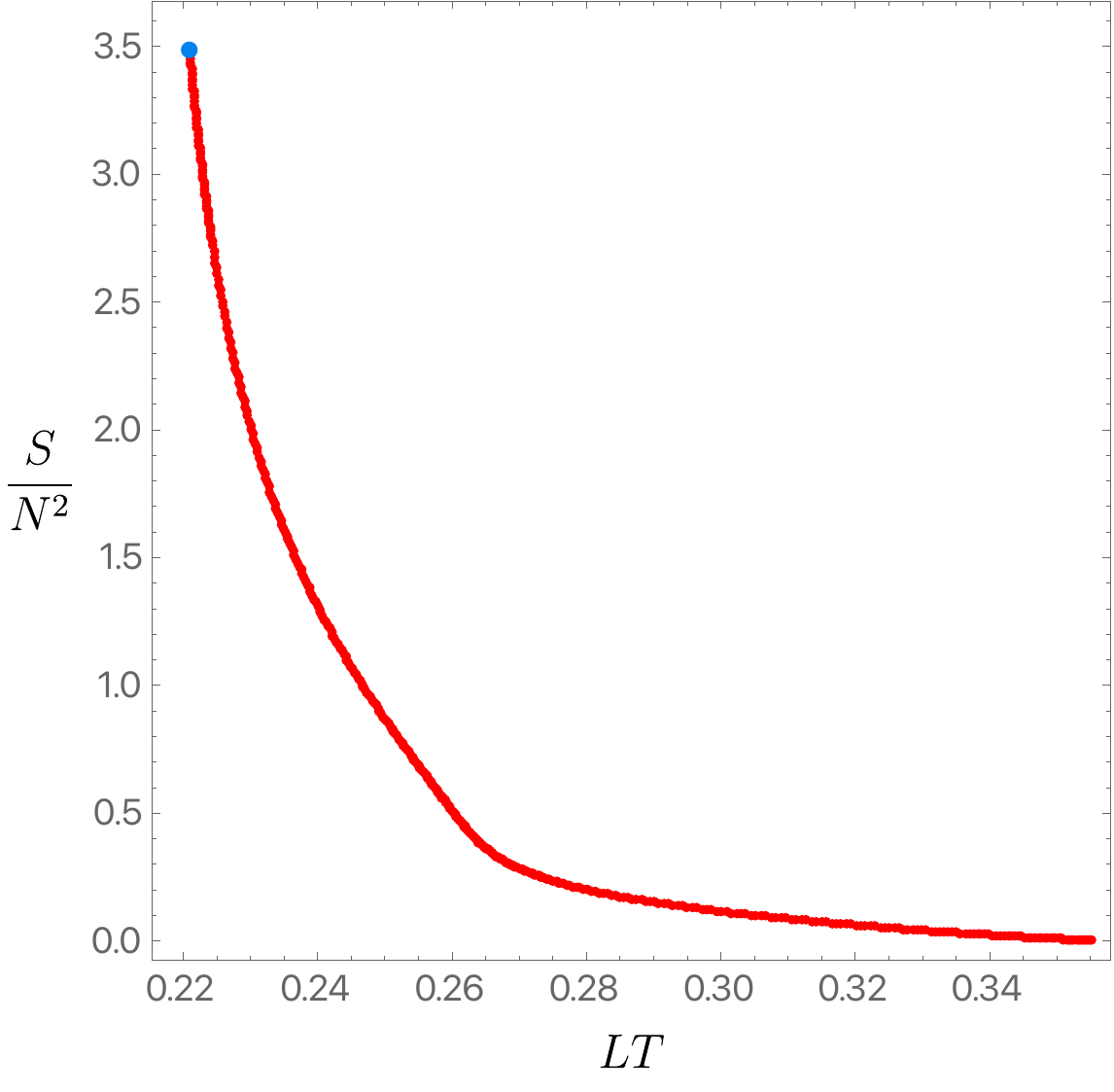}
\vspace{-1cm}
\caption{Entropy as a function of the temperature.}
\label{fig:entropyT}
\end{figure}

In Figs.~\ref{fig:MuE}--\ref{fig:entropyT}, we present representative numerical results for relevant thermodynamic quantities of the HBHs. As before, we fix $J/N^2 = 0.05$, and show the plots for three different values of the charge: $QL/N^2=0.15$ (left panels), $QL/N^2=0.6$ (middle panels) and $QL/N^2=0.7$ (right panels). These plots represent, for the aforementioned specific values of $J$ and $Q$, the otherwise generic qualitative behaviour of HBH solutions with charge $Q<Q_c(J)$ (left panels), $Q$ slightly above $Q_c(J)$ (middle panels) and $Q>Q_c(J)$ (right panels), where $Q_c(J)$ (see footnote~\ref{foot:Qc}) is a critical charge expected to be present in the system from the thermodynamic model analysis of Section~\ref{sec:toymodel} and whose existence was confirmed in~\cite{Markeviciute:2018yal, Markeviciute:2018cqs}.
 In some of these plots, the star ($\star$) denotes the extrapolated BPS limit, and the blue disk is the merger point between the hairy and CLP families of black holes (which coincides with the point obtained using the independent linear onset numerical code of Section~\ref{sec:Onset3Q}). Some plots also have a vertical dashed line that signals the BPS limit: either  $E=E_\BPS$ or  $T=0$, depending on which quantity is plotted. The red points always describe the HBH family. 
%%%%%%%%%%%%%%%%%%
%%%%%%%%%%%%%%%%%%
%%%%%%%%%%%%%%%%%%
%%%%%%%%%%%%%%%%%%

In Figs.~\ref{fig:MuE}--\ref{fig:entropyE}, the horizontal axis is always the dimensionless energy $E L/N^2$ and the vertical axis is the chemical potential $\mu$, horizon angular velocity $\O_H\,L$, entropy $S/N^2$, respectively. We see that the HBH moves from the merger point with CLP black hole (blue disk) to the BPS point (black $\star$). We make the following observations from these plots. First we note that in these figures, all three plots (i.e. for the three different values of charge) are qualitatively similar. Secondly, from Figs.~\ref{fig:MuE}--\ref{fig:OmegaE}, we see that $\mu \to 1^+$ and $\O_H L \to 1^-$ as $E \to E_\BPS$ (see $\star$ black point and dashed vertical line). Thirdly, we see that Fig.~\ref{fig:entropyE} displays a piece of the CLP black hole family (green curve) with the given $J$ and $Q$, starting at its extremal configuration (green diamond with $T_\CLP=0$ and $S_\CLP\neq 0$) and extending to higher values of energy with increasingly higher entropy (there is no upper bound on $E$ of CLP as best seen in Fig.~\ref{figJ0p05:phasediagram}). The blue disk on this curve pinpoints the merger point with the HBH family with the same dimensionless $J$ and $Q$ (red points). At this merger point, the two BH solutions have the same $S(E)$ and $\frac{\partial S}{\partial E}\big|_{J, Q}$ because this is a second-order phase transition. The HBH family then extends to lower energies till $E=E_\BPS$ while also decreasing its entropy. There is an intermediate window of energies where both CLP and HBHs coexist with the same $(E, Q, J)$ (but different $S,\mu,\O_H$) but, for smaller energies (although above $E_\BPS$), only HBHs do exist. The inset plots in these figures display $\Delta S/N^2$ as a function of $E L/N^2$ in the window where CLP (horizontal green line) and HBHs (red points) coexist, where $\Delta S$ is the entropy difference between the HBH and the CLP BH with the same $(E, Q, J)$. We conclude that the latter always has higher entropy for a given $(E, Q, J)$ where CLP and HBHs coexist. In other words, the HBH dominates the microcanonical ensemble for all values of $(E,Q,J)$ where hairy and CLP BHs coexist. Fourthly, Fig.~\ref{fig:entropyE} also shows (or in the case of the left panel, suggests) that the entropy of the HBH falls to zero in the BPS limit, $E \to E_\BPS$ (see black $\star$  and dashed vertical line). If this is indeed the case (see further discussion below where we zoom-in the region around the relevant $\star$ point), this means that the BPS limit of the hairy black hole is {\it singular} since the entropy and thus horizon radius do vanish. That is to say, the BPS limit of HBHs is a {\it naked singularity} instead of being a supersymmetric black hole with scalar hair. This is an extremely important finding. In particular, (for small $(E, Q, J)$ in the case of the left panel) it violates the expectations from the thermodynamic model (\cite{Bhattacharyya:2010yg} and Section~\ref{sec:toymodel}) and perturbative (Section~\ref{sec:hbh-pert}) analyses. This important matter will be discussed further below.   
%%%%%%%%%%%%%%%%%%
%%%%%%%%%%%%%%%%%%
%%%%%%%%%%%%%%%%%%
We now move to a discussion of Figs.~\ref{fig:temperatureE}--\ref{fig:entropyT}. Here, we finally observe a sharp distinction in the qualitative behaviour of HBHs with $Q<Q_c(J)$ and $Q>Q_c(J)$. In Fig.~\ref{fig:temperatureE}, we plot the dimensionless temperature of HBHs as a function of their dimensionless energy. We see that if $Q>Q_c(J)$ (middle and right panels), the temperature of the HBH family starts at a finite value at the merger blue point and ends at {\it finite} value at the BPS limit $E\to E_\BPS$ (see vertical dashed line; for the given $\{J, Q\}$ in these plots, the BPS temperature is higher than the one at the merger but this is not necessarily the case for other $\{J, Q\}$ choices). For sufficiently large $Q>Q_c(J)$, this evolution is such that the temperature increases monotonically as $E$ decreases and approaches $E\to E_\BPS$ (right panel). In contrast, for $Q\gtrsim Q_c(J)$ (middle panel), the curve has local maxima and minima before reaching the finite value in the BPS limit (it could be that the number of local extrema gets very large as $Q\to Q_c(J)$). The key conclusion for $Q> Q_c(J)$ is that in the BPS limit one has $E\to E_\BPS$, $\O_H L\to 1^-$, $\mu\to 1^+$ but $T\to T_\BPS\neq 0$. (So, for $Q> Q_c(J)$, the only solutions with $T=0$ are extremal CLP BHs when they exist). This behaviour might sound unfamiliar, but it is qualitatively similar to the static single-charge hairy black hole solution of another sector ($Q_1=Q_2=0$, $Q_3=Q$) of $U(1)^3$ supergravity constructed in~\cite{Dias:2022eyq}, for which the limiting temperature is $TL=1/\pi$. On the other hand, for $Q<Q_c(J)$ (left panel of  Fig.~\ref{fig:temperatureE}), we see a sharply distinct behaviour: the temperature at the merger blue disk is still finite {\it but} ends up being zero in the BPS limit $E\to E_\BPS$ (see dashed vertical line). So, the key conclusion for $Q< Q_c(J)$ is that in the BPS limit one has $E\to E_\BPS$, $\O_H L\to 1^-$, $\mu\to 1^+$ and, this time, $T\to T_\BPS= 0$. 

This substantial difference in the qualitative behaviour of certain thermodynamic quantities of HBHs with  $Q<Q_c(J)$ and  $Q>Q_c(J)$ is also observed in the plot of the entropy as a function of temperature (Fig.~\ref{fig:entropyT}). For $Q>Q_c(J)$ (middle and right panels), the entropy and temperature start finite at the merger blue point. Still, the entropy vanishes (see horizontal dashed line) in the BPS limit at higher temperature, namely when $E\to E_\BPS$, $\O_H L\to 1^-$, $\mu\to 1^+$ {\it but} $T\to T_\BPS\neq 0$ (again, in the middle panel, note the oscillatory behaviour when $Q$ is just slightly above $Q_c(J)$). On the other hand, the behaviour is substantially distinct for  $Q<Q_c(J)$ (left panel): the entropy and temperature start again finite at the merger blue point but the entropy vanishes (as we will fully confirm below) in the BPS limit, this time, at zero temperature, namely when $E\to E_\BPS$, $\O_H L\to 1^-$, $\mu\to 1^+$ {\it and} $T\to T_\BPS= 0$ (see vertical dashed line). The entropy decreases dramatically in a very small vicinity of $T=0$. This is one of the reasons it is so difficult to find the full evolution of this family of solutions in the `late stages' as it approaches $T\to 0$. The sharp transition, when $Q$ increases from $Q<Q_c(J)$ towards $Q>Q_c(J)$ observed in Figs.~\ref{fig:temperatureE}--\ref{fig:entropyT} (but also in Figs.~\ref{fig:epsilonT},~\ref{fig:vevT},~\ref{fig:VevApsiE}), in the qualitative behaviour of HBH solutions was previously identified and well documented in~\cite{Markeviciute:2018yal, Markeviciute:2018cqs}.

The reader should note that the numerical findings for the $Q<Q_c(J)$ case are {\it not} consistent with the conclusions extracted from the thermodynamic model (\cite{Bhattacharyya:2010yg} and Section~\ref{sec:toymodel}) and perturbative (Section~\ref{sec:hbh-pert}) analyses (which are valid for small dimensionless $(E, Q, J)$; thus they only capture the properties of the $Q<Q_c(J)$ case\footnote{Although the thermodynamic model is still capable of suggesting the existence of a critical charge $Q_c(J)$ because the static supersymmetric soliton of the non-interacting mixture (on `top' of which we `place' a small rotating CLP BH) is regular only up to a critical charge $Q_c(J=0)$.}). Indeed, the latter analyses predict that the entropy is {\it finite} in the BPS limit. This contradiction between the analytical and numerical analyses raises a puzzle that must be addressed. 

\begin{figure}[H]
\centering
\includegraphics[width=0.48\textwidth]{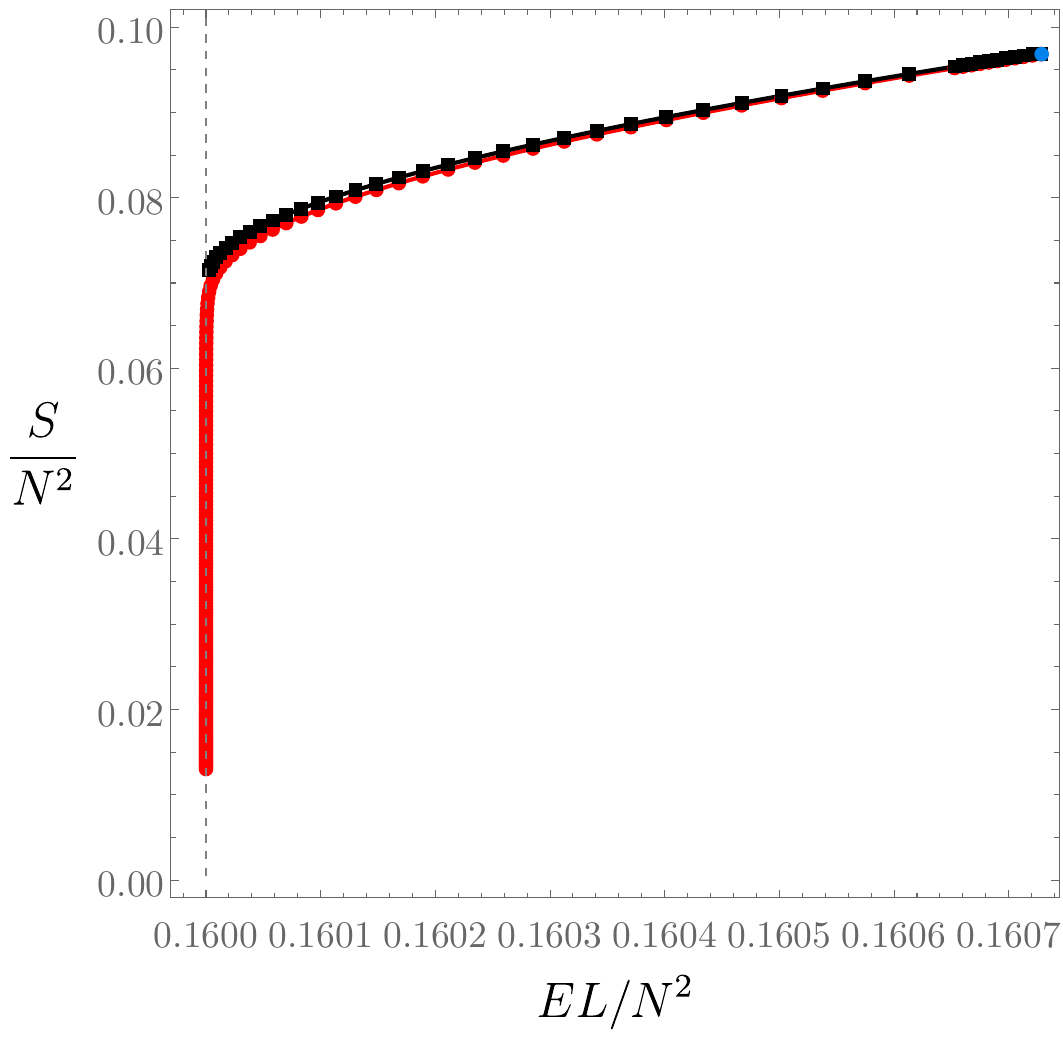}~\includegraphics[width=0.48\textwidth]{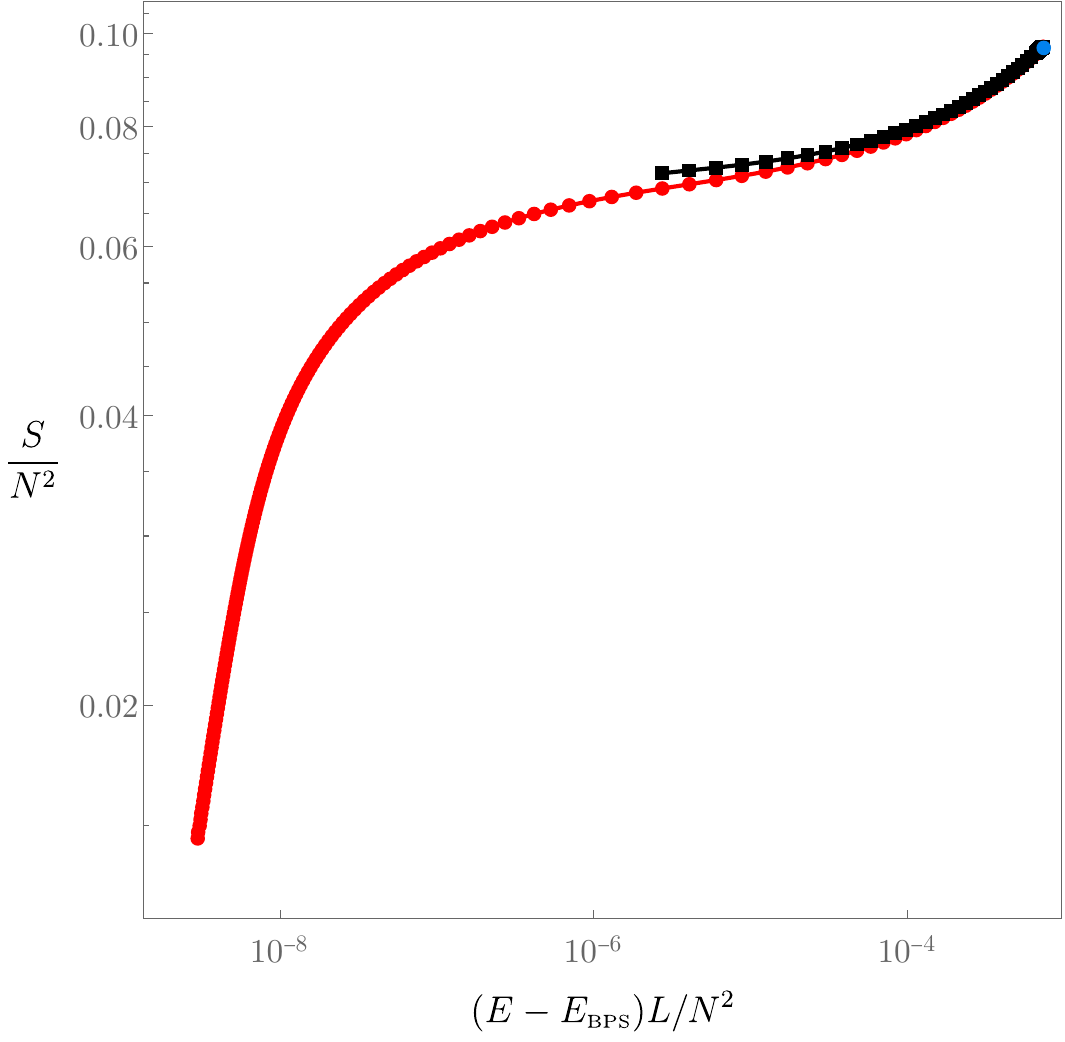}
\vspace{-0.25cm}
\caption{Hairy BHs with $ J /N^2=0.005$ and $Q L/N^2=0.05<Q_c(J)$. {\bf Left panel:} 
entropy as a function of the dimensional energy. Red disks are the numerical data while the black squares describe the perturbative result \eqref{hbh_pert_thermo}. The blue disk is the merger point between the hairy and the CLP families and the dashed vertical line signals $E=E_\BPS$.
{\bf Right panel:} log-log plot for entropy $S/N^2$ as a function of $E-E_\BPS$.}
\label{fig:entropyE-NumPert1}
\end{figure}

\begin{figure}[H]
\centering
\includegraphics[width=0.48\textwidth]{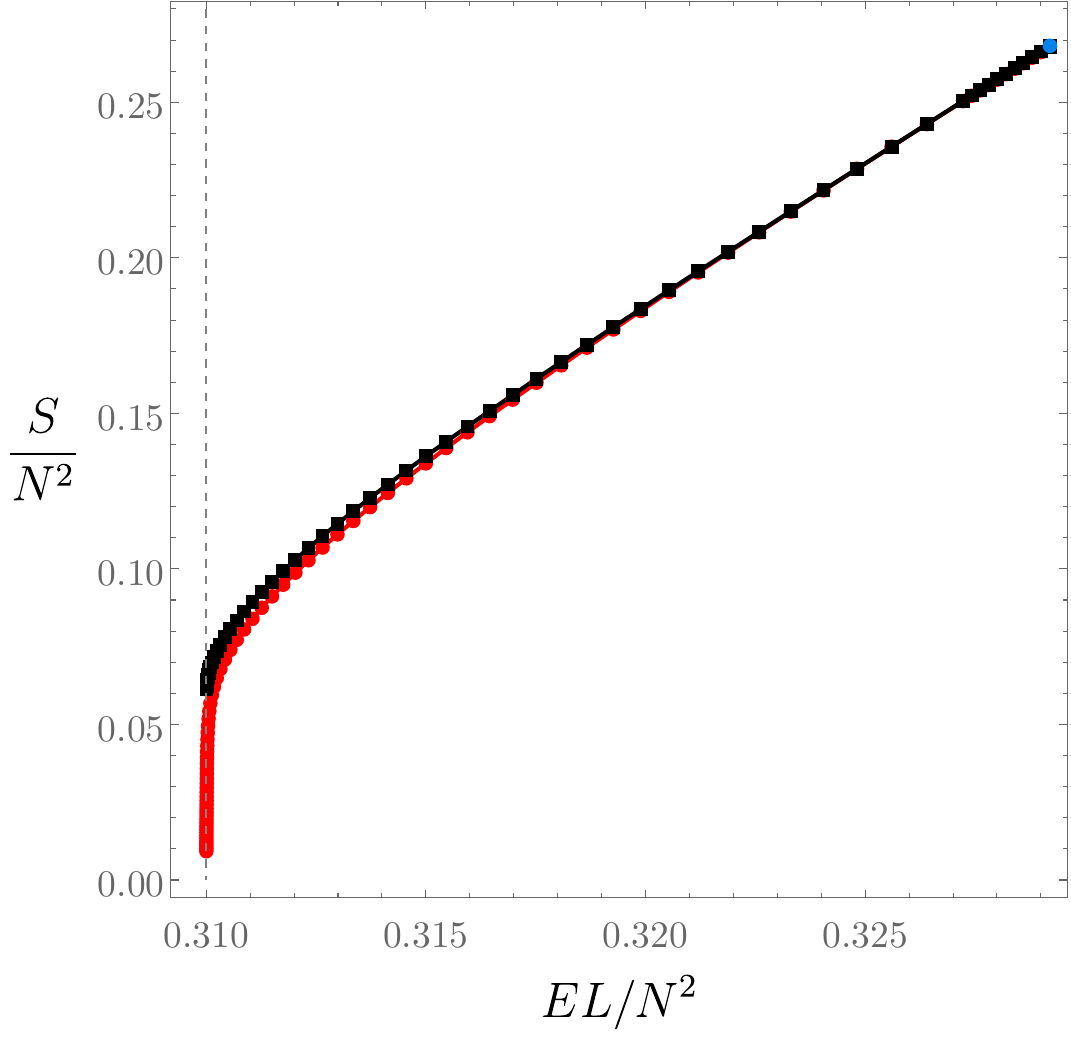}~\includegraphics[width=0.48\textwidth]{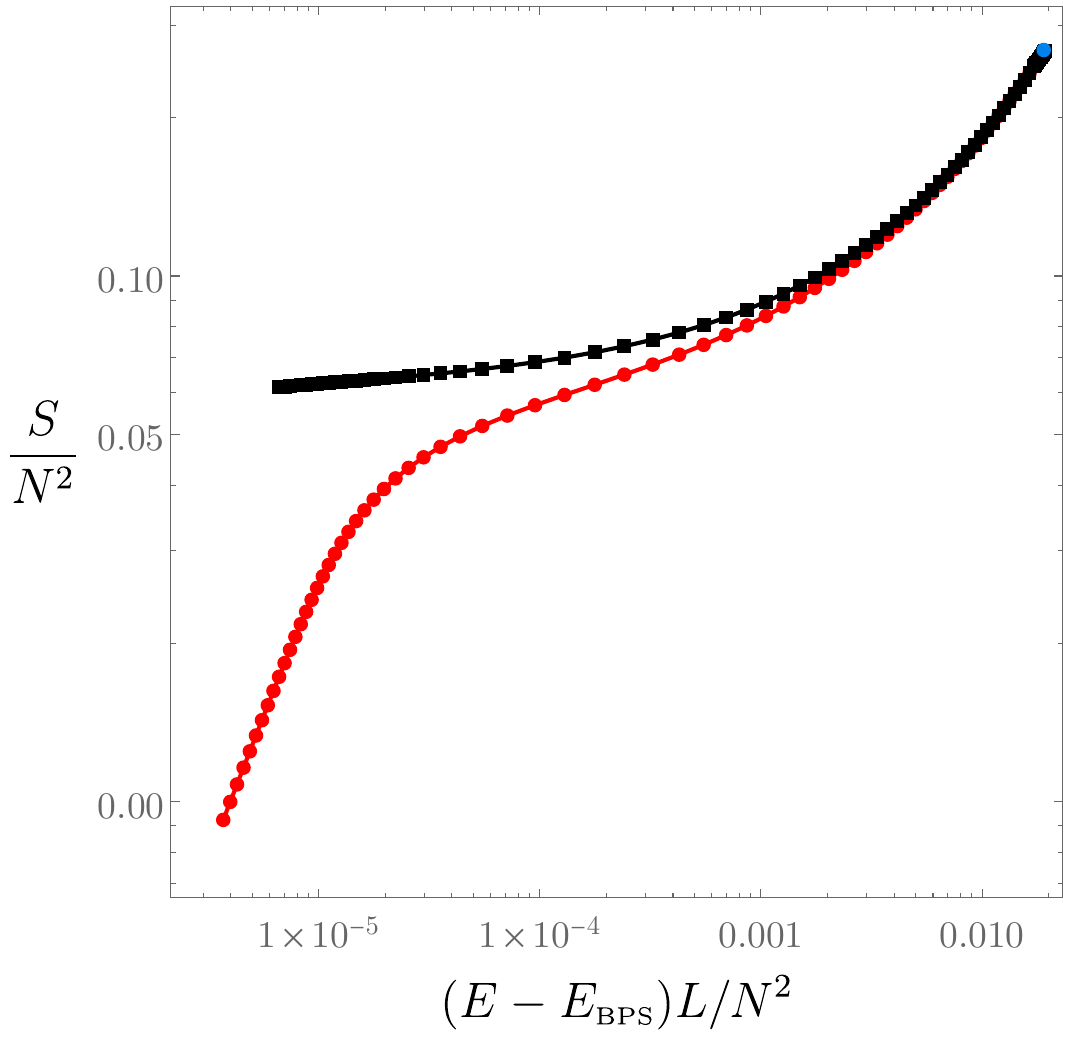}
\vspace{-0.25cm}
\caption{Similar to Fig.~\ref{fig:entropyE-NumPert1} but this time for hairy black holes with $ J /N^2=0.005$ and $Q L/N^2=0.1<Q_c(J)$.}
\label{fig:entropyE-NumPert2}
\end{figure}

Before doing this, let us compare our numerical findings in more detail with the numerical findings of~\cite{Markeviciute:2018yal, Markeviciute:2018cqs}. Our results for the $Q>Q_c(J)$ case are in perfect agreement with those of~\cite{Markeviciute:2018yal, Markeviciute:2018cqs}. In particular, we agree that the entropy of HBHs vanishes in the BPS limit (where $T_\BPS\neq 0$). However, there is a {\it disagreement} in the main physical conclusion in the $Q<Q_c(J)$ case. More concretely, the numerical data collected in~\cite{Markeviciute:2018yal, Markeviciute:2018cqs} suggested that the BPS limit of a $Q<Q_c(J)$ HBH should be a {\it regular} supersymmetric HBH, as conjectured by the thermodynamic model of ~\cite{Bhattacharyya:2010yg} (see also our Section~\ref{sec:toymodel}). This particular conclusion of~\cite{Markeviciute:2018yal, Markeviciute:2018cqs} is at odds with ours. For the numerical data collected in~\cite{Markeviciute:2018yal, Markeviciute:2018cqs}, we observe a perfect match between our results and theirs. The key difference is that we have {\it extended} the data collection to {\it smaller temperatures}, which reveals an unexpected and intricate behaviour of the system as it approaches $T=0$  (that, as far as we are aware, is very rare). In more detail,~\cite{Markeviciute:2018yal, Markeviciute:2018cqs} stop collecting data at $T L\sim \CO{(10^{-3})}$. For two orders of magnitude (i.e., for $10^{-3} \lesssim T L \lesssim 10^{-1}$), $S$ is a monotonic function of $T$ with approximately constant slope. Naturally, this led~\cite{Markeviciute:2018yal, Markeviciute:2018cqs}  to use this data to extrapolate the system's behaviour at $T=0$, which yields a finite entropy in the BPS limit. The data collection of~\cite{Markeviciute:2018yal, Markeviciute:2018cqs}  stopped at $T L\sim \CO{(10^{-3})}$ because $i)$ this is typically considered to be a very small value, $ii)$ it is very difficult to generate numerical data for smaller $T$, and $iii)$ the outcome of the extrapolation above was in agreement with the non-interacting thermodynamic analysis of~\cite{Bhattacharyya:2010yg}. Altogether, there was no reason to push further the challenging numerical computations. It turns out that if we extend the data collection for even smaller $T$ than~\cite{Markeviciute:2018yal, Markeviciute:2018cqs} the slope of $S$ changes and becomes higher as will be more accurately described below (see Figs.~\ref{fig:entropyE-NumPert1}--\ref{fig:entropyE-NumPert2}).

In our case, however, we benefit from the luxury of having the thermodynamic quantities \eqref{hbh_pert_thermo} that emerge from the perturbative analysis that we performed in Section~\ref{sec:hbh-pert}. When we compare the perturbative curve with the numerical curve, e.g. for $S(E)$, the agreement is very good for a large range of energies (for small $E, Q, J$), but, very surprisingly, this agreement clearly becomes increasingly bad as the temperature gets closer to zero (although $E$ is decreasing). This remarkable, unexpected behaviour (and associated contradiction/puzzle) motivated us to extend the numerical computation for even lower values of $T$, namely till $T L\sim \CO{(10^{-7})}$, i.e. four orders of magnitude closer to $T=0$ than~\cite{Markeviciute:2018yal, Markeviciute:2018cqs} attained. 
This discussion is best illustrated analysing immediately the outcome of our exercise. 
%%%%%%%%%%%%%%%%%%
%%%%%%%%%%%%%%%%%%

%%%%%%%%%%%%%%%%%%
%%%%%%%%%%%%%%%%%%

In Figs.~\ref{fig:entropyE-NumPert1}--\ref{fig:entropyE-NumPert2} we take an HBH with  $ J /N^2=0.005$ (i.e., one order of magnitude lower than in Figs.~\ref{fig:epsilonE}--\ref{fig:entropyT}) and $Q L/N^2=0.05<Q_c(J)$ (Fig.~\ref{fig:entropyE-NumPert1}) or $Q L/N^2=0.1<Q_c(J)$ (Fig.~\ref{fig:entropyE-NumPert2}) and, in both figures, we plot the entropy as a function of the dimensionless energy for a HBH family that merges with the CLP BH family at the blue disk (right side of plots). 
We display both the numerical data that we collected (red disks) and the entropy (black squares) that we obtain from the perturbative formulae \eqref{hbh_pert_thermo}. In both figures, the perturbative result is a very good (if not excellent) approximation to the numerical data in a very large window of energies starting at the blue merger disk. However, in the left panel, one observers that very close to $E=E_\BPS$  (where $T\to 0$; see vertical dashed line), the numerical entropy falls sharply in a region near $E_\BPS$ where the change of $E$ is extremely small. On the other hand, this fall-off is much less substantial in the case of the black square perturbative curve. This feature is amplified in the right panel (of both Figs.~\ref{fig:entropyE-NumPert1}--\ref{fig:entropyE-NumPert2}), where we still display the entropy of the red numerical and black perturbative curves but, this time, 1) we plot $E-E_\BPS$ in the horizontal axis and 2) we use a log-log plot. Again, the perturbative curve proves to be a very good approximation in a wide window that starts at the blue merger point but, clearly, the perturbative curve stops being a good approximation to the exact numerical data as $E-E_\BPS\to 0$ (actually, for too small energies the perturbative formula breaks down and this is why it does not extend so close to $E_\BPS$ as the red numerical curve). This happens for the two charges displayed in  Figs.~\ref{fig:entropyE-NumPert1}--\ref{fig:entropyE-NumPert2} (which are selected illustrative cases). In particular, note that if we  extrapolate the perturbative black curve, it indicates that the entropy should attain a {\it finite} value at $E= E_\BPS$. This is in agreement with the perturbative BPS entropy \eqref{BPS_limit_pert_S}. However, the extrapolated red numerical data indicates that, in reality, on has $S= 0$ at $E= E_\BPS$ (where we also have $T=0, \mu=1$ and $\O_H L=1$).

The following conclusion is now inevitable. The non-interacting thermodynamic model of~\cite{Bhattacharyya:2010yg} (and of our Section~\ref{sec:toymodel}) and the perturbative analysis of Section~\ref{sec:hbh-pert} predict that, for $Q<Q_c(J)$, the 2-parameter BPS limit of the non-extremal 3-parameter HBH family should be a {\it regular} (i.e., with finite $S_\BPS$) supersymmetric HBH. Nevertheless, the exact numerical solution -- when stretched to values remarkably close to the BPS limit -- unequivocally demonstrates that the BPS limit of hairy black holes is a {\it singular} (i.e. it has $S_\BPS=0$ and the scalr field and curvature invariants diverge at the horizon). This is the main finding of our paper. In particular, this demonstrates that the perturbation theory of Section~\ref{sec:hbh-pert} must be breaking down as $E\to E_\BPS$, although here we are at even smaller values of $E$ (at fixed $J,Q$) than at the merger (recall that we expect our perturbation theory to be valid only for small $E,Q,J$). Yet, it must be breaking down in a very subtle way only for $E-E_\BPS\ll 1$   since Figs.~\ref{fig:entropyE-NumPert1}--\ref{fig:entropyE-NumPert2}
also undoubtedly demonstrate that it is a excellent approximation for a very wide range of energies (and $Q,J$) where HBHs exist, as long as we are not extremely close to the BPS configuration (and, of course if $(E,Q,J)$ is small). Very roughly, we can say that our perturbation theory fails only if we are in the $\sim 1\%$ (say) 3-dimensional region around the 2-dimensional BPS surface. This is very unique and puzzling and certainly deserves identifying the root-cause. We do this analysis in subsection~\ref{sec:PerturbativeFailing}. 

\begin{figure}[H]
\centering
\includegraphics[width=0.48\textwidth]{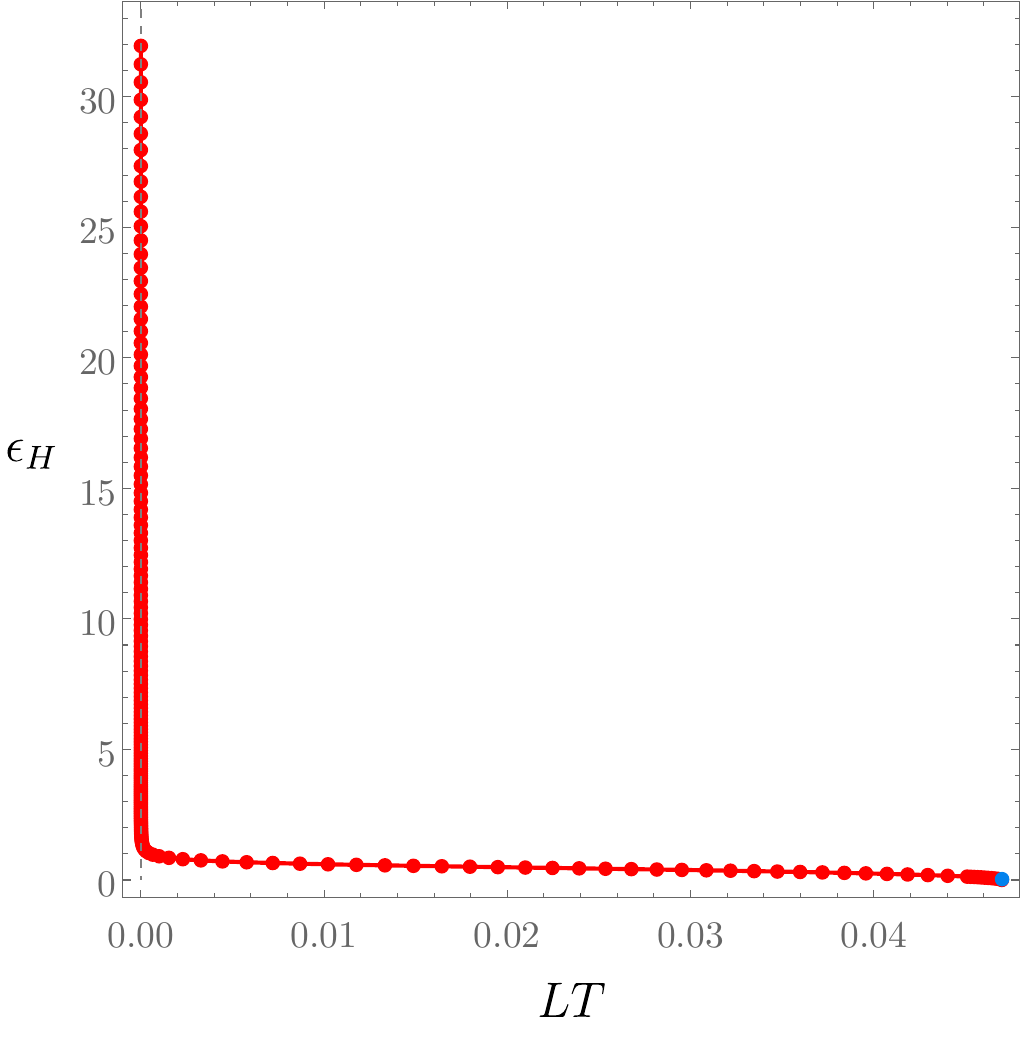}~\includegraphics[width=0.48\textwidth]{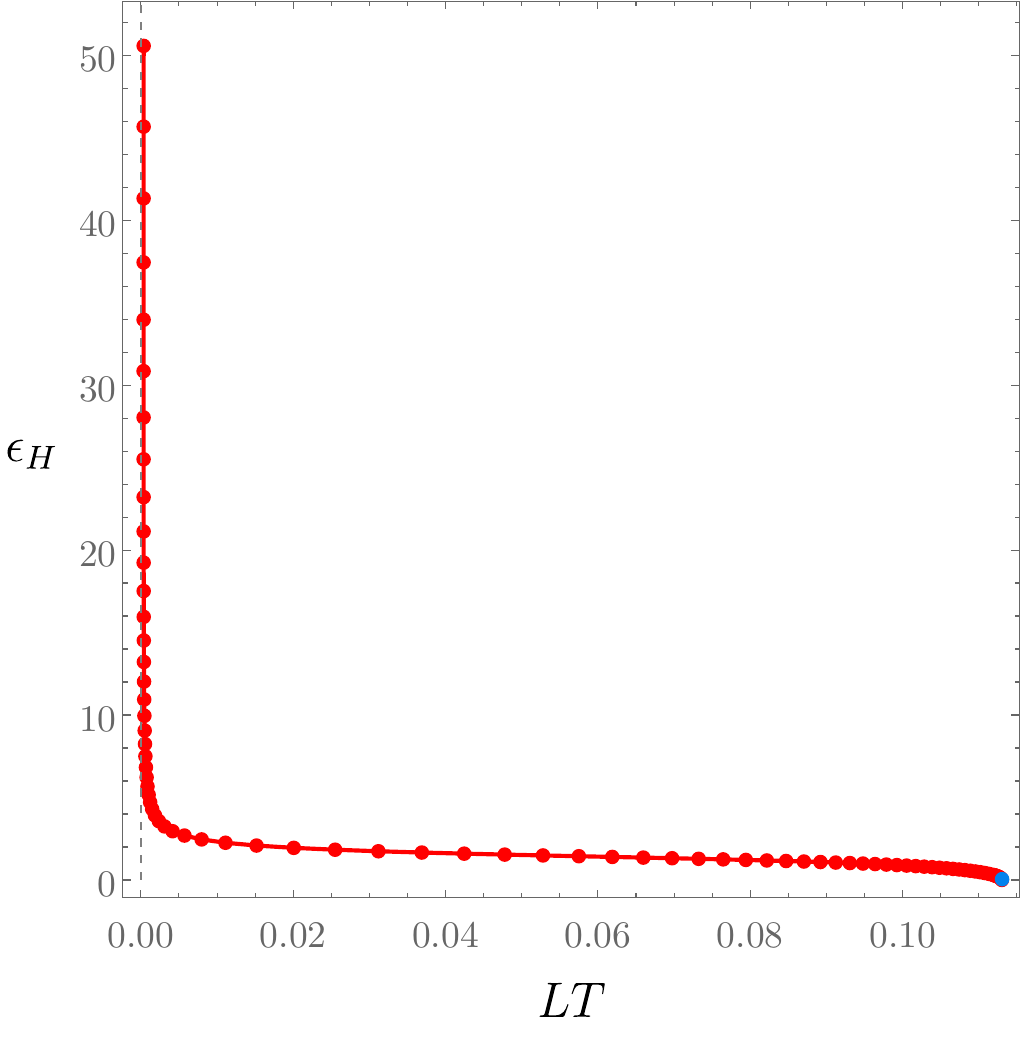}
\vspace{-0.25cm}
\caption{{\bf Left panel:} Hairy BHs with $ J /N^2=0.005$ and $Q L/N^2=0.05<Q_c(J)$. {\bf Right panel:} Hairy BHs with $ J /N^2=0.005$ and $Q L/N^2=0.1<Q_c(J)$.
In both panels, we plot the value of the scalar field $\Phi$ at the horizon as a function of the dimensional energy. The blue disk is the merger point between the hairy and the CLP families and the dashed vertical line signals $E=E_\BPS$.
}
\label{fig:eHdiverge}
\end{figure}

As an additional test of our findings, in Section~\ref{sec:susy-analysis}, we will directly search for supersymmetric HBH solutions by solving the BPS equations of the system. Here as well, we will find {\it no} evidence for the existence of a {\it regular} supersymmetric black hole, in particular for $Q<Q_c(J)$. Supersymmetric hairy black hole solutions seem to be unavoidably singular and we will argue that the fundamental reason for this is because the value $\e_H$ of the scalar field $\Phi$ diverges at the horizon in the BPS limit. This implies that curvature invariants, such as the Kretschmann scalar, also diverge at $r=r_+$. This is  illustrated in the left panel of Figs.~\ref{fig:epsilonE} or~\ref{fig:epsilonT} when $J/N^2=0.05$ and for $Q L/N^2$ or, even more unequivocally, in Fig.~\ref{fig:eHdiverge} for $J/N^2=0.005$ and $QL/N^2=0.05$ (left panel) and $QL/N^2=0.1$ (right panel) (these are thus the same solutions of Figs.~\ref{fig:entropyE-NumPert1}--\ref{fig:entropyE-NumPert2}). Altogether, the BPS limit of a HBHs is a {\it singular} supersymmetric BHs.
%%%%%%%%%%%%%%%%%%
%%%%%%%%%%%%%%%%%%

The main finding of our paper has a far-reaching consequence. As discussed in the Introduction, one could envisage a scenario for the phase diagram of the ($S_3$-invariant) $U(1)^3$ gauged supergravity theory \eqref{action} where 
the 2-parameter family of hairy supersymmetric black holes of \eqref{action} would exist as regular solutions and reduce to the 1-parameter Gutowski-Reall (GR) black hole family~\cite{Gutowski:2004ez} when the scalar field vanishes (recall that the GR solution is the special case $Q_1 = Q_2 = Q_3=Q$ and $J_1 = J_2=J$ of the Kunduri-Lucietti-Reall supersymmetric black hole~\cite{Kunduri:2006ek}).  
If this scenario were to hold true, this would solve a long standing puzzle since it would finally identify supersymmetric solutions with $\Delta_{\hbox{\tiny KLR}}\neq 0$, where the latter charge constraint is defined in \eqref{charge_constraint} (in the present context, with $Q_1 = Q_2 = Q_3=Q$ and $J_1 = J_2=J$). Unfortunately, we have however demonstrated that the sector of the theory that we studied does not have regular supersymmetric HBHs that reduce to the regular supersymmetric GR BH when the scalar field vanishes. The BPS limit of HBHs has zero entropy as well as divergent scalar field and divergent curvature invariants at the horizon, $r=r_+$.

%%%%%%%%%%%%%%%%%%%%%%%%%
\subsection{Revisiting the BPS limit of perturbation theory  \label{sec:PerturbativeFailing}}
%%%%%%%%%%%%%%%%%%%%%%%%%%%

In the previous subsection we have gathered exact numerical data that strongly suggests that the BPS limit of hairy black holes is a {\it singular} supersymmetric limit with $E\to E_\BPS$, $\mu\to1$, $\O_HL\to 1$, $TL\to 0$, and $ S\to 0$. This is in sharp disagreement -- as best illustrated in Figs.~\ref{fig:entropyE-NumPert1}--\ref{fig:entropyE-NumPert2} -- with the expectations of the perturbative and non-interacting thermodynamic analyses (of Section~\ref{sec:hbh-pert} and of~\cite{Bhattacharyya:2010yg}/Section~\ref{sec:toymodel}, respectively) which predict that the BPS limit of HBHs should have finite entropy. Thus, we must revisit our pertubative analyses to identify why it is failing as we approach the BPS limit.

To do this, we define a new parameter $\d$ by the equation
\begin{equation}
\begin{split}
\label{BPS_lim_analytics}
E - E_\BPS &= \frac{3}{4} \left( 2-\g-2\sqrt{1-\g} \right) y_+^4 +\CO(y_+^6,\e^2 y_+^4), \qquad 0 \leq \g \leq 1, \\
&\equiv  \frac{3}{4} \left( 1 - \sqrt{1-\delta} \, \right)^2 y_+^4  , \qquad 0 < \d \leq 1,
\end{split}
\end{equation}
where the RHS of the first line follows from \eqref{BPS_limit}, namely $E_\BPS=3Q+2J/L$, and we used the perturbative expressions \eqref{hbh_pert_thermo} for $(E,Q,J)$.
The BPS limit  then corresponds to take $\d \to 0$. Using \eqref{BPS_lim_analytics}, we can solve for $\g$ in terms of $\d$. This yields, to leading order, 
$\d = \g + \CO(y_+^2,\g  y_+^2, \g \e^2)$.
With this, we can rewrite all the perturbative thermodynamic quantities in terms of $\d$ and, in principle, analyse the BPS limit $\d \to 0$ of the hairy black hole. 

But there is a first-principles problem with this analysis. Clearly, the perturbative analysis of Section~\ref{sec:hbh-pert} relies on a double expansion in $y_+\ll 1$ and $\e \ll 1$ whereby the $\g$ parameter -- introduced in \eqref{introduceGamma} -- is implicitly assumed to be $\CO(1)$. However, we are interested on studying the $\delta\sim\g\sim 0$ limit of the system. This is at odds with the fact that, to get \eqref{hbh_pert_thermo}, we took Taylor expansions in $y_+ \ll 1$ and $\e \ll 1$ assuming that these two expansion parameters were not of the same order as $\g$ (and thus $\d$). That is, we implicitly assumed that $\d\gg y_+$ and $\d\gg\e$. It follows that analysing the $\delta\sim\g\to 0$ limit of \eqref{hbh_pert_thermo} is a delicate issue and we must proceed with caution. Note that this also applies to the analysis associated to \eqref{gamma_BPS}--\eqref{BPS_limit_pert_S} which must be revisited and scrutinized.

To proceed, we express the HBH thermodynamics \eqref{hbh_pert_thermo} in terms of $\delta$. For the discussion that will follow, it is enough to focus our attention in one of the thermodynamic quantities, temperature say, and the exact coefficients of the expansion are not relevant. It suffices to note that for small $\d$  the temperature of the black hole behaves as   
\begin{equation}
\begin{split}
\label{T_hbh_pert_delta}
&T L = \bigg( \CO(\d) y_+ + \CO(\d)  y_+^3 +\CO(\d \ln \d) y_+^5+ \CO \left( \d (\ln \d)^2 \right) y_+^7+\CO(y_+^9) \bigg) \\
& +  \e^2 \bigg( \CO(\d) y_+ + [ - 1/\pi + \CO(\d\ln \d) ] y_+^3 + \CO(y_+^5) \bigg) + \e^4  \bigg( \CO(\d) y_+ + \CO(y_+^3) \bigg) + \CO(\e^6). 
\end{split}
\end{equation}
In this expansion, a given sub-leading term should be (considerably) smaller that its precedent. From the successive terms in the $\e^0$ (and in the $\e^2$) contribution in \eqref{T_hbh_pert_delta}, one sees that this requires that  $|\!\ln \d| y_+^2 \ll 1$, i.e.,
\begin{equation}\label{validityPT}
|\!\ln \d| \ll y_+^{-2}\,.
\end{equation}
The fact that one has  
${\underset{\d \rightarrow 0}{\lim}\, \ln \d \to \infty}$
means that, if we are at {\it fixed} $y_+\ll 1$,  our perturbation theory breaks downs before $\d$ can vanish to yield the desired BPS limit. This explains why our exact numerical data is increasingly less well approximated by the perturbative expressions~\eqref{hbh_pert_thermo} as the BPS limit is approached:  see Figs.~\ref{fig:entropyE-NumPert1}--\ref{fig:entropyE-NumPert2}. It also explains why we {\it cannot} trust one of the (supposedly) main predictions  of perturbative analysis of Section~\ref{sec:hbh-pert}, namely that the entropy is finite in the BPS limit. That is, it reveals why the naive perturbative prediction is not matching the numerical findings of Section~\ref{sec:NumericalResults} where we clearly see that $S\to 0$ in the BPS limit: see again Figs.~\ref{fig:entropyE-NumPert1}--\ref{fig:entropyE-NumPert2}.   

But this is not the whole story because so far we have not yet fully analysed the contributions that are proportional to the scalar condensate $\e$ (which hairy black holes certainly contain, including in their BPS limit). This time, it is important to focus our attention in the first term $ \CO(\d) y_+$ of the first line of~\eqref{T_hbh_pert_delta} and compare it the second term $-\e^2 y_+^3$ of the second line (again, similar contributions are present in other thermodynamic quantities \eqref{hbh_pert_thermo}). We see that, because the latter has a negative contribution to $T$, to have a non-negative temperature one must necessarily have
\begin{equation}\label{T0hairyBHs}
\delta\gtrsim \CO(\e^2 y_+^2)\,.
\end{equation}
Note that since 
$|\!\ln (\e^2 y_+^2) | \ll y_+^{-2}$, this requirement is within the regime of validity~\eqref{validityPT} of the perturbative expansion. 
So, \eqref{T0hairyBHs} indicates that, if we assume that $y_+$ is {\it fixed} during the process, hairy black holes should reach the zero temperature configuration as one approaches $\delta\sim \CO(\e^2 y_+^2)$. That is to say, this analysis suggests that hairy black holes would never reach $E=E_\BPS$ (which corresponds to $\delta=0$)  because $\delta$ is capped from below at $\delta\sim \CO(\e^2 y_+^2)$ where the HBH system reaches $T=0$. In other words, there would be a small gap between the $T=0$ boundary of HBH and the BPS curve in Fig.~\ref{figJ0p05:phasediagram}.

In the arguments of the previous two paragraphs, $y_+$ is assumed to be fixed while we try to approach $\delta=0$. On the other hand, if we do not force the system to be at fixed $y_+$ as we approach the BPS limit, there is a sense in which we can rescue the perturbative and non-interacting thermodynamic analyses and reinterpret them to get expectations that do match the fact that the exact numerical results, in the BPS limit $\delta\to 0$, strongly suggests that one reaches a singular supersymmetric HBH with $E\to E_\BPS$, $\mu\to 1$, $\O_HL\to 1$, $TL\to 0$, and $S\to 0$.  
Indeed, we see that the condition~\eqref{validityPT} for our  perturbation theory to be valid can be satisfied when $\delta \to 0$ if (and only if) we {\it simultaneously} send $y_+\to 0$. In this `double' BPS limit one finds, from the perturbative expressions~\eqref{hbh_pert_thermo} and from the BPS limit~\eqref{BPS_limit_pert}-\eqref{BPS_limit_pert_S} of the non-interacting thermodynamic model, that one has $E\to E_\BPS$, $\mu\to1$, $\O_HL\to 1$, $T L\to 0$, and $S\to 0$. That is, the BPS limit of the hairy black holes should be a {\it singular} supersymmetric hairy black hole (this time, in agreement with the numerical findings of Section~\ref{sec:NumericalResults}). However, this reinterpretation is a hand-waved one since Figs.~\ref{fig:entropyE-NumPert1}--\ref{fig:entropyE-NumPert2} unequivocally demonstrate that perturbation theory simply breaks down close to the BPS limit, in agreement with~\eqref{validityPT}.

%%%%%%%%%%%%%%%%%%%%%%%%%%%%%%%%%%
\section{Solutions of the BPS equations}
\label{sec:susy-analysis}
%%%%%%%%%%%%%%%%%%%%%%%%%%%%%%%%%%

In Section~\ref{sec:MainResults}, we have gathered overwhelming evidence that the 2-parameter BPS limit of the 3-parameter family of hairy black holes is a {\it singular} supersymmetric black hole with $E\to E_\BPS$, $\mu\to1$, $\O_HL\to 1$, $TL\to 0$, and $S\to 0$ (with diverging scalar field and curvature invariants at $r=r_+=0$), except in the strict limit where the scalar field vanishes and one gets the 1-parameter family of Gutowski-Reall supersymmetric black holes~\cite{Gutowski:2004ez} with finite entropy. We have also seen that this is {\it naively} in contradiction with the outcome of the non-interacting thermodynamic (\cite{Bhattacharyya:2010yg} and Section~\ref{sec:toymodel}) and perturbative (Section~\ref{sec:hbh-pert}) analyses (which predict a finite entropy in the BPS limit) although, in Section~\ref{sec:PerturbativeFailing}, we understood that this is because these analyses breakdown (very) close to the 2-dimensional BPS surface. In this section, to collect bullet-proof evidence for the main claim of this paper, we provide a third argument that supports the conclusion above: we search directly for BPS solutions (instead of studying the $T\to 0$ limit of $T\neq 0$ hairy black holes). More precisely, we study the near-horizon behaviour of the BPS equations that govern supersymmetric solutions (necessarily with $E= E_\BPS$, $\mu= 1$, $\O_HL= 1$, $T= 0$) and show that, with certain reasonable assumptions, no supersymmetric hairy black hole solution exists.

%%%%%%%%%%%%
\subsection{BPS equations}
%%%%%%%%%%%%%

The action \eqref{action} is the bosonic part of a larger $\CN=2$ supersymmetric theory~\cite{Liu:2007rv}, which is a consistent truncation of $\CN=8$ gauged supergravity~\cite{Gunaydin:1985cu,Pernici:1985ju,Kim:1985ez,Khavaev:1998fb,Cvetic:1999xp,Cvetic:1999xx,Cvetic:2000eb,Ciceri:2014wya,Baguet:2015sma} (the latter is a consistent dimensional reduction of Type IIB supergravity on AdS$_5\times S^5$ along the $S^5$). Supersymmetric solutions in this system can be constructed by solving the Killing spinor or BPS equations~\cite{Liu:2007rv}. To describe these equations, we will need to move to a different set of coordinates. Given the functions $g$ and $h$ in the ansatz \eqref{3Q:ansatz}, we start by defining the function $H(x)$ via the following differential equation and boundary condition
\begin{equation}
\begin{split}
\label{Hx_def}
H'(x) = \frac{1}{\sqrt{ g \left( \sqrt{H(x)} \right) h \left( \sqrt{H(x)}\right) } }  , \qquad H(0) = \frac{r_+^2}{L^2} \equiv y_+^2.
\end{split}
\end{equation}
Next, we change coordinates by setting
\begin{equation}
\begin{split}
r = L \sqrt{H(x)} . 
\end{split}
\end{equation}
In the new coordinates, the ansatz \eqref{3Q:ansatz} takes the form
\begin{equation}
\begin{split}
\label{susy_gauge}
\dt s^2 &= - f \dt t^2 + L^2 \frac{\dt x^2}{4hfH}  + L^2 H \left[ \frac{1}{4} \dt \O_2^2  + h \left( \dt \psi + \frac{1}{2} \cos \t \dt \phi  - w \dt t \right)^2 \right] , \\
A &= A_t \dt t + A_\psi \left( \dt \psi + \frac{1}{2} \cos \t \dt \phi \right)  , \qquad \Phi = \Phi^\dagger = 2 \sinh \varphi . 
\end{split}
\end{equation}
The boundary condition in \eqref{Hx_def} implies that, in these new coordinates, the horizon is located at $x=0$.

The BPS equations for the system \eqref{action} and with the ansatz \eqref{susy_gauge} were derived in~\cite{Liu:2007rv}. It was shown that supersymmetric solutions satisfy the following algebraic identities
\begin{equation}
\begin{split}
\label{susy_rel}
f &= \frac{x^2 \eta}{h H^2}  , \quad~ h = \eta - \frac{4\o^2}{H^3} , \quad~ w = \frac{2 x \o}{L h H^3} , \quad~ A_t  = \frac{x}{H} , \quad~ A_\psi = 2 L \left( U + \frac{\o}{H} \right),
\end{split}
\end{equation}
where the algebraically independent functions $\{U,\eta,\o,\varphi,H\}$ satisfy the following set of differential equations
\begin{subequations} \label{BPSeqns}
\begin{align}
\label{eq1}  ( x U ) ' &= - H \cosh \varphi ,  \\
\label{eq2} [ x^2 ( \eta - 1 ) ]' &= - 6 x U \cosh \varphi , \\
\label{eq3} ( x^{-2} \o )' &= 3 H ( H \cosh \varphi + 2 U ) / ( 2 x^3 ) , \\
\label{eq4} \varphi' &=  2U \sinh \varphi / ( \eta x )  , \\
\label{eq5}  ( \eta x H' - \eta H + 4 U^2  )' &= 2  \o' \cosh\varphi  - H^2 \sinh^2 \varphi / x . 
\end{align}
\end{subequations}

%%%%%%%%%%%%%%%%%%%%%
\subsection{Near-horizon analysis of the BPS equations}
%%%%%%%%%%%%%%%%%%%%%%%%

In the rest of this section, we analyse the structure of solutions to the equations \eqref{susy_rel}-\eqref{BPSeqns} near the horizon, i.e. near $x=0$.\footnote{Since the metric ansatz \eqref{susy_gauge} was obtained via a coordinate transformation of \eqref{3Q:ansatz}, it is clear that the horizon is at $x=0$. However, we can also check that in \eqref{susy_gauge}, the horizon for BPS solutions must be located at $x=0$ (without making reference to \eqref{3Q:ansatz}). To see this, suppose the horizon is at $x=x_\CH\neq0$. Then, by definition we must have $f(x_\CH) =x_\CH^2 h(x_\CH)^{-1} y_+^{-4} \eta(x_\CH) =0$. Since $h(x_\CH)> 0$ and $r_+ > 0$, this implies $\eta(x_\CH)=0$. This then immediately leads to a contradiction since $h(x_\CH) = - 4r_+^{-6}\o(x_\CH)^2  \leq 0$. Consequently, the horizon is located at $x_\CH=0$} We are particularly interested in black hole solutions with smooth horizons. All such solutions satisfy\footnote{Note that we employ little-$o$ notation: If $f(x) = o(g(x))$, then $\lim\limits_{x\to0} (f/g) = 0$.}
\begin{equation}
\begin{split}\label{assumptions}
H(x) = y_+^2 + o(1) , \qquad h(x) = b^2 + o(1) , \qquad \eta(x) = o(x^{-2}) , \qquad \o(x) = o(x^{-1}) . 
\end{split}
\end{equation}
The first condition is necessary since we are interested in solutions with a horizon, i.e., non-vanishing entropy. In other words, we are specifically looking for regular supersymmetric black hole solutions, not solitonic ones (or even the singular supersymmetric BHs with $S=0$). The second condition in \eqref{assumptions}  imposes smoothness of the horizon. The horizon topology is warped $S^3$ with $b$ being the warping factor. The third requirement in \eqref{assumptions}  ensures that $f$ vanishes on the horizon (as it should, by definition). Finally, the fourth requirement is enforced by consistency with the first three. In this section, we will prove that if the assumptions \eqref{assumptions} hold, then:
\begin{enumerate}[leftmargin=*]
\item[(1)] If $\varphi$ is finite on the horizon, then $\varphi=0$ everywhere in spacetime.
\item[(2)] $\varphi$ cannot diverge monotonically at the horizon.
\end{enumerate}
The main ingredient in our proof will be the following theorem: Let $f:\mrr_+ \to \mrr$ and $g:\mrr_+ \to \mrr$ be monotonic functions in an open neighbourhood of $x=0$. Then,
\begin{equation}
\begin{split}
f'(x) = o(g'(x)) \quad \implies \quad f = \begin{cases}
C + o(g(x)-g(0)) &\text{if $g$ is bounded.} \\
o(g(x)) & \text{if $g$ is unbounded.}
\end{cases}
\end{split}
\end{equation}
We prove this theorem in Appendix~\ref{app:theorem_proof}.

The consequence of (1) and (2) are the following. (1) states that if $\varphi$ is finite on the black hole horizon, it must vanish everywhere, reducing our theory to minimal supergravity. This is the theory in which Gutowski and Reall (GR) first discovered the eponymous supersymmetric black hole~\cite{Gutowski:2004ez}. Our analysis recovers the GR black hole as a solution, as expected. (1) and (2) allow us to reject a large class of asymptotic behaviours near the horizon. Quite importantly, this gives credence to our conjecture that a {\it regular} supersymmetric hairy black hole described by \eqref{3Q:ansatz} does not exist since, as clearly illustrated in Fig.~\ref{fig:eHdiverge}, the scalar field $\Phi$ (and thus $\varphi$) diverges at the horizon. Consequently, curvature invariants such as the Kretschmann scalar also diverge. Of course, this does not rule out a solution in which the charged scalar diverges while oscillating infinitely, e.g. $x^{-1} \sin(1/x)$. In principle, it is still possible that such exotic hairy black holes exist as solutions to our system.

Next, we prove (1) and then (2).

\paragraph{Proof of (1):} In this case, we assume that $\varphi$ behaves as
\begin{equation}
\begin{split}
\label{ass_1}
\varphi = \varphi_0 + o(1),
\end{split}
\end{equation}
with $\varphi_0$ being finite. Since the BPS equations have a $\varphi \to -\varphi$ symmetry, we can assume, without loss of generality, that $\varphi_0 \geq 0$. Near $x=0$, \eqref{eq1} simplifies to 
\begin{equation}
\begin{split}
\label{U_sol_susy}
(x U)'  = - y_+^2 \cosh \varphi_0 + o(1)  \quad \implies \quad U = - \frac{C}{x} - y_+^2 \cosh \varphi_0 + o(1). 
\end{split}
\end{equation}
We first assume that $C \neq 0$. Then, near $x=0$, \eqref{eq2} simplifies to
\begin{equation}
\begin{split}
[ x^2 ( \eta - 1 ) ]' &= 6 C  \cosh \varphi_0 + o(1) \quad \implies \quad  \eta = \frac{C'}{x^2}  + \frac{6 C \cosh \varphi_0}{x} +  o(x^{-1}) .
\end{split}
\end{equation}
The third condition in \eqref{assumptions} implies that $C' = 0$. We next turn to \eqref{eq3} which simplifies near $x=0$ to
\begin{equation}
\begin{split}
( x^{-2} \o )' &=  -  \frac{3 C y_+^2}{x^4} + o(x^{-4}) \quad \implies \quad \o = \frac{C y_+^2}{x} + o(x^{-1}) , \\
\end{split}
\end{equation}
The fourth condition in \eqref{assumptions} now implies that $C = 0$. We now go back to \eqref{U_sol_susy} and set $C=0$. Redoing our analysis in these conditions, we find
\begin{equation}
U = - y_+^2 \cosh \varphi_0 + o(1) , \quad \eta =  1 + 3  y_+^2 \cosh^2 \varphi_0 + o(1) , \quad \o= \frac{3}{4} y_+^4 \cosh \varphi_0 + o(1) . 
\end{equation}
With this, \eqref{eq4} near $x=0$ simplifies to
\begin{equation}
\begin{split}
\varphi' =  -  \frac{y_+^2 \sinh ( 2 \varphi_0 )}{1 + 3  y_+^2 \cosh^2 \varphi_0} \frac{1}{x} + o(x^{-1} ) ~\implies~ \varphi =  -  \frac{y_+^2 \sinh ( 2 \varphi_0 )}{1 + 3  y_+^2 \cosh^2 \varphi_0} \ln x + o(\ln x) . \\
\end{split}
\end{equation}
The logarithmic divergence at $x=0$ is inconsistent with \eqref{ass_1}, unless we have $\varphi_0=0$. Using this, we reconsider \eqref{eq4}. Near $x=0$, the differential equation now reduces to
\begin{equation}
\begin{split}
x \varphi' &=  -  \frac{2y_+^2}{1 + 3  y_+^2} \varphi \quad \implies \quad \varphi = K x^{- \frac{2y_+^2}{1+3y_+^2} } . 
\end{split}
\end{equation}
Since the power of $x$ is negative, this result contradicts our assumption that $\varphi$ is finite on the horizon unless the integration constant $K=0$. However, $K$ parameterises the first deviation of $\varphi$ away from the horizon. Setting this to zero implies that $\varphi$ must vanish \emph{everywhere} in the spacetime. This completes the proof of (1).

We additionally note that when $\varphi=0$, the remaining functions behave as
\begin{equation}
\begin{split}
U &= - y_+^2 + o(1) , \quad~ \eta =  1 + 3  y_+^2 + o(1) , \quad~ \o= \frac{3}{4} y_+^4 + o(1) , \quad~ H = y_+^2 + o(1) .
\end{split}
\end{equation}
This is precisely the near-horizon behaviour of the Gutowski-Reall  black hole~\cite{Gutowski:2004ez}.

\paragraph{Proof of (2):} In this case, we can assume that $\varphi \to +\infty$ as $x \to 0$ without loss of generality. Near $x=0$, \eqref{eq1} simplifies to
\begin{equation}
\begin{split}
\label{U_form_2}
( x U ) ' = - \frac{y_+^2}{2} e^\varphi \quad \implies \quad U = \frac{C}{x} - \frac{y_+^2}{2x} \rho + o ( x^{-1} \rho) , \qquad \rho' = e^\varphi . 
\end{split}
\end{equation}
Let us now consider three cases:
\begin{itemize}[leftmargin=*]
\item $\rho$ diverges monotonically as $x\to0$: In this case, $C/x$ is the subdominant term in $U$. Using this, \eqref{eq2} simplifies near $x=0$ to
\begin{equation}
\begin{split}
[ x^2 ( \eta - 1 ) ]' = \frac{3}{4} y_+^2 (\rho^2)'  \quad \implies \quad \eta   = \frac{3 y_+^2 \rho^2}{4 x^2} + o(x^{-2}\rho^2).
\end{split}
\end{equation}
This violates \eqref{assumptions} so we reject it as a possibility.

\item $\rho=o(1)$ near $x=0$ ($\rho' = e^\varphi$ diverges):\footnote{If $\rho=\rho_0 +o(1)$ near $x=0$, then we can simply redefine the function $\rho \to \rho - \rho_0$ so that $\rho_{new} = o(1)$.} In this case, $C/x$ is the dominant term in $U$. \eqref{eq2} then simplifies to
\begin{equation}
\begin{split}
[ x^2 ( \eta - 1 ) ]' &= - 3 C \rho'  \quad \implies \quad \eta = \frac{C'}{x^2} - \frac{3 C \rho}{x^2} + o(x^{-2}\rho) . 
\end{split}
\end{equation}
The third condition in \eqref{assumptions} implies that $C'=0$. \eqref{eq3} then simplifies near $x=0$ to
\begin{equation}
\begin{split}
\o &= - \frac{Cy_+^2}{x}  + o(x^{-1}), .
\end{split}
\end{equation}
where we used the fact that $\rho'$ must diverge slower $1/x$ as $x\to0$. The fourth condition in \eqref{assumptions} then implies that $C=0$. Setting $C=0$ in \eqref{U_form_2} and redoing our analysis, we find
\begin{equation}
\begin{split}
U = - \frac{y_+^2\rho}{2x}  + o ( \rho/x) , \qquad \eta =  \frac{3y_+^2\rho^2}{4x^2} + o(\rho^2/x^2) 
\end{split}
\end{equation}
where we used the fact that $\rho$ must vanish slower than $x$ as $x \to 0$. This implies that $\eta$ diverges as $x \to 0$. Using this, we find that \eqref{eq3} simplifies to
\begin{equation}
\begin{split}
\o &= \frac{3 y_+^4}{4} x^2 \int \dt x \frac{( x^{-2} \rho )'}{x} + o \left( x^2 \int \dt x \frac{( x^{-2} \rho )'}{x} \right) .
\end{split}
\end{equation}
Note that since $\rho$ vanishes slower than $x$ as $x \to 0$, $\o$ diverges as $x\to0$.
We now use these results to evaluate $h$ as given in \eqref{susy_rel}:
\begin{equation}
\begin{split}
h &= \eta - \frac{4\o^2}{H^3} \\
&= \frac{3y_+^2}{4} \left[ \frac{\rho^2}{x^2} - 3 \left( x^2 \int \dt x \frac{( x^{-2} \rho )'}{x} \right)^2 \right] + o \left( x^2 \int \dt x \frac{( x^{-2} \rho )'}{x} \right)+ o(\rho^2/x^2) 
\end{split}
\end{equation}
The second condition in \eqref{assumptions} requires that $h$ be finite as $x \to 0$. However, both  terms inside the square brackets are separately divergent in this limit. Consistency with \eqref{assumptions} then requires that the two divergences must cancel each other. This requires that 
\begin{equation}
\begin{split}
\frac{\rho^2}{x^2} - 3 \left( x^2 \int \dt x \frac{( x^{-2} \rho )'}{x} \right)^2 = 0 \,.
\end{split}
\end{equation}
This implies a differential equation for $\rho$, which we can solve
\begin{equation}
\begin{split}
( x^{-3} \rho )' = \pm \sqrt{3} \frac{( x^{-2} \rho )'}{x} \quad \implies \quad \rho = K\, x^{\frac{1}{2} ( 3 \mp \sqrt{3} ) } .
\end{split}
\end{equation}
The requirement that $\rho=o(1)$ and $\rho' \to \infty$ implies that we take the upper sign above. It follows that this case could only work if $\rho$ diverges as a power law. In this case, we then find
\begin{equation}
\begin{split}
U &= - \frac{K\, y_+^2}{2} x^{\a-1} + o (x^{\a-1}) , \\
\eta &=  \frac{3K^2y_+^2}{4} x^{2(\a-1)} + o(x^{2(\a-1)}) , \\
\o &= \frac{3 K y_+^4}{4} \frac{\a-2}{\a-3} x^{\a-1}  + o(x^{\a-1}) , \\
e^\varphi &= K \, \a x^{\a-1}  + o(x^{\a-1}).  
\end{split}
\end{equation}
where $\a = \frac{1}{2} ( 3 - \sqrt{3} ) \approx 0.633975$. However, if we plug in this form into \eqref{eq4} and expand near $x \to 0$, we find
\begin{equation}
\begin{split}
( e^{-\varphi} )' =  U / ( x \eta ) \quad \implies \quad \frac{1-\a}{K\,  \a} x^{-\a} =  - \frac{2}{3 K} x^{-\a}  + o(x^{-\a}). 
\end{split}
\end{equation}
It is immediately clear that this equation does not hold for $\a = \frac{1}{2} ( 3 - \sqrt{3} )$. We have, therefore, reached a contradiction, and we reject this possibility.
\end{itemize}

To conclude, in this section we looked for a supersymmetric hairy black hole by solving directly the BPS equations (more precisely by studying the near-horizon behaviour of the solutions). We find evidence that supersymmetric black holes can only exist if the scalar field at the horizon is not finite. Consequently, the pullback of curvature invariants to the horizon also diverge. That is to say, the analysis of the BPS equations is consistent with the main exact numerical results of Section~\ref{sec:NumericalResults}: when hairy black holes approach the BPS limit, they become {\it singular} because the scalar field  and curvature invariants diverge at the horizon.

%%%%%%%%%%%%%%%%%%%%%%%%%%%%%%%%%%%%%%%%%%%%%%%%%%%
\section{Summary and Comments}
\label{sec:conclusion}
%%%%%%%%%%%%%%%%%%%%%%%%%%%%%%%%%%%%%%%%%%%%%%

The phase diagram of asymptotically AdS$_5\times S^5$ black hole solutions of Type IIB supergravity plays a fundamental role in the AdS/CFT correspondence~\cite{Maldacena:1997re, Gubser:1998bc, Witten:1998qj, Aharony:1999ti} and is expected to be very rich (starting with the solutions in~\cite{Witten:1998qj}). In this paper,  we have contributed to our understanding of the phase diagram of a sector of the full theory.

More concretely,  we started by using the fact that a dimensional reduction of Type IIB supergravity along the $S^5$ yields 5D $\CN=8$ gauged supergravity~\cite{Gunaydin:1985cu, Pernici:1985ju},  which is a consistent reduction of the full Type IIB supergravity on AdS$_5\times S^5$~\cite{Gunaydin:1985cu, Pernici:1985ju, Kim:1985ez, Khavaev:1998fb, Cvetic:1999xp, Cvetic:1999xx, Cvetic:2000eb, Ciceri:2014wya, Baguet:2015sma} (the full non-linear reduction ansatz\"e for the AdS$_5\times S^5$ compactification of IIB supergravity was worked out in~\cite{Baguet:2015sma}). The 10D Type IIB fields $\{g_{ab}, \Phi, C, B_\2, C_\2, C_\4 \}$ are equivalently encoded in the 5D spectrum of gauged $\CN=8$ supergravity whose field content consists of one graviton, 15 $SO(6)$ gauge fields, 12 two-form gauge potentials in the $6+\widebar{6}$ representations of $SO(6)$, 42 scalars in the $1+1+20^\prime+10+\overline{10}$ representations of $SO(6)$ and the fermionic superpartners. This theory contains too many fields and constructing generic solutions thereof is practically impossible. A further consistent truncation of this theory, known as $SO(6)$ gauged supergravity~\cite{Cvetic:2000nc}, is obtained by setting (in the bosonic sector) all fields except the graviton, 15 $SO(6)$ gauge fields and the $20^\prime$ scalars, to zero. This theory is obtained by a dimensional reduction of the the $SL(2,\mrr)$-invariant sector of Type IIB supergravity, which retains only the 10D graviton and self-dual 5-form field strength. This still has too many fields to solve for. Fortunately, there is a further (and final) consistent truncation which breaks the $SO(6)$ gauge group down to its $U(1)^3$ Cartan subgroup. The bosonic fields of this truncation, known as $U(1)^3$ gauged supergravity~\cite{Liu:2007rv}, are the graviton, two neutral real scalar fields $\{\varphi_1, \varphi_2\}$\footnote{It is often convenient to replace the two real scalar fields $\{\varphi_1, \varphi_2\}$ with 3 real scalars $\{X_1,X_2,X_3\}$ subject to the constraint $X_1X_2 X_3=1$.}, and 3 complex scalar fields $\{\Phi_1,\Phi_2,\Phi_3 \}$ that are charged under three $U(1)$ gauge field potentials $\{A^1_{(1)},A^2_{(1)},A^3_{(1)}\}$. For reference, when the charged scalar fields $\Phi_{1,2,3}$ vanish, 
$U(1)^3$ gauged supergravity reduces to the STU model~\cite{Behrndt:1998jd}, and when all the scalar fields are set to zero (including the neutral scalars) and the gauge fields are set equal, the theory reduces to minimal gauged supergravity~\cite{Gunaydin:1983bi}. Black hole solutions of STU model are fully known~\cite{Behrndt_1999, Cvetic:2004ny, CVETIC2004273, Gutowski:2004ez, Gutowski:2004yv, Chong:2005hr, Chong:2005da, Chong:2006zx, Cvetic:2005zi,Kunduri:2006ek, MEI200764, Wu:2011zzh, Wu:2011gq}.\footnote{The most general non-extremal black hole solution with $\Phi_{1, 2, 3}=0$~\cite{Wu:2011gq} has 6 conserved charges: the energy $E$,  three $U(1)$ electric charges $\{Q_1, Q_2, Q_3\}$,  and two independent angular momenta $\{J_1, J_2\}$ along the two independent rotation planes of AdS$_5$ with $SO(4)$ symmetry. In the holographic dictionary,  the dual thermal states in $\CN=4$ SYM have $SU(4)\cong SO(6)$ $R$-charge given by the weight vector $(Q_1, Q_2, Q_3)$ and chemical potentials $\{\mu_1, \mu_2, \mu_3\}$ given by the sources of $\{A^1_{(1)}, A^2_{(1)}, A^3_{(1)}\}$~\cite{Cvetic:1999xp}. On the other hand,  $(J_1, J_2)$ is proportional to a weight vector of the four dimensional rotation group $SO(4)$. In the dual CFT language,  one usually works with $J_L \equiv J_1 + J_2$ and $J_R \equiv J_1 - J_2$,  which are proportional to the weights with respect to the two $SU(2)$ factors in $SO(4) \sim SU(2)_L \times SU(2)_R$~\cite{Kunduri:2006ek}. In this paper,  we were `only' interested on the most general non-extremal solution of~\cite{Wu:2011gq} with  arbitrary $Q_1=Q_2=Q_3$ and $J_1=J_2$) found by Cveti\v{c}-L\"u-Pope~\cite{CVETIC2004273}.}

In the present paper, we focused our attention on the $S_3$-invariant sector of $U(1)^3$ gauged supergravity, which sets $\Phi_{1,2,3} \equiv \Phi$, $A_\1^{1,2,3}\equiv A$, and $\varphi_{1,2} \equiv 0$. This theory is described by action~\eqref{action}. The most general `bald' ($\Phi=0$) black holes of this theory with $Q_{1,2,3}\equiv Q$ and equal angular momenta $J_1=J_2$ were found by Cveti\v{c}-L\"u-Pope~\cite{CVETIC2004273} (when $J=0$ this is the  Behrndt-Cveti\v{c}-Sabra solution with $Q_{1,2,3}\equiv Q$ ~\cite{Behrndt_1999}; see also~\cite{Dias:2022eyq}).\footnote{\label{foot:uplift} We reinforce that $\CN=8$, $SO(6)$, and $U(1)^3$ gauged supergravity are consistent truncations in the sense that every solution of 5D supergravity lifts to a solution of 10D Type IIB supergravity~\cite{Gunaydin:1985cu, Pernici:1985ju, Kim:1985ez, Khavaev:1998fb, Cvetic:1999xp, Cvetic:1999xx, Cvetic:2000nc, Cvetic:2000eb, Ciceri:2014wya, Baguet:2015sma}. More precisely, a solution of 5D $SO(6)$ gauged supergravity (which consists of $20^\prime$ scalar fields $T_{ij}$, 15 $SO(6)$ gauge fields $(A_a)_{ij}$ and a 5D metric $g_{ab}$) uplifts to a $SL(2,\mrr)$-invariant solution of 10D Type IIB supergravity which retains only the 10D metric ${\hat g}_{mn}$ and self-dual 5-form field ${\hat H}_{pqrst}$. Explicitly, the 10D solution is given by~\cite{Cvetic:2000nc}
\begin{align*}
&{\hat g}_{mn} \dt x^m \dt x^n = \D^{1/2} g_{ab} \dt x^a \dt x^b +  L^2  \D^{-1/2} \s^\text{T}_\1 T \s_\1 , \qquad \s_\1 = D \mu T^{-1}  , \\
&{\hat H}_\5 = {\hat G}_\5 +  {\hat \star} {\hat G}_\5 \quad \text{where} \quad {\hat G}_\5 = \frac{1}{L}  U \e_\5 + L \s^\text{T}_\1 \w \star D T \mu + \frac{L^2}{2} \s^\text{T}_\1 \w  \star F_\2 \w  \s_\1  .
\end{align*}
Here, $\mu$ is a 6D vector with unit-norm ($\mu^\text{T} \mu = 1$) that describes the standard embedding of $S^5 \hookrightarrow \mrr^6$, $\D \equiv \mu^\text{T} T \mu$, $U \equiv 2  \mu^\text{T} T^2 \mu - \D \Tr\,T$, $F_\2 \equiv \dt A_\1 + L^{-1}  A_\1 \w A_\1$, $D T \equiv \dt T + L^{-1}   [ A_\1 ,  T ]$, and $D \mu \equiv \dt \mu + L^{-1}  A_\1 \mu$. Solutions of $U(1)^3$ gauged supergravity can be extended to solutions of $SO(6)$ gauged supergravity (see~\cite{Dias:2022eyq}) and therefore, to those of 10D Type IIB supergravity.} These CLP back holes are to be seen as the `Kerr-Newmann--AdS$_5$ black hole' solution of the theory (it exactly reduces to the Reissner-Nordstr\"om--AdS$_5$ when $J=0$ solution since the Chern-Simons term in the action~\eqref{action} vanishes in this case). One of our aims was to study (in perturbation theory and numerically) the spectrum of hairy black holes of the theory with $\Phi\neq 0$, thus complementing the non-interacting thermodynamic model of~\cite{Bhattacharyya:2010yg} and the numerical studies of~\cite{Markeviciute:2018cqs, Markeviciute:2018yal}. Hairy black holes are a 3-parameter family of solutions that we can take to be the energy $E$, charge $Q_{1,2,3}\equiv Q$ and angular momentum $J_{1,2}\equiv J$ (there are other physical observables like the VEVs of the scalar and gauge fields that are related to these charges). We find exact agreement with the numerical data of~\cite{Markeviciute:2018cqs, Markeviciute:2018yal}, if we restrict to the parameter space region that was analysed there. In this paper, we extend the analysis of~\cite{Markeviciute:2018cqs, Markeviciute:2018yal} to much lower temperatures. Additionally, for most of the parameter space, as long as we are not too close to the BPS surface (more below), the perturbative analysis that we perform in Section~\ref{sec:hbh-pert} matches our exact numerical results of Sections~\ref{sec:numerics}~$\&$~\ref{sec:NumericalResults} and of~\cite{Markeviciute:2018cqs, Markeviciute:2018yal}. In particular, we find that there exists a critical curve $Q_c(J)$ at the BPS surface $E_\BPS(Q,J)$. Hairy black holes  approach the BPS surface quite differently depending on whether they reach the BPS limit below or above this critical curve $Q_c(J)$. Precise distinctions between the two cases are best summarized in Figs.~\ref{fig:epsilonT},~\ref{fig:vevT},~\ref{fig:entropyT} and associated discussions in Section~\ref{sec:NumericalResults}. Here, it suffices to highlight that, in the BPS limit, the temperature of hairy black holes approaches zero (a finite value) if they reach the BPS surface below (above) the critical curve $Q_c(J)$.
%%%%%%%%%%%%%
\begin{figure}[h]
\centering
\includegraphics[width=0.45\textwidth]{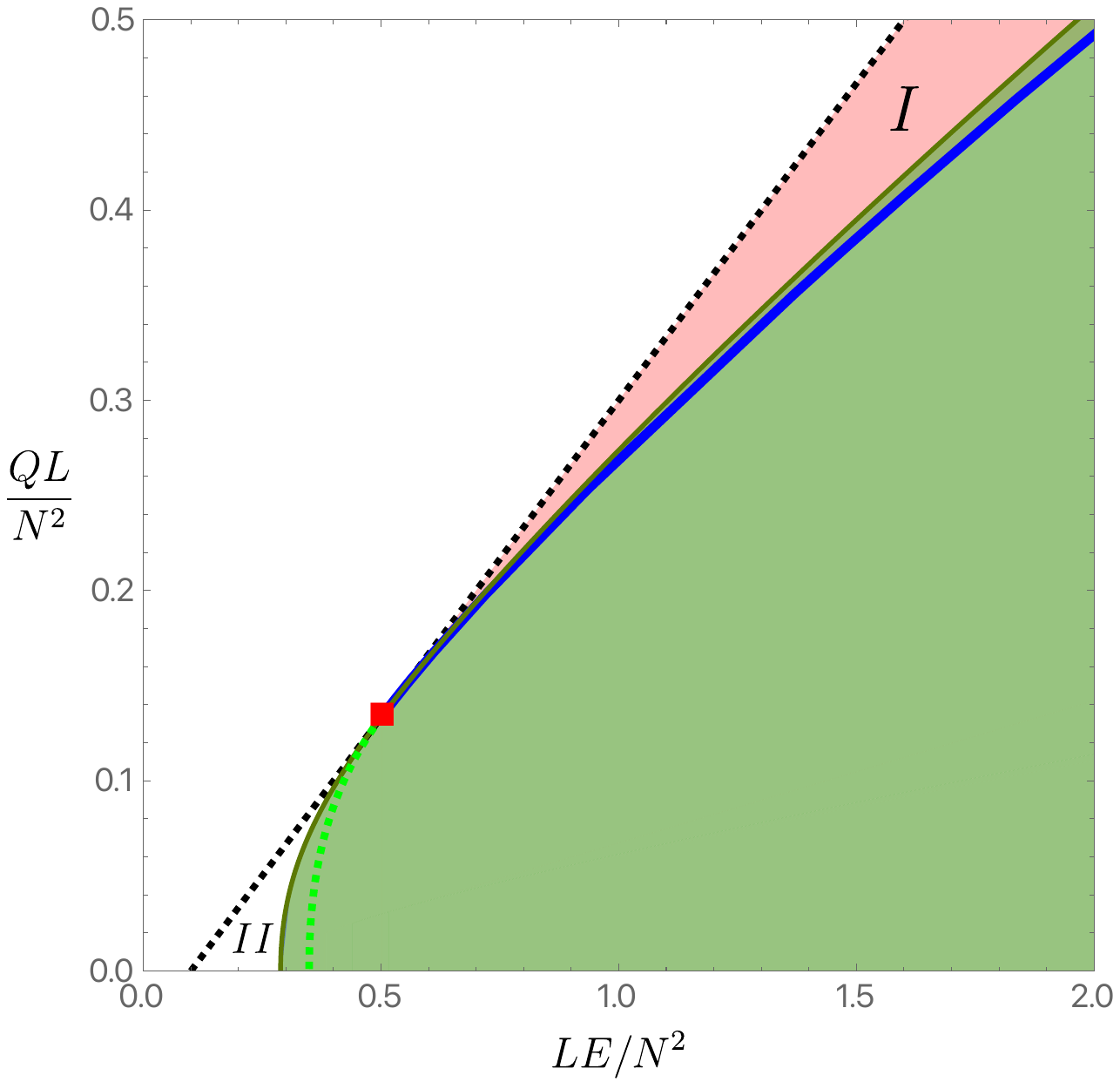} \quad
\includegraphics[width=0.45\textwidth]{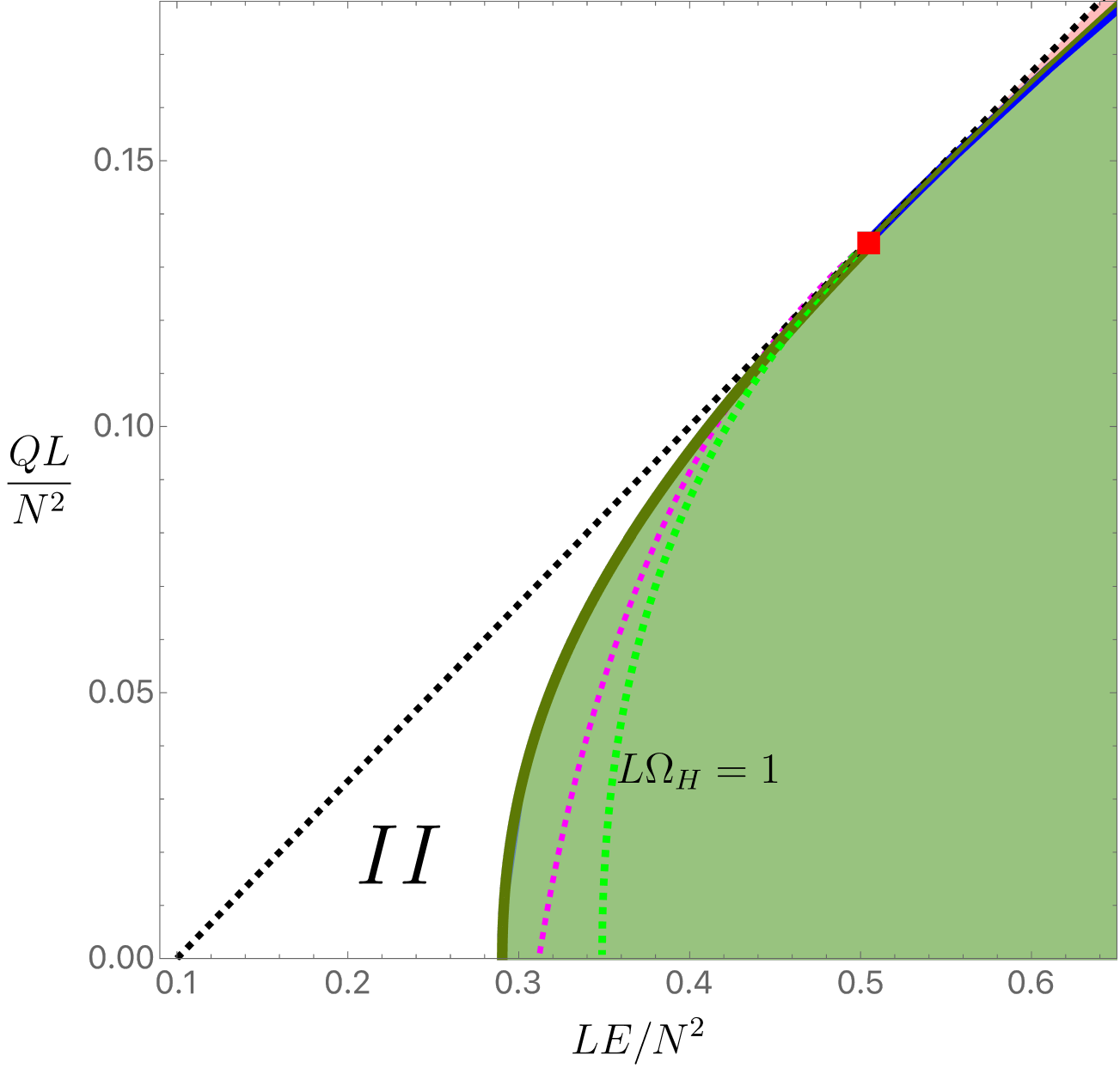}
\caption{Phase diagram charge vs energy for fixed $ J /N^2$. The green area describes regular CLP black holes, with the upper dark-green boundary of this region being the extremal CLP with $T=0, S \neq 0$. The black dashed line describes the BPS line with $Q= \frac{1}{3}\left(E-2 J/L \right)$.   
The blue curve describes the merger line CLP and hairy black holes. 
It meets both the BPS and extremal CLP curves at the red square  %$(E, Q)\frac{L}{N^2}\simeq (0.502733, 0.134244)$ 
which describes the Gutowski-Reall (GR) supersymmetric BH. 
Hairy BHs exist in region $I$ above the blue merger line all the way up to the BPS line, including in the light red region where CLP BHs do not exist. The light green curve that starts at the GR BH and extends to $Q=0$ has $\O_H L=1$. The plot on the right panel is a zoom in plot that focus on white region $II$. The magenta dashed curve is explained in the text.}
\label{fig:phasediagramConc}
\end{figure}
%%%%%%%%%%%%%

In a phase diagram $(E,Q,J)$ of stationary solutions of the theory, the 3-parameter family of hairy black holes exists in a very wide volume (very much like the CLP black hole family). More concretely, the 3-parameter family of CLP black holes exists in a volume delimited by the surface $Q=0$ and the extremal surface, where their temperature $T$ vanishes and their entropy is finite (these are regular extremal CLP black holes). On the other hand, hairy black holes exist in a volume whose boundaries are the merger surface with the CLP family of black holes (where the scalar hairy vanishes) and the BPS surface $E=E_\BPS=3Q+2J$. Hairy and CLP black holes coexist in a wide volume delimited by the merger surface and CLP's extremal surface. This discussion of the phase diagram is more clear if we fix the angular momentum $J$ and display the 2-dimensional phase diagram as sketched in~Fig.~\ref{fig:phasediagramConc} (this figure is for $J/N^2=0.05$ but the qualitative features are universal, i.e. independent of the particular value of $J\neq 0$ that is fixed).\footnote{If $J=0$ the dashed black BPS line starts at $E=Q=0$ and the square point is at $E=Q=0$ but it no longer describes a regular sumpersymmetric Gutowski-Reall black hole. The dark-green extremal CLP and blue merger curves still meet at this point.} In this figure, CLP black holes exist in the green shaded area with upper boundary (dark-green curve) given by the extremal CLP family with $T=0$ and $S\neq 0$. CLP black holes become unstable to the condensation of the scalar field $\Phi$ of the theory~\eqref{action} at the blue curve (Section~\ref{sec:Onset3Q}): CLP black holes above this blue onset curve (and below CLP's extremal curve) are unstable. This instability onset curve is also the merger line between CLP and hairy black holes in the limit where the latter reach $\Phi=0$. Hairy black holes exist above this blue merger line and the dashed black BPS curve (the value $\e_H$ of $\Phi$ at the horizon and the  VEV $\langle \CO_\Phi \rangle$  of the scalar field are both zero at the merger line and then increase as we move away from it as shown in Figs.~\ref{fig:epsilonE}-\ref{fig:vevT}). This includes the green-shaded area between the merger and the extremal CLP lines (where hairy black holes coexist with CLP black holes), but also the light-red shaded area (region $I$) where hairy black holes are the only stationary black hole solutions of the theory. The BPS (dashed black), extremal CLP (dark-green) and merger (blue) lines all meet at the red square which describes the  Gutowski-Reall supersymmetric black hole of the theory~\cite{Gutowski:2004ez} (for the particular value of $J$). Recall that the latter is a 1-parameter family of black holes (and thus a point in Fig.~\ref{fig:phasediagramConc} where we fix $J$).   

Quite importantly, for values of $(E,Q,J)$ where hairy black holes co-exist with CLP black holes, the former always have the highest entropy, as best illustrated in Fig.~\ref{fig:entropyE}. That is to say, hairy black holes dominate the microcanonical ensemble. It follows that, in a time evolution process where we keep $E,Q,J$ fixed and we perturb a CLP black hole that sits in the unstable region, it is natural to expect that the system will evolve towards a hairy black hole with the same $E,Q,J$. If the hairy black hole is not the endpoint of the instability it should be at least a metastable state of the evolution.  

Furthermore, we find that our hairy black holes fill an important gap in our understanding of the gauge/gravity duality of the system. In its weakest form (low energies) this is the duality between Type IIB supergravity in global AdS$_5\times S^5$ with radius $L$ and $N$ units of self-dual $F_5$ flux on $S^5$ and $\mathcal{N}=4$ SYM on the Einstein Static Universe with gauge group $SU(N)$. In this context, CLP and hairy black holes with Hawking temperature $T$ and chemical potentials $(\mu,\O_H)$ are dual to thermal states of SYM with temperature $T$ and chemical potentials $(\mu,\O_H)$. Now, from the SYM perspective, there is no reason not to have thermal states in the light-red region $I$ of Fig.~\ref{fig:entropyE}. But CLP black holes do not exist in this region and thus they could not be the required dual gravitational solutions. As Fig.~\ref{fig:entropyE} shows, the hairy black holes are the missing gravitational solutions (there could be others, not known so far) since they exist above the blue merger line including in the light-red region $I$ and all the way up to the dashed black BPS curve.

Our main motivation to revisit 
the numerical construction of hairy black holes performed in~\cite{Markeviciute:2018cqs, Markeviciute:2018yal} is related to the BPS limit of the hairy black holes. In this limit, solutions of the light-red region $I$ of Fig.~\ref{fig:entropyE} approach the dashed black BPS curve. As discussed in the Introduction (Section~\ref{sec:intro}), there is a longstanding open problem in $U(1)^3$ gauged supergravity (or more generically, in $SO(6)$ gauge supergravity) and associated gauge/gravity duality. Namely, from the SYM perspective the supersymmetric thermal states should have two fugacities, i.e. they should be described by two parameters. However, the only known supersymmetric black hole of the theory is the Gutowski-Reall black hole~\cite{Gutowski:2004ez}, which is a 1-parameter solution ($J$, say; and thus this solution is the red square in Fig.~\ref{fig:entropyE} where we fix $J$). Indeed, recall that the solutions of~\cite{Gutowski:2004ez} satisfy $E=E_\BPS=3Q+2J$, which suggests that it is a 2-parameter family, but, in addition, it also satisfies the charge constraint $\Delta_{\hbox{\tiny KLR}}=0$ in \eqref{charge_constraint} with $Q_{1,2,3}\equiv Q$ and $J_{1,2}$. At the end of the day, the Gutowski-Reall solution is `simply' a 1-parameter family of supersymmetric black holes. Therefore, there is a missing gravitational parameter. This problem has also received some attention recently due to developments in the microstate counting of $\ads_5$ black holes via the SYM index \cite{Hosseini:2018tha, Choi:2018hmj, Benini:2018ywd, Cabo-Bizet:2018ehj, Larsen:2019oll, Benini:2020gjh}.\footnote{In AdS/CFT, black hole microstates can be counted by evaluating the partition function of the CFT, which is incredibly hard due to it being strongly coupled. However, it can be shown that the partition function in SYM is related to the Witten index with complex fugacities. The index can be evaluated at strong coupling.} In $S_3$-invariant sector considered in this paper, the SYM index also depends on only one parameter, which matches the parameter in the GR black hole exactly. The potential existence of new supersymmetric (hairy) black holes with \emph{two} parameters instead of one is an interesting possibility that we explored in this paper. To this end, we were motivated by the conjecture of~\cite{Bhattacharyya:2010yg}, which proposed that the missing parameter should be describing scalar hair (i.e. it should be the expectation value of the dual operator to the scalar field). If so, hairy black holes should exist and, in particular, the system should have a regular supersymmetric hairy black hole that (when we fix $J$) would start at the red square in Fig.~\ref{fig:entropyE} and extend upwards along the dashed black BPS curve. In the full $(E,Q,J)$ phase diagram, supersymmetric hairy black holes would be described by a surface sitting at the BPS surface and with a `lower' boundary described by the Gutowski-Reall curve. Ref.~\cite{Bhattacharyya:2010yg} supported its conjecture with a non-interacting thermodynamic model (that does not use the equations of motion of the theory besides when using information about the CLP black holes) whereby the hairy black hole is constructed as a mixture of a small CLP black hole that is placed on top of the supersymmetric soliton of the theory.  The distribution of energy, charge and angular momentum among the two components is such that they are in thermodynamic equilibrium and entropy is maximized. Interestingly, this model (or an effectively equivalent one that we discuss in Section~\ref{sec:toymodel}) indeed predicts that, in the BPS limit, hairy black holes should have $E\to E_\BPS$, $\mu\to 1$, $\O_HL\to 1$, $TL\to 0$, and {\it finite} entropy. The numerical data collected by~\cite{Markeviciute:2018cqs, Markeviciute:2018yal} up to temperatures as low as $TL\sim\CO(10^{-3})$ supported this conjecture of~\cite{Bhattacharyya:2010yg}.

However, there is a twist to this story. As described in section~\ref{sec:susy-analysis}, motivated by the above predictions, we attempted
to find the supersymmetric hairy black hole by solving directly the BPS equations but did not find evidence for the existence of such a solution. This fact forced us to revisit the previous perturbative and numerical results, and was the main motivation for the present paper. We started by questioning whether the non-interacting thermodynamic model of~\cite{Bhattacharyya:2010yg} (and of our Section~\ref{sec:toymodel}) is indeed a good one. If it is, it should produce the correct leading order thermodynamics of the system. To check if this is the case, in Section~\ref{sec:hbh-pert} we constructed hairy black holes within perturbation theory (with a matching asymptotic expansion) with a double expansion in  the horizon radius and scalar condensate amplitude (i.e. for small $E,Q,J$). We concluded that the leading order terms in the BPS limit of the perturbative thermodynamics \eqref{hbh_pert_E}--\eqref{hbh_pert_S} indeed recovers the thermodynamic model results~\eqref{MQJ_mix} of~\cite{Bhattacharyya:2010yg}. Not only does this seems to confirm the thermodynamic model but if further reinforces the associated conjecture of~\cite{Bhattacharyya:2010yg}. From this viewpoint, supersymmetric hairy black holes should exist and be regular with finite entropy.

This perturbative analysis does not solve the aforementioned puzzle. Therefore, in Section~\ref{sec:numerics}, we decided to find the exact numerical hairy black holes, much like in~~\cite{Markeviciute:2018cqs, Markeviciute:2018yal}. But this time we developed a numerical code to stretch the data collection to {\it much lower} temperatures; typically 4 orders of magnitude lower,  from $T L\sim \CO(10^{-3})$ down to $T L\sim \CO(10^{-7})$, say. This proved to be truly enlightening, as best illustrated in Figs.~\ref{fig:entropyE-NumPert1}--\ref{fig:entropyE-NumPert2}. Collecting data down to small temperatures, but not too small (say, `only' down to $TL\sim 10^{-3}$), it does seem that an extrapolation of the available numerical data to $T=0$ indeed yields a BPS hairy black hole with finite entropy, in agreement with~\cite{Markeviciute:2018cqs, Markeviciute:2018yal} and thus~\cite{Bhattacharyya:2010yg}. But this is misleading because if we stretch the code to collect data to even lower temperatures (say, down to $TL\sim 10^{-7}$), we find that the entropy starts falling dramatically in the very last stages of the $T\to 0$ approach, as unequivocally displayed in Figs.~\ref{fig:entropyE-NumPert1}--\ref{fig:entropyE-NumPert2}. With the full data displayed in Figs.~\ref{fig:entropyE-NumPert1}--\ref{fig:entropyE-NumPert2}, an extrapolation of it down to $T=0$ now finds that the BPS limit of the hairy black holes has, after all, {\it vanishing} entropy.  More precisely, in the supersymmetric limit, the hairy black hole family approaches
$E\to E_\BPS$, $\mu\to 1^+$, $\O_HL\to 1^-$, $T\to 0$, and $S\to 0$. Not less importantly, we further find that the supersymmetric hairy black holes are {\it singular} because their entropy vanishes and the value $\e_H$ of the scalar field at the horizon, and thus invariant curvatures at the horizon, do diverge as clearly shown in Fig.~\ref{fig:eHdiverge}. This proves that the conjecture of~\cite{Bhattacharyya:2010yg} does {\it not} hold and this is the main finding of our study. Our analysis assumes that the solution studied in this paper represents the relevant phase within the STU model at very low temperatures. However, it remains to be verified that our hairy black holes are \emph{stable} against potential condensations involving the remaining fields in the STU model, which for simplicity we have set to zero (e.g. $\varphi_r$, $\Phi_1-\Phi_2$, $\Phi_2-\Phi_3$ or $\Phi_1-\Phi_3$).

To have further support for our main finding, in Section~\ref{sec:susy-analysis} we looked for a supersymmetric hairy black hole by solving directly the BPS equations (more concretely, we analysed the near-horizon behaviour of the solutions). We concluded that supersymmetric hairy black holes can only exist if the scalar field at the horizon is not finite (thus, the pullback of the curvature to the horizon also diverges). That is to say, the analysis of the BPS equations is consistent with the main exact numerical results of Section~\ref{sec:NumericalResults}: when hairy black holes approach the BPS limit, they become {\it singular} because the scalar field and curvature invariants diverge at the horizon. Of course, the Gutowski-Reall black holes are regular supersymmetric solutions because the scalar field vanishes everywhere. 

It follows from the above findings that a scalar condensate observable (e.g. the expectation value of the dual operator to the scalar field) cannot be the missing gravitational parameter that would extend the radius of existence of supersymmetric black holes from the 1-parameter Gutowski-Reall curve to the 2-parameter BPS surface delimited by the previous curve (or, in Fig.~\ref{fig:phasediagramConc} where we fix $J$, from the red square to the whole dashed BPS curve above it).

The new numerical data collected down to extremely small values of the temperature further shows that the perturbative analysis of Section~\ref{sec:hbh-pert} fails very, very, close to the BPS surface, although it is a very good approximation (if not excellent) otherwise. Again, this is best illustrated in Fig.~\ref{fig:onsetNumPert}, Figs.~\ref{fig:entropyE-NumPert1}--\ref{fig:entropyE-NumPert2} and associated discussions. Since the leading order contribution of this perturbative analysis gives the non-interacting thermodynamic model of~\cite{Bhattacharyya:2010yg}/Section~\ref{sec:toymodel}, this also means that the latter fails to predict the correct properties of the BPS system. This failure so close to the BPS limit requires an explanation. As discussed in detail in Section~\ref{sec:PerturbativeFailing}, it indeed turns out that the perturbative analysis (and associated non-interacting limit) breaks down for extremely small temperatures although it is valid otherwise (as long as we further work at small $(E,Q,J)$). We should emphasize that this failure of the perturbative analysis, and associated non-interacting thermodynamic model, very close to the BPS limit (or extremal limit where the temperature vanishes) is surprising in the sense that, in other systems where these methods can be employed~\cite{Basu:2010uz,Bhattacharyya:2010yg,Dias:2011tj,Dias:2016pma,Dias:2011at,Stotyn:2011ns,Dias:2013sdc,Cardoso:2013pza,Dias:2015rxy,Dias:2022eyq}, they proved to give very good approximations even close to extremality.

An inspection of Fig.~\ref{fig:phasediagramConc} raises one question that has not yet been addressed. As expected from the gauge/gravity correspondence of the system, we have concluded that hairy black holes fill the region that extends from the blue merger line all the way up to the dashed BPS curve (where strictly speaking there is no regular hairy black hole). This includes the light-red region $I$ where CLP black holes do not exist. What about in the white region $II$ (around the bottom-left `corner') between the dark-green 
extremal CLP curve and the dashed BPS curve? From the AdS$_5$/CFT$_4$ correspondence, there should exist black holes also in this white region $II$ because, from the SYM perspective, there is no reason not to have thermal states with these $(E,Q,J)$. But these cannot be the hairy black holes constructed in~\cite{Markeviciute:2018cqs, Markeviciute:2018yal} and in the current paper. So what could they be? We conjecture that the following occurs. In Fig.~\ref{fig:phasediagramConc}, as best seen in its right panel where we do a zoom in of white region $II$, we also display a light-green curve that starts at the Gutowski-Reall red square and extends down to $Q=0$ (in a full phase diagram $(E,Q,J)$, where we do not fix $J$, this would be a surface). This curve describes CLP black holes with $\O_H L=1$, i.e. it can be obtained from the CLP thermodynamics~\eqref{BH_charges}--\eqref{BH_thermo} by setting $\O_H L=1$. CLP black holes to the left (right) of this curve have $\O_H L>1$ ($\O_H L<1$). On general grounds, spinning black holes with $\O_H L>1$ should be unstable to superradiant instabilities~\cite{Hawking:1999dp, Kunduri:2006qa, Dias:2011at, Stotyn:2011ns, Dias:2013sdc, Cardoso:2013pza, Dias:2015rxy, Green:2015kur, Ishii:2018oms, Ishii:2020muv}. Actually, we already stated that the CLP black holes of our theory are to be seen as the Kerr-Newmann--AdS$_5$ black holes of the theory. Thus, although superradiant instabilities of CLP where never studied, it seems safe to borrow knowledge from  superradiant studies of Kerr-AdS in five or even four dimensions available in the literature~\cite{Hawking:1999dp, Cardoso:2004hs, Kunduri:2006qa, Dias:2011at, Dias:2013sdc, Cardoso:2013pza, Dias:2015rxy, Green:2015kur}. We will do so in what follows.

In Section~\ref{sec:Onset3Q}, we studied CLP instabilities that, roughly, one can say that have a `charged' nature. In the sense that the associated scalar condensation instability emerges because the CLP black holes above the blue merger line in Fig.~\ref{fig:phasediagramConc} have chemical $\mu>1$ and the near-horizon effective mass of the scalar field violates~\cite{Gubser:2008px, Dias:2010ma, Basu:2010uz, Bhattacharyya:2010yg, Dias:2011tj, Dias:2022eyq} the 2-dimensional Breitlohner-Freedmann  bound~\cite{Breitenlohner:1982bm, Breitenlohner:1982jf}. Accordingly, such instabilities `first appear' in the $m=0$ sector of perturbations, where $m$ is the azimuthal quantum number, because the presence of spin is not particularly relevant. On the other hand, to study superradiant instabilities of rotational nature (in black holes with $\O_H L>1$) one must consider perturbations of the type $\Phi(t,r,\psi,\theta,\phi)\sim e^{-i\o t}e^{im\psi}\Phi(r)Y(\theta,\phi)$, this time, with integer $m\geq 1$. Borrowing lessons from~\cite{Dias:2011at,Dias:2013sdc,Cardoso:2013pza,Dias:2015rxy}, we can expect that the onset curve of the $m=1$ superradiant instability will look like the magenta dashed curve {\it sketched} in the right panel of Fig.~\ref{fig:phasediagramConc}: it is to the left of the light-green $\O_HL=1$ curve and starts at the red square. CLP black holes to the left (right) of this $m=1$ onset curve are unstable (stable) to $m=1$ modes. We then expect (but not necessarily for the low $m$-modes)  the onset curves with $m\geq 2$ to be in-between the magenta and light-green curves with increasingly higher $m$ onset curves closer to the $\O_H L=1$ curve (which should be reached from the left in the strict $m\to \infty$ limit); see e.g. Fig.~3 and 6 of~\cite{Cardoso:2013pza} to have an idea of how the onset curves might look like. From the general relativity studies~\cite{Dias:2011at, Stotyn:2011ns, Dias:2015rxy, Ishii:2018oms, Ishii:2020muv}, one further expects that the $m$-onset curves should be merger lines between the CLP family and a new family of hairy black holes, with winding number $m$, that were coined `{\it black resonators}' in~\cite{Dias:2015rxy}. Such (hairy) black resonators distinguish from the hairy black holes in the light-red region $I$ on a key aspect.  Unlike on the hairy black holes constructed in the present paper, for the black resonators $\partial_t$ and $\partial_\psi$ are {\it not} Killing vector fields, but the linear combination $\partial_t+\O_H \partial_\psi$ that generates the horizon is a Killing vector field. Thus, the black resonators are not time-independent neither axisymmetric but they are time periodic. For values of $(E,Q,J)$ where they co-exist, black resonators should have higher entropy than CLP black holes, i.e. they should dominate the microcanonical ensemble. We conjecture that such (hairy) black resonators (for each $m$) will start at the $m$-merger curve (e.g. the dashed magenta curve for $m=1$ sketched in Fig.~\ref{fig:phasediagramConc}) and extend to its left all the way to the BPS black dashed curve (this time, below the red square). Thus, there is a region (till the extremal CLP dark-green curve) where they are conjectured to co-exist with CLP black holes but they are also conjectured to fill the white region $II$ in Fig.~\ref{fig:phasediagramConc}. We further conjecture that black resonators approach the BPS curve with $E\to E_\BPS$, $\mu\to 1^-$, $\O_H L\to 1^+$, and $TL\to 0$. Without performing the actual computation, it remains unclear whether this BPS will have finite entropy or whether this limit will be regular. Note that hairy black holes in region $I$ reach the BPS limit with $\mu=1$ from above and $\O_H L=1$ from below, while the black resonators of region $II$ should reach the BPS limit with $\mu=1$ from below and $\O_H L=1$ from above.

In spite of our best efforts, our findings and conjectures have not identified the gravitational missing parameter that would describe regular supersymmetric black holes that extend the 1-parameter family of Gutowski-Reall black holes to the whole 2-dimensional BPS surface in the 3-dimensional parameter space $(E,Q,J)$. What could this missing parameter be, or what could these regular supersymmetric black holes be? This question has, in fact, been addressed in recent work \cite{Kim:2023sig, Choi:2024xnv} and we briefly comment on these recent results and contrast them with our current explorations.

In our work, we studied the instability of the CLP black hole to the perturbations of the charged scalar $\Phi=\Phi_{1,2,3}$ (Section~\ref{sec:Onset3Q}) and found that for some values of $E$, $Q$, and $J$, the CLP black hole is unstable and the scalar field condenses. The ``endpoint'' of this condensation process is the hairy black hole that we constructed in the present paper. When this hairy black hole solution is uplifted to 10D Type IIB supergravity (see footnote~\ref{foot:uplift}), the charged scalar appears as a Kaluza-Klein (KK) mode on the $S^5$. Of course, since $\Phi$ is not the only KK mode on the $S^5$, it reasonable to ask whether a similar instability can be seen for some of the other KK modes of the theory. This is precisely what was studied in \cite{Kim:2023sig, Choi:2024xnv}. However, instead of analyzing the linearized instability in AdS by solving the bulk linearized EoM\footnote{In \cite{Ezroura:2024xba}, the spectrum of CLP black holes was analyzed from a bulk perspective, revealing that not all instabilities can be attributed to those arising from simple charged scalar fields with specific mass and charge.} (as we have done in section~\ref{sec:Onset3Q}), \cite{Kim:2023sig, Choi:2024xnv} studies the instability on the SYM side of the AdS/CFT correspondence. To understand their results, we start by noting that the charged scalar $\Phi$ of this theory is dual to the operator $\sum_{i=1}^6 \Tr[(X^i)^2]$ in SYM, where $X^i$ are the six scalar fields of SYM. At leading order in large $N$, this operator has $\D=2$. More general KK modes are dual to $\frac{1}{2}$-BPS operators of the form $\CO_{n,\ell,m} = (\p^2)^n \p_{\mu_1} \cdots \p_{\mu_\ell} \Tr[X^{(i_1} \cdots X^{i_m)}]$ which, at leading order in large $N$ has $\D = 2n+\ell+m$.\footnote{All the fields of $\CN=8$ gauged supergravity are dual to the operators with $m=2$.} The condensation instability is studied by analyzing the contribution of such operators to the SYM index. It was shown that CLP black holes are afflicted by two types of instabilities.

When $\O_H L > 1$ (left of the green curve in Fig.~\ref{fig:phasediagramConc}), the modes with $\ell \sim N$ (recall that $N$ is large) are unstable. The bulk solution dual of this operator is a quantum gas of rotating gravitons that lives in a disk of radius $r \sim \sqrt{N} L$, has charges $E,J = \CO(N^2)$, and has entropy $S = \CO(N)$. Note that this graviton gas is supersymmetric. As in the present paper, the ``endpoint'' of this instability is expected to be a `hairy' black hole solution, which \cite{Kim:2023sig, Choi:2024xnv} referred to as a \emph{Grey Galaxy}.\footnote{We argued that the region to the left of the green curve in Fig.~\ref{fig:phasediagramConc}, in particular the white region $II$, is filled with several $m$-mode black resonators: for each $m$ there is a black resonator and they should all be singular in the BPS limit. On the other hand, the grey galaxy is a mixture of gravitons with several $m$ that is regular at the BPS surface.}
 It comprises a CLP black hole surrounded by the graviton gas. At large $N$, the gas and the black hole are widely separated and can be treated as a non-interacting mix similar to what we did in section~\ref{sec:toymodel}. However, unlike our construction in section~\ref{sec:toymodel}, this is \emph{not} a toy model for the grey galaxy, but an exact solution!\footnote{We emphasize that this happens because the graviton gas is extremely far away from the centre of the CLP black hole, so its backreaction is subleading in $\sqrt{N}$. For the hairy black holes studied in the present paper, the scalar hair is ``large'' throughout the entire spacetime and its backreaction onto the CLP black hole is not subleading. In fact, in the BPS limit, the scalar field diverges on the horizon, and its backreaction is so strong that it destroys the black hole completely, and the limiting solution has zero entropy. This sort of intricate feature is entirely missing for the grey galaxies.} These grey galaxies start existing to the left of the green curve in Fig.~\ref{fig:phasediagramConc} and completely fill up region $II$. Moreover, they reach the BPS curve, and the limiting supersymmetric grey galaxy is a Gutwoski-Reall black hole surrounded by the graviton gas.

When $\mu > 1$ (left of the blue curve in Fig.~\ref{fig:phasediagramConc}), the modes with $m \sim N$ are unstable. The bulk dual of these operators is the well-known dual giant graviton solution \cite{Grisaru:2000zn, McGreevy:2000cw}, which is a special case of the more general LLM solution \cite{Lin:2004nb}. The thermodynamics of the dual giant graviton is similar to the graviton gas discussed in the previous paragraph in that they have a radius $r \sim \sqrt{N} L$, charges $E,Q = \CO(N^2)$, and entropy $S = \CO(N)$. The dual giants are supersymmetric. The endpoint of the instability is a new solution, which \cite{Kim:2023sig, Choi:2024xnv} refers to as the \emph{Dual Dressed Black Holes} (DDBH). As before, because the dual giants and the CLP black hole at the center are widely separated, their thermodynamic properties are reproduced via a non-interacting model similar to section \ref{sec:toymodel}, i.e., they can be treated as a non-interacting mix. However, unlike the quantum graviton gas of the previous paragraph, the dual giant gravitons are classical and they backreact non-trivially onto the CLP black hole. The full backreacted solution, if it exists, should interpolate between the LLM geometry at large distances and the CLP black holes in the interior. Despite this backreaction, one expects (precisely because the dual giants are so far away from the black hole) that the non-interacting thermodynamic model reproduces the exact thermodynamics of the full DDBH solution. The DDBHs start existing to the left of the blue curve in Fig.~\ref{fig:phasediagramConc} and fill up region $I$. As before, they reach down to the BPS curve, and the limiting supersymmetric DDBH is the GR black hole surrounded by one, two, or three dual giant gravitons (depending on how many of the chemical potentials $\mu_K$ are greater than 1). The hairy black hole constructed in this paper have $\mu = \mu_{1,2,3} > 1$, so they are unstable to the nucleation of three dual giant gravitons.

Even with the new solutions constructed in \cite{Kim:2023sig, Choi:2024xnv}, there are still vast portions of the phase space for which we have no known $\CO(N^2)$ entropy growth, and one must continue to search for more general solutions to achieve a full understanding of the phase space of black hole solutions in AdS/CFT.\footnote{For instance, for $J_1=J_2=0$, there are no known solutions with $\CO(N^2)$ entropy and the methods of \cite{Kim:2023sig, Choi:2024xnv} do not allow us construct new solutions either. A similar argument also extends to the case $J_1 \neq J_2$, where there are whole finite-measure regions in phase space where solutions are yet to be constructed.} This issue has also been recently addressed in \cite{Armas:2024dtq} which searched for more general black brane solutions with $\ads_m \times S^n$ asymptotics for $(m,n)=(5,5), (4,7),(7,4)$ using the blackfold approach.\footnote{Solutions with $(m,n)=(5,5)$ are embedded in Type IIB supergravity whereas those with $(m,n)=(7,4)$ and $(4,7)$ are embedded in $M$-theory.} The basic idea here is to approximate near-horizon solutions by long-wavelength deformations of a black $p$-brane solution, which can be constructed in a hydrodynamic expansion. The background $p$-brane solution is either known exactly or can be constructed in a perturbative expansion in $r_+/L$ where $r_+$ is the horizon size of the $p$-brane. The deformations of the $p$-brane is constructed in a perturbative expansion in $r_+/\ell$, where $\ell$ is the characteristic wavelength of the deformation. The overall solution is then constructed in a double perturbative expansion in the above-mentioned parameters. The far-field solution is constructed by solving the supergravity EoM, just like we have done here. The full solution is then constructed by the method of matched asymptotics. All in all, the complete solution is obtained in a double-perturbative expansion in the size of the black $p$-brane and the wavelength of its deformations. The qualitative structure of the solution is rather similar to the perturbative solution that we have constructed in this paper, and it seems very likely that our hairy black hole solutions can be understood in this blackfold language. We leave an exploration of this for future work.

%%%%%%%%%%%%%%%%%%%%%%%%%%%%%%%%%%%%%%%%%%%%%%%%%%%
\section*{Acknowledgements}
%%%%%%%%%%%%%%%%%%%%%%%%%%%%%%%%%%%%%%%%%%%%%%%%%%%

We would like to thank Diksha Jain and Shiraz Minwalla for many useful conversations. The authors acknowledge the use of the IRIDIS High Performance Computing Facility, and associated support services at the University of Southampton, in the completion of this work. O.C.D. acknowledges financial support from the STFC “Particle Physics Grants Panel (PPGP) 2020” Grant No. ST/X000583/1. P.M. and J.E.S.'s work have been partially supported by STFC consolidated grant ST/T000694/1. P.M. is supported by the European Research Council (ERC) under the European Union’s Horizon 2020 research and innovation programme (grant agreement No 852386).

%%%%%%%%%%%%%%%%%%%%%%%%%%%%%%%%%%%%%%%%%%%%%%%%%%%
%%%%%%%%%%%%%%%%%%%%%%%%%%%%%%%%%%%%%%%%%%%%%%%%%%%
\appendix
%%%%%%%%%%%%%%%%%%%%%%%%%%%%%%%%%%%%%%%%%%%%%%%%%%%
%%%%%%%%%%%%%%%%%%%%%%%%%%%%%%%%%%%%%%%%%%%%%%%%%%%

%%%%%%%%%%%%%%%%%%%%%%
\section{Differential Equations in Perturbation Theory}
\label{app:perturbative_eq}
%%%%%%%%%%%%%%%%%%%%%%

In this appendix, we list the differential equations obtained at each order in $\e$ and $y_+$ in the far-field, intermediate-field and near-field region. We set $L=1$ in this appendix. All functions are expanded in powers of $\e$ as
\begin{equation}
\begin{split}\label{main-exp}
f(r) &=  \sum_{n=0}^\infty \e^{2n} f_{(2n)}(r) , \qquad g(r) =  \sum_{n=0}^\infty \e^{2n} g_{(2n)}(r) , \qquad h(r) = \sum_{n=0}^\infty \e^{2n} h_{(2n)}(r) , \\
w(r) &=  \sum_{n=0}^\infty \e^{2n} w_{(2n)}(r) , \qquad A(r) = \sum_{n=0}^\infty \e^{2n} A_{(2n)}(r) , \qquad B(r) = \sum_{n=0}^\infty \e^{2n} B_{(2n)}(r) , \\
\Phi(r) &= \sum_{n=0}^\infty \e^{2n+1} \Phi_{(2n+1)}(r) .
\end{split}
\end{equation}

\subsection{Far-Field Equations}
\label{app:far_field_eq}

In the far-field region, we expand all the functions as
\begin{equation}
\begin{split}\label{main-exp-far}
f_{(2n)}(r) &= \sum_{k=0}^\infty y_+^{2k} f^\far_{(2n,2k)}(r)  , \qquad g_{(2n)}(r) = \sum_{k=0}^\infty y_+^{2k} g^\far_{(2n,2k)}(r)  , \\
h_{(2n)}(r) &=  \sum_{k=0}^\infty y_+^{2k} h^\far_{(2n,2k)}(r)  , \qquad w_{(2n)}(r) = \sum_{k=0}^\infty y_+^{2k} w^\far_{(2n,2k)}(r)  , \\
A_{(2n)}(r) &= \sum_{k=0}^\infty y_+^{2k} A^\far_{(2n,2k)}(r)  , \qquad B_{(2n)}(r) =  \sum_{k=0}^\infty y_+^{2k} B^\far_{(2n,2k)}(r)  , \\
\Phi_{(2n-1)}(r) &= \sum_{k=0}^\infty y_+^{2k} \Phi^\far_{(2n-1,2k)}(r)  . 
\end{split}
\end{equation}
To bring the equations into a simple form, we write
\begin{equation}
\begin{split}
g^\far_{(2n,2k)}(r) &= X^\far_{(2n,2k)}(r) +  \frac{1}{3} \left( r [ h^\far_{(2n,2k)}(r)]' + h^\far_{(2n,2k)}(r) \right) , \\
h^\far_{(2n,2k)}(r) &= \frac{1}{r^2} \left( Y^\far_{(2n,2k)}(r) + 3 f^\far_{(2n,2k)}(r) \right) .
\end{split}
\end{equation}
The equations at each order take the form
\begin{equation}
\begin{split}\label{odd-eq}
P_{(2n-1,2k)}^{\far,\Phi}(r) &= \left( \frac{r^3}{1+r^2}  [ (1+r^2) \Phi^\far_{(2n-1,2k)}(r) ]' \right)' , \\
P_{(2n,2k)}^{\far,A}(r) &= \left( r^3 [ A^\far_{(2n,2k)}(r) ]' \right)' , \\
P_{(2n,2k)}^{\far,B}(r) &=  \left( \frac{1+r^2}{r^3} [ r^2 B^\far_{(2n,2k)}(r) ]' \right)' , \\
P_{(2n,2k)}^{\far,w}(r) &= \left( r^5 [ w^\far_{(2n,2k)}(r)]' \right)' , \\
P_{(2n,2k)}^{\far,X}(r) &= [X^\far_{(2n,2k)}(r)]' , \\
P_{(2n,2k)}^{\far,Y}(r) &=  [ r^2 Y^\far_{(2n,2k)}(r) ]' , \\
P_{(2n,2k)}^{\far,f}(r) &= \left( \frac{1+r^2}{r^5} [ r^2 f^\far_{(2n,2k)}(r) ]' \right)' .
\end{split}
\end{equation}
The sources $P$ appearing on the LHS of these equations are all fixed by the solution at lower order in perturbation theory, and consequently can be treated as known functions. Each of these equations can easily be solved by integrating the sources. The integration constants are fixed by the boundary conditions \eqref{asymp_exp} and by matching with the intermediate-field solution, as described in~\cite{Bhattacharyya:2010yg, Dias:2022eyq}.

\subsection{Intermediate-Field Equations}
\label{app:int_field_eq}

In the intermediate-field region, we expand all functions as
\begin{equation}
\begin{split}\label{main-exp-int}
f_{(2n)}(r_+y) &= \sum_{k=0}^\infty y_+^{2k} f^\intt_{(2n,2k)}(y)  , \qquad g_{(2n)}(r_+y) = \sum_{k=0}^\infty y_+^{2k} g^\intt_{(2n,2k)}(y)  , \\
h_{(2n)}(r_+y) &= \sum_{k=0}^\infty y_+^{2k} h^\intt_{(2n,2k)}(y)  , \qquad w_{(2n)}(r_+y) = \sum_{k=0}^\infty y_+^{2k} w^\intt_{(2n,2k)}(y)  , \\
A_{(2n)}(r_+y) &= \sum_{k=0}^\infty y_+^{2k} A^\intt_{(2n,2k)}(y)  , \qquad B_{(2n)}(r_+y) = y_+^2 \sum_{k=0}^\infty y_+^{2k} B^\intt_{(2n,2k)}(y)  , \\
\Phi_{(2n+1)}(r_+y) &= \sum_{k=0}^\infty y_+^{2k} \Phi^\intt_{(2n+1,2k)}(y) . 
\end{split}
\end{equation}
To bring the equations into a simpler form, we write
\begin{equation}
\begin{split}
g^\intt_{(2n,2k)}(y) &= X^\intt_{(2n,2k)}(y) + \frac{1}{3} [ y h^\intt_{(2n,2k)}(y) ]'  + \frac{1}{2y} [ y^2 Y_{(2n,2k)}(y) ]' , \\
f^\intt_{(2n,2k)}(y) &= Y_{(2n,2k)}^\intt(y)  + \frac{2}{3y^2} [ h_{(2n,2k)}^\intt(y) - 3 A_{(2n,2k)}^\intt(y) ] , \\
B_{(2n,2k)}^\intt(y) &= \sqrt{1-\g} Z_{(2n,2k)}^\intt(y) + \frac{1}{6} y^5 [ w_{(2n,2k)}^\intt (y) ]' . 
\end{split}
\end{equation}
The equations at each order take the form
\begin{equation}
\begin{split}
\label{hbh_equations_int}
P_{(2n-1,2k)}^{\intt,\Phi}(y)  &= \left( \frac{( y^2 - 1 )^2}{y} [ \Phi^\intt_{(2n-1,2k)}(y) ] ' \right)' ,  \\
P_{(2n,2k)}^{\intt,X}(y) &= X^\intt_{(2n,2k)}(y) , \\
P_{(2n,2k)}^{\intt,Y}(y) &= [ y^3 [ Y^\intt_{(2n,2k)}(y) ]' ]' ,  \\
P_{(2n,2k)}^{\intt,h}(y) &= \left( \frac{[ (y^2-1)^2 h_{(2n,2k)}^\intt(y) ]' }{y(y^2-1)^2} \right)' , \\
P_{(2n,2k)}^{\intt,A}(y) &= [ y^3 [ A_{(2n,2k)}^\intt(y) ]' ]' , \\
P_{(2n,2k)}^{\intt,Z}(y) &= [ Z^\intt_{(2n,2k)}(y) ]' , \\
P_{(2n,2k)}^{\intt,w}(y) &= \left( \frac{(y^2-1)^6}{y^7} \left( \frac{y^7 [ w_{(2n,2k)}^\intt(y) ]'  }{(y^2-1)^2}  \right)' \right)' .
\end{split}
\end{equation}
The integration constants here are fixed by matching with the far-field solution and the near-field solution.

\subsection{Near-Field Equations}
\label{app:near_field_eq}

In the near-field region, we expand all functions as
\begin{equation}
\begin{split}\label{main-exp-near}
f_{(2n)}\left(r_+ + \frac{z r_+^3}{L^2}\right) &= y_+^4 \sum_{k=0}^\infty y_+^{2k} f^\near_{(2n,2k)}(z)  , \\
g_{(2n)}\left(r_+ + \frac{z r_+^3}{L^2}\right) &= \sum_{k=0}^\infty y_+^{2k} g^\near_{(2n,2k)}(z)  , \\
h_{(2n)}\left(r_+ + \frac{z r_+^3}{L^2}\right) &= \sum_{k=0}^\infty y_+^{2k} h^\near_{(2n,2k)}(z)  , \\
w_{(2n)}\left(r_+ + \frac{z r_+^3}{L^2}\right) &= y_+^2 \sum_{k=0}^\infty y_+^{2k} w^\near_{(2n,2k)}(z)  , \\
A_{(2n)}\left(r_+ + \frac{z r_+^3}{L^2}\right) &= y_+^2 \sum_{k=0}^\infty y_+^{2k} A^\near_{(2n,2k)}(z)  , \\
B_{(2n)}\left(r_+ + \frac{z r_+^3}{L^2}\right) &= y_+^2 \sum_{k=0}^\infty y_+^{2k} B^\near_{(2n,2k)}(z)  , \\
\Phi_{(2n+1)}\left(r_+ + \frac{z r_+^3}{L^2}\right) &= \sum_{k=0}^\infty y_+^{2k} \Phi^\near_{(2n+1,2k)}(z)  . 
\end{split}
\end{equation}
To bring the equations into a simpler form, we write
\begin{equation}
\begin{split}
g^\near_{(2n,2k)}(z) &= X^\near_{(2n,2k)}(z)  + [A^\near_{(2n,2k)}(y)]' + \frac{1}{3} \left( h^\near_{(2n,2k)}(z) + ( 2 z + \g ) [ h^\near_{(2n,2k)}(z) ]' \right) \\
&\qquad \qquad + \frac{3+4z+5\g}{(4(2z+\g)^2} f^\near_{2n,2k-2}(z) - \frac{3-4z+\g}{8(2z+\g)} [ f^\near_{2n,2k-2}(z)]' , \\
A^\near_{(2n,2k)}(z) &= Y^\near_{(2n,2k)}(z) + \frac{1}{2(2z+\g)} f^\near_{(2n,2k)}(z) , \\
B^\near_{(2n,2k)}(z) &= \sqrt{1-\g} \left( Z^\near_{(2n,2k)}(z) + \frac{2}{3} h^\near_{(2n,2k)}(z) - \frac{2 z + \g}{12} [ h^\near_{(2n,2k)}(z)]' \right. \\
&\left. \qquad \qquad - \frac{1}{4} X^\near_{(2n,2k)}(z)  - \frac{1}{4} [ Y^\near_{(2n,2k)}(z) ]' + \frac{1}{6} [ W^\near_{(2n,2k)}(z)]'  \right. \\
&\left. \qquad \qquad - \frac{3+48z+13\g}{48(2z+\g)^2} f^\near_{(2n,2k-2)}(z) + \frac{3+48z+\g}{96(2z+\g)} [ f^\near_{(2n,2k-2)}(z)]' \right) , \\
w^\near_{(2n,2k)}(z) &= \sqrt{1-\g} \left( W^\near_{(2n,2k)}(z) + \frac{3}{4(2z+\g)} f^\near_{(2n,2k)}(z) \right) .
\end{split}
\end{equation}
The equations at each order take the form
\begin{equation}
\begin{split}
\label{hbh_equations_near}
P_{(2n-1,2k)}^{\near,\Phi}(z) &= [ z (z+\g) [ \Phi^\near_{(2n-1,2k)}(z) ] ' ]' ,  \\
P_{(2n,2k)}^{\near,X}(z) &= X^\near_{(2n,2k)}(z)  , \\
P_{(2n,2k)}^{\near,h}(z) &= \left( z ( z + \g ) ( 2 z + \g )^2 \left( \frac{h^\near_{(2n,2k)}(z)}{2z+\g} \right)' \right)' , \\
P_{(2n,2k)}^{\near,Z}(z) &= [ Z^\near_{(2n,2k)}(z) ]'  , \\
P_{(2n,2k-2)}^{\near,f}(z) &= \left( \frac{f^\near_{(2n,2k-2)}(z)}{ 2z+\g } \right)'' , \\
P_{(2n,2k)}^{\near,Y}(z) &= [ (2z+\g)^2 [ Y^\near_{(2n,2k)}(z) ]' ]'  , \\
P_{(2n,2k)}^{\near,W}(z) &= \left( z ( z + \g ) ( 6 z^2 + 6 z \g + \g^2 )^2 \left( \frac{[ W^\near_{(2n,2k)}(z) ]'}{ 6 z^2 + 6 z \g + \g^2  }  \right)' \right)' , \\
\end{split}
\end{equation}
The integration constants here are fixed by requiring regularity on the horizon $z=0$ and matching with the intermediate-field solution.

Note that the near-field equations mixes the functions appearing at $\CO(y_+^{2k})$ and those appearing at $\CO(y_+^{2k-2})$. This is due to the residual coordinate freedom in the near-field perturbative expansion, as noted in~\cite{Bhattacharyya:2010yg}.

\section{Proof of Theorem}
\label{app:theorem_proof}

Let $f:\mrr_+ \to \mrr$ and $g:\mrr_+ \to \mrr$ be monotonic functions in an open neighbourhood of $x=0$. Then,
\begin{equation}
\begin{split}
f'(x) = o(g'(x)) \quad \implies \quad f(x) = \begin{cases}
C + o(g(x)-g(0)) &\text{if $g$ is bounded.} \\
o(g(x)) & \text{if $g$ is unbounded.}
\end{cases}
\end{split}
\end{equation}

\paragraph{Proof}

Without loss of generality, we assume that $g$ is an increasing function for all $x \in (0,\ve)$ for some $\ve > 0$ (if it is a decreasing function, we can replace $g(x) \to - g(x)$ everywhere). We then have $g'(x) > 0$ for all $x \in (0,\ve)$. $f'(x)= o(g'(x))$ means that for all $c>0$, there exists a $\d_c \in (0,\ve)$ such that
\begin{equation}
\begin{split}\label{prop_1}
|f'(x)| < c g'(x) \quad \forall  \quad x \in (0,\d_c) .
\end{split}
\end{equation}
The main result we will use is that for all $0 \leq x_1 \leq x_2 \leq \d_c$, we have
\begin{equation}
\begin{split}\label{prop_2}
|f(x_2) - f(x_1)| &= \left| \int_{x_1}^{x_2}  \dt t f'(t) \right| \leq \int_{x_1}^{x_2} \dt t | f'(t) |  = c \int_{x_1}^{x_2}  \dt t g'(t) < c [ g(x_2) - g(x_1) ] .
\end{split}
\end{equation}
We consider three separate cases. 
\begin{enumerate}[leftmargin=*]
\item $g,g'=\CO(1)$ as $x \to 0$: In this case, $g'$ is bounded, so $f'$ is also bounded due to \eqref{prop_1}. Thus, $\lim_{x\to0} f$ exists, and we define this as $f(0)$. Then setting $x_1=0$ and $x_2=x$ in \eqref{prop_2}, we find
\begin{equation}
\begin{split}
|f(x)-f(0)| < c | g(x) - g(0) | \quad \implies \quad f(x) = f(0) + o ( g(x) - g(0) ) . 
\end{split}
\end{equation}
\item $g = \CO(1)$ and $g' \to \infty$ as $x \to 0$: In this case, since $g$ is bounded, \eqref{prop_2} implies that $f$ is also bounded. Since it is monotonic, it has a well-defined limit of $0$ and the proof of Case I trivially extends here. Note that this is the only case where the requirement that $f$ is monotonic is necessary.

\item $g\to-\infty$ and $g' \to \infty$ as $x \to 0$: In this case, we take $x_2=\d$ and $x_1=x$ in \eqref{prop_2} and use one extra step 
\begin{equation}
\begin{split}
|f(x)| \leq |f(x)-f(\d)| + |f(\d)| < c g(\d) - c g(x) + |f(\d)| . 
\end{split}
\end{equation}
Taking $x$ to be sufficiently small so that $g(x) < 0$ and $c g(\d) + |f(\d)|<c |g(x)|$, we find
\begin{equation}
\begin{split}
|f(x)| <  2c | g(x) | \quad \implies \quad f(x) = o(g(x)) . 
\end{split}
\end{equation}
\end{enumerate}
This completes the proof.

%%%%%%%%%%%%%%%%%%%%%%%%%%%%%%%%%%%%%%%%%%%%%%%%%%%%%%
\bibliography{refsSO6sugra}{}
\bibliographystyle{JHEP}

\end{document}